\documentclass[journal]{IEEEtran}

\usepackage{ifpdf}

\usepackage{cite}

\ifCLASSINFOpdf
  \usepackage[pdftex]{graphicx}
\else
  \usepackage[dvips]{graphicx}
\fi

\usepackage{amsmath}

\usepackage{algorithmic}

\usepackage{array}

\ifCLASSOPTIONcompsoc
  \usepackage[caption=false,font=normalsize,labelfont=sf,textfont=sf]{subfig}
\else
\usepackage[caption=false,font=footnotesize]{subfig}
\fi
\usepackage{nomencl}
\makenomenclature
\usepackage{stfloats}

\ifCLASSOPTIONcaptionsoff
    \usepackage[nomarkers]{endfloat}
    \let\MYoriglatexcaption\caption
 \renewcommand{\caption}[2][\relax]{\MYoriglatexcaption[#2]{#2}}
\fi

\usepackage{amssymb}
\usepackage{url}
\usepackage{tikz}
\usetikzlibrary{arrows,decorations,backgrounds,shadows,plotmarks, positioning, calc,shapes, patterns,chains,intersections,external,decorations.pathreplacing,calligraphy}
\usepackage[american,europeaninductors,smartlabels,EFvoltages,europeanresistors]{circuitikz}
 \usepackage{xcolor}
	\definecolor{UCLgreenD}{rgb}{0.333,0.314,0.145}    
	\definecolor{UCLgreenM}{rgb}{0.561,0.600,0.243}    
	\definecolor{UCLgreenB}{rgb}{0.710,0.741,0.000}    
	\definecolor{UCLgreenL}{rgb}{0.733,0.773,0.573}    
	\definecolor{UCLredD}{rgb}{0.396,0.114,0.196}    
	\definecolor{UCLredM}{rgb}{0.576,0.153,0.173}    
	\definecolor{UCLredB}{rgb}{0.835,0.000,0.196}    
	\definecolor{UCLredL}{rgb}{0.878,0.235,0.192}    
	\definecolor{UCLpurpleD}{rgb}{0.294,0.220,0.298}       
	\definecolor{UCLpurpleM}{rgb}{0.314,0.027,0.471}       
	\definecolor{UCLpurpleB}{rgb}{0.675,0.078,0.353}       
	\definecolor{UCLpurpleL}{rgb}{0.7764,0.690,0.737}      
	\definecolor{UCLblueD}{RGB}{0,61,76}        
	\definecolor{UCLblueM}{RGB}{0,40,85}        
	\definecolor{UCLblueB}{RGB}{0,151,169}      
	\definecolor{UCLblueL}{RGB}{141,185,202}    
	\definecolor{UCLblueC}{RGB}{164,219,232}    
	\definecolor{UCLblueIOE}{RGB}{50,85,164}    
	\definecolor{UCLyellow}{RGB}{246,190,0}     
	\definecolor{UCLorange}{RGB}{234,118,0}   
	\definecolor{UCLgrey}{RGB}{140,130,121}   
	\definecolor{UCLbrown}{RGB}{78,54,41}  
	\definecolor{UCLGRAY}{RGB}{140, 130, 121}
\definecolor{PCBgreen}{RGB}{0,45,4}
\definecolor{UCLGreenD}{RGB}{85, 80, 37}
\definecolor{UCLGreenM}{RGB}{143, 153, 62}
\definecolor{UCLGreenB}{RGB}{181, 189, 0}
\definecolor{UCLGreenL}{RGB}{187, 197, 146}
\makeatletter
\newcommand*{\rom}[1]{\expandafter\@slowromancap\romannumeral #1@}
\makeatother
\begin{document}
%
\title{An Isolated Gate Driver for Multi-Active Bridges with Soft Switching}
\author{
Ferdinand Grimm, Pouya Kolahian, John Wood, Richard Bucknall and Mehdi Baghdadi
}

\markboth{
}%
{Shell \MakeLowercase{\textit{et al.}}: Bare Demo of IEEEtran.cls for IEEE Journals}

\maketitle

\begin{abstract}
The design of gate drivers is an important topic in power converter topologies that can help reduce switching losses and increase power density.
Gate driving techniques that offer zero-voltage switching and/or zero current switching have recently been successfully proposed for different modular multilevel converters such as the cascaded H bridge. 
Previous papers on other multilevel converters such as the multi-active bridge, however, do not sufficiently assess the topics of gate driver design for this topology.
This work presents a novel isolated gate driver architecture tailored to the multi-active bridge topology. 
Zero voltage switching is then achieved using two multi-winding transformers. 
The advantages of the proposed topology are not only a reduction of switching losses but also reduced component count.
 The topology is evaluated on a prototype using experimental results. 
 It was shown using simulation and experiments that the proposed topology has a high efficiency while providing compact power packaging. 
 Especially for converters with many levels, the proposed topology is therefore advantageous compared to existing solutions.
\end{abstract}


%
\IEEEpeerreviewmaketitle

\mbox{}
\nomenclature{gnd}{Gate driver reference voltage.}
\nomenclature{GND}{Multi-active bridge reference voltage.}
\nomenclature{u}{Input variable.}
\nomenclature{\(\mathbf{b}\)}{Input vector.}
\nomenclature{\(\mathbf{A}\)}{System matrix.}
\nomenclature{\(\mathbf{x}\)}{State-space variable.}
\nomenclature{\(\mathbf{x}^{\textrm{ss}}\)}{Steady state of the state-space variable.}
\nomenclature{\(x_i\)}{$i$-th entry of the vector $\mathbf{x}$.}
\nomenclature{\(\dot{\mathbf{x}}\)}{First time derivative of $\mathbf{x}$.}
\nomenclature{\(\ddot{\mathbf{x}}\)}{Second time derivative of $\mathbf{x}$.}
\nomenclature{\(M\)}{Number of modules of the multi-active bridge.}
\nomenclature{\(i_{\textrm{GS}}\)}{Current charging the gate-source capacitance.}
\nomenclature{\(i_m\)}{Magnetization current of the gate driver transformer.}
\nomenclature{\(L_s\)}{Stray inductance of the gate driver transformer.}
\nomenclature{\(R_c\)}{Conduction resistance of the gate driver.}
\nomenclature{\(R_{\textrm{ON}}\)}{ON-state resistance of the gate drive MOSFET.}
\nomenclature{\(R_{\textrm{eq}}\)}{Equivalent resistance of the gate driver.}
\nomenclature{\(V_{\textrm{DC}}\)}{DC-voltage of the multi-active bridge.}
\nomenclature{\(V_{\textrm{S}}\)}{Supply voltage of the multi-active bridge.}
\nomenclature{\(V_{\textrm{gd}}\)}{Supply voltage of the gate driver.}
\nomenclature{\(V_{\textrm{GS}}\)}{Voltage between the Gate and Source terminals on the multi-active bridge circuit.}
\nomenclature{\(V_{\textrm{mid}}\)}{Midpoint voltage of the multi-active bridge circuit.}
\nomenclature{\(t_{\textrm{rise}}\)}{Timing constant.}
\nomenclature{\(t_{\textrm{high}}\)}{Timing constant.}
\nomenclature{\(t_{\textrm{fall}}\)}{Timing constant.}
\nomenclature{\(t_{\textrm{zero}}\)}{Timing constant.}
\nomenclature{\(t_{\textrm{low}}\)}{Timing constant.}
\nomenclature{\(S_i\)}{Signal that is applied to switch $i$.}
\nomenclature{\(n\)}{Turns number of the gate driver transformer windings.}
\nomenclature{\(N\)}{Turns number of the multi-active bridge circuit transformer windings.}
\nomenclature{\(L_{m}\)}{Magnetization inductance of the gate driver.}
\nomenclature{\(L_{M}\)}{Magnetization inductance of the multi-active bridge circuit.}
\nomenclature{\(C_{\textrm{dc}}\)}{DC-link capacitance of the gate driver.}
\nomenclature{\(C_{\textrm{DC}}\)}{DC-link capacitance of the multi-active bridge circuit.}
\nomenclature{\(C_{\textrm{GS}}\)}{Gate-Source capacitance.}
\nomenclature{\(C_{\textrm{eq}}\)}{Equivalent gate-source capacitance of the multi-active bridge.}
\nomenclature{\(C_{\textrm{DS}}\)}{Drain-Source capacitance.}
\nomenclature{\(q\)}{Switch on the gate driver.}
\nomenclature{\(Q\)}{Switch on the multi-active bridge circuit.}

\printnomenclature

\section{Introduction}
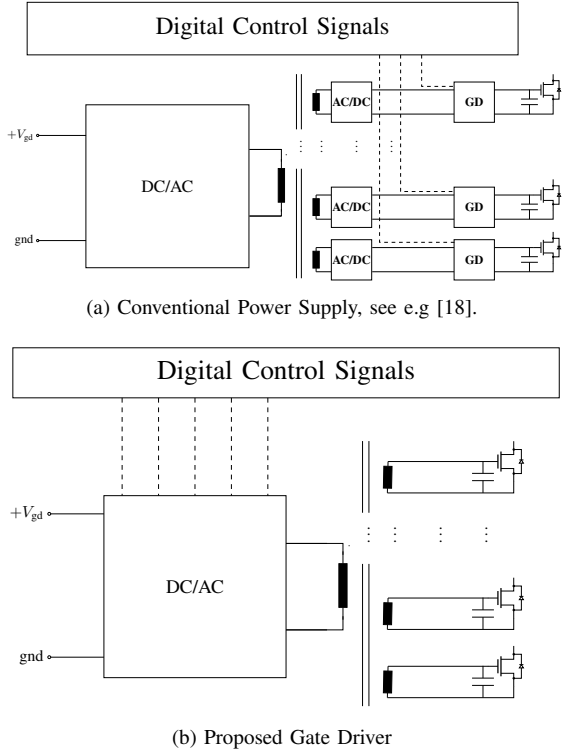
\begin{figure}[!htbp]
    \centering
    	\ctikzset{bipoles/thickness=1}
	\ctikzset{bipoles/length=0.8cm}
	\ctikzset{tripoles/thyristor/height=.8}
	\ctikzset{tripoles/thyristor/width=1}
	\ctikzset{bipoles/diode/height=.375}
	\ctikzset{bipoles/diode/width=.3}
    \ctikzset{inductors/scale=.75}
	\tikzstyle{block} = [draw,fill=white, rectangle, minimum height=1cm, minimum width=6em]
	\tikzstyle{sum} = [draw, fill=white, circle, node distance=1cm]
	\tikzstyle{pinstyle} = [pin edge={to-,thin,black}]
	\subfloat[Conventional Power Supply, see e.g  \cite{GDT2}.]{
	\resizebox{0.85\linewidth}{!}{
		\begin{circuitikz}[
		longL/.style = {european inductor, inductors/scale=0.75,
 inductors/width=1.6, inductors/coils=9}]
 \ctikzset{chips/scale=1.5}
 \draw (-2,15) rectangle (12,16.5);
 \node at (5,15.75){\huge{Digital Control Signals}};
 \draw[dashed] (9.25,15) -- ++ (0,-0.9) -- ++ (0.95,0);
 \draw[dashed] (8.65,15) -- ++ (0,-3.9) -- ++ (1.5,0);
 \draw[dashed] (8.05,15) -- ++ (0,-5.35)-- ++ (2.15,0);
		\draw
		(13,14)   node[nmos,bodydiode](igbt4a){};
		\draw
		(13,11) node[nmos,bodydiode](igbt7a){}
		(13,9.5) node[nmos,bodydiode](igbt8a){}
		; 
	    \node [
		rectangle,draw,
		minimum width=1.15cm,
		minimum height=1.15cm,
		] (igbt8a-Box1) at (7.25,9.15) {};
		\node at (igbt8a-Box1.center) {\textbf{AC/DC}};
		\node [
		rectangle,draw,
		minimum width=1.15cm,
		minimum height=1.15cm,
		] (igbt8a-Box2) at (10.75,9.15) {};
		\node at (igbt8a-Box2.center) {\textbf{GD}};
	    \node [
		rectangle,draw,
		minimum width=1.15cm,
		minimum height=1.15cm,
		] (igbt7a-Box1) at (7.25,10.65) {};
		\node at (igbt7a-Box1.center) {\textbf{AC/DC}};
		\node [
		rectangle,draw,
		minimum width=1.15cm,
		minimum height=1.15cm,
		] (igbt7a-Box2) at (10.75,10.65) {};
		\node at (igbt7a-Box2.center) {\textbf{GD}};
		\node [
		rectangle,draw,
		minimum width=1.15cm,
		minimum height=1.15cm,
		] (igbt4a-Box1) at (7.25,13.65) {};
		\node at (igbt4a-Box1.center) {\textbf{AC/DC}};
		\node [
		rectangle,draw,
		minimum width=1.15cm,
		minimum height=1.15cm,
		] (igbt4a-Box2) at (10.75,13.65) {};
		\node at (igbt4a-Box2.center) {\textbf{GD}};
        \draw (igbt7a.B)--++(-1.115,0); 
        \draw (igbt7a.B) ++(-2.265,0)--++(-2.335,0);
        \draw (igbt7a.B) ++(-5.775,0) -- ++(-0.425,0) coordinate (I7AB){};
        \draw (igbt8a.B)--++(-1.115,0); 
        \draw (igbt8a.B) ++(-2.265,0)--++(-2.335,0);
        \draw (igbt8a.B) ++ (-5.775,0) --++(-0.425,0) coordinate (I8AB){};
        \draw (igbt4a.B)--++(-1.115,0); 
        \draw (igbt4a.B) ++(-2.265,0)--++(-2.335,0);
        \draw (igbt4a.B) ++ (-5.775,0) --++(-0.425,0) coordinate (I4AB){};
        \draw (igbt7a.E) |- ++ (-1.675,-0.25);
        \draw (igbt7a.E)++(-2.825,-0.25) --++(-2.335,0);
        \draw (igbt7a.E)++(-6.335,-0.25) --++(-0.425,0) coordinate (I7AE){};
        \draw (igbt8a.E) |- ++ (-1.675,-0.25);
        \draw (igbt8a.E)++(-2.825,-0.25) --++(-2.335,0);
        \draw (igbt8a.E)++(-6.335,-0.25) --++(-0.425,0) coordinate (I8AE){};
        \draw (igbt4a.E) |- ++ (-1.675,-0.25);
        \draw (igbt4a.E)++(-2.825,-0.25) --++(-2.335,0);
        \draw (igbt4a.E)++(-6.335,-0.25) --++(-0.425,0) coordinate (I4AE){};
        \draw (igbt8a.E) ++ (-0.675,-0.25) to[C] ++(0,0.7);
        \draw (igbt7a.E) ++ (-0.675,-0.25) to[C] ++(0,0.7);
        \draw (igbt4a.E) ++ (-0.675,-0.25) to[C] ++(0,0.7);
        \draw
        (I4AE) to[L] (I4AB);
        \draw
        (I7AE) to[L] (I7AB);
        \draw
        (I8AE) to[L] (I8AB);
        \node at (6.25,12.00) (I2AE2){};
        \node at (6.25,14.75) (I1AB2){};
        \node at (6.25,10.00) (I1AE3){};
        \node at (6.25,12.75) (I2AB3){};
        \node[left of = I2AE2,node distance = 0.43cm] (GDIC1){};
        \node[left of = I2AE2,node distance = 0.57cm] (GDIC2){};
        \node[below of = GDIC1,node distance = 3.5cm] (GDIC11){};
        \node[below of = GDIC2,node distance = 3.5cm] (GDIC21){};
        \draw
        (GDIC11) -- ++ (0,3.25) node (GDIC12){};
        \draw
        (GDIC21) -- ++ (0,3.25) node (GDIC22){};
        \node[above of = GDIC12, node distance = 1cm] (GDIC13){};
        \node[above of = GDIC22, node distance = 1cm] (GDIC23){};
        \draw
        (GDIC13) -- ++ (0,1.7);
        \draw
        (GDIC23) -- ++ (0,1.7);
        \node[left of = I1AE3] (GDPMaux){};
        \node[left of = I2AB3] (GDPPaux){};
        \node[above of = GDPMaux,node distance=0.4cm] (GDPM){};
        \node[below of = GDPPaux,node distance=0.6cm] (GDPP){};
        \draw
        (GDPP) to[longL] (GDPM);
        \node[below of = GDPP, node distance=0.25cm] (GDPaux){};
        \node at (5.815, 12.5) {$\mathbf{\vdots}$};
        \node at (6.4, 12.5) {$\mathbf{\vdots}$};
        \node at (7.4, 12.5) {$\mathbf{\vdots}$};
        \node at (8.4, 12.5) {$\mathbf{\vdots}$};
		\draw
		(-1.7,12.7) node [ocirc]{} node [left] {{$+V_{\textrm{gd}}$}}  -- (-0.32,12.7)
		(-1.7,9.7) node [ocirc]{} node [left] {{gnd}}  -- (-0.32,9.7);
		\coordinate (V4) at (4.34,12.3);
	    \coordinate (V6) at (4.34,10.4);
		\draw (V6) -- ++(0.9,0);
		\draw (V4) -- ++(0.9,0);
		\draw (GDPP)++(0,-0.2) |- (V4);
		\draw (GDPM)++(0,0.2) |- (V6);
 		\node [
 		rectangle,draw,
 		minimum width=4.65cm,
 		minimum height=4.65cm,
 		] (DC-Box) at (2,11.25) {\Large{DC/AC}};
		\end{circuitikz}
	}
}
\\
	\subfloat[Proposed Gate Driver]{
	\resizebox{0.85\linewidth}{!}{
		\begin{circuitikz}[longL/.style = {european inductor, inductors/scale=0.75,
 inductors/width=1.6, inductors/coils=9}]
 \ctikzset{chips/scale=1.5}
 \draw (-2,15.4) rectangle (10,16.5);
 \node at (4,15.95) {\LARGE{Digital Control Signals}};
 \draw[dashed] (0.4,15.4)  -- ++ (0,-2.15);
 \draw[dashed] (1.2,15.4) -- ++ (0,-2.15);
 \draw[dashed] (2,15.4) -- ++ (0,-2.15);
 \draw[dashed] (2.8,15.4) -- ++ (0,-2.15);
 \draw[dashed] (3.6,15.4) -- ++ (0,-2.15);
		\draw
		(9,14)   node[nmos,bodydiode](igbt4a){};
		\draw
		(9,11) node[nmos,bodydiode](igbt7a){}
		(9,9.5) node[nmos,bodydiode](igbt8a){}
		; 
        \draw
        (igbt7a.B) -- ++ (-0.2,0) to++(-2,0) coordinate (I7AB){};
        \draw
        (igbt8a.B) |- ++ (-0.2,0)to++(-2,0) coordinate (I8AB){};
        \draw
        (igbt4a.B) |- ++ (-0.2,0)to++(-2,0) coordinate (I4AB){};
        \draw
        (igbt7a.E) |- ++ (-2.775,-0.25) coordinate (I7AE){};
        \draw
        (igbt8a.E) |- ++ (-2.775,-0.25) coordinate (I8AE){};
        \draw
        (igbt4a.E) |- ++ (-2.775,-0.25) coordinate (I4AE){};
        \draw (igbt8a.E) ++ (-0.675,-0.25) to[C] ++(0,0.7);
        \draw (igbt7a.E) ++ (-0.675,-0.25) to[C] ++(0,0.7);
        \draw (igbt4a.E) ++ (-0.675,-0.25) to[C] ++(0,0.7);
        \draw
        (I4AE) to[L] (I4AB);
        \draw
        (I7AE) to[L] (I7AB);
        \draw
        (I8AE) to[L] (I8AB);
        \node at (6.25,12.00) (I2AE2){};
        \node at (6.25,14.75) (I1AB2){};
        \node at (6.25,10.00) (I2AE3){};
        \node at (6.25,12.75) (I1AB3){};
        \node[left of = I2AE2,node distance = 0.43cm] (GDIC1){};
        \node[left of = I2AE2,node distance = 0.57cm] (GDIC2){};
        \node[below of = GDIC1,node distance = 3.5cm] (GDIC11){};
        \node[below of = GDIC2,node distance = 3.5cm] (GDIC21){};
        \draw
        (GDIC11) -- ++ (0,3.25) node (GDIC12){};
        \draw
        (GDIC21) -- ++ (0,3.25) node (GDIC22){};
        \node[above of = GDIC12, node distance = 1cm] (GDIC13){};
        \node[above of = GDIC22, node distance = 1cm] (GDIC23){};
        \draw
        (GDIC13) -- ++ (0,1.7);
        \draw
        (GDIC23) -- ++ (0,1.7);
        \node[left of = I2AE3] (GDPMaux){};
        \node[left of = I1AB3] (GDPPaux){};
        \node[above of = GDPMaux,node distance=0.4cm] (GDPM){};
        \node[below of = GDPPaux,node distance=0.6cm] (GDPP){};
        \draw
        (GDPP) to[longL] (GDPM);
        \node[below of = GDPP, node distance=0.25cm] (GDPaux){};
        \node at (5.815, 12.5) {$\mathbf{\vdots}$};
        \node at (6.4, 12.5) {$\mathbf{\vdots}$};
        \node at (7.4, 12.5) {$\mathbf{\vdots}$};
        \node at (8.4, 12.5) {$\mathbf{\vdots}$};
		\draw
		(-1.2,12.85) node [ocirc]{} node [left] {{$+V_{\textrm{gd}}$}}  -- (-0.0,12.85)
		(-1.2,9.7) node [ocirc]{} node [left] {{gnd}}  -- (-0.0,9.7);
		\coordinate (V4) at (4.00,12.2);
	    \coordinate (V6) at (4.00,10.3);
		\draw (V6) -- ++(0.9,0);
		\draw (V4) -- ++(0.9,0);
		\draw (GDPP)++(0,-0.2) |- (V4);
		\draw (GDPM)++(0,0.2) |- (V6);
 		\node [
 		rectangle,draw,
 		minimum width=4.0cm,
 		minimum height=4.0cm,
 		] (DC-Box) at (2,11.25) {\large{DC/AC}};
		\end{circuitikz}
	}
	}
    \caption{Overview of different isolated auxiliary supply concepts for MOSFET switches. Dashed lines show data signals from the controller. The conventional approach, shown in (a), generates an AC signal from the voltage supply which is then distributed to all MOSFETs using a multi-winding transformer. At the MOSFETs, the voltage is rectified and serves as a supply for the gate drivers connected to each MOSFET. The gate driver then generates the signals that are required to switch the MOSFET. The switching state is determined by a controller run e.g. on an FPGA. The FPGA communicates with each MOSFET directly. In the proposed approach, shown in (b), the voltage supply is converted to an AC signal as well. Similar to the conventional approach, the AC signal is then distributed to the MOSFETs using a multi-winding transformer. At the MOSFET however, the multi-winding transformer is directly connected to the gate and source terminals to enable the switching. Since there is no gate driver in between, the controller is directly connected to the primary side DC/AC part, switching all MOSFETs synchronously.} 
    \label{fig:GD1}
\end{figure}

 Recent advances in multi-winding transformer research have produced interesting novel approaches for power converters derived from the multi-winding transformer technology \cite{MABMIMO}.
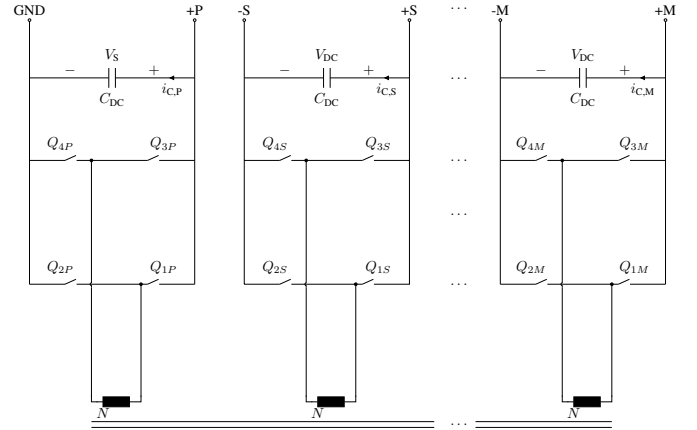
\begin{figure}[!htbp]
    \centering
    	\ctikzset{bipoles/thickness=1}
	\ctikzset{bipoles/length=0.8cm}
	\ctikzset{tripoles/thyristor/height=.8}
	\ctikzset{tripoles/thyristor/width=1}
	\ctikzset{bipoles/diode/height=.375}
	\ctikzset{bipoles/diode/width=.3}
	\tikzstyle{block} = [draw,fill=white, rectangle, minimum height=1cm, minimum width=6em]
	\tikzstyle{sum} = [draw, fill=white, circle, node distance=1cm]
	\tikzstyle{pinstyle} = [pin edge={to-,thin,black}]
	\resizebox{0.5\textwidth}{!}{
		\begin{tikzpicture}
        \draw (7.9,7.6)--(7.9,7.45) node (T1C1) {} to[L] ++ (-1.2,0) node (T1C2) {}  --++ (0,0.15);
        \draw (13.1,7.6)--(13.1,7.45) node (T1B1) {} to[L] ++ (-1.2,0)  node (T1B2) {}  --++ (0,0.15);
        \draw (19.3,7.6)--(19.3,7.45) node (T1A1) {} to[L] ++ (-1.2,0)  node (T1A2) {}  --++ (0,0.15);
		\node[below right of = T1A2, node distance=0.4cm] {$N$};
		\node[below right of = T1B2, node distance=0.4cm] {$N$};
		\node[below right of = T1C2, node distance=0.4cm] {$N$};
        \draw (6.7,6.97) -- (15,6.97);
        \draw (6.7,6.83) -- (15,6.83);
        \draw (16,6.97) -- (19.3,6.97);
        \draw (16,6.83) -- (19.3,6.83);
		\draw
		(17.35,10.7) node(igbt6a){$Q_{2M}$}
		(19.85,10.7) node(igbt6b){$Q_{1M}$}
		(17.35,13.7) node(igbt4a){$Q_{4M}$}
		(19.85,13.7) node(igbt4b){$Q_{3M}$}; 
		\draw
		(16.6,13.3)
		to[nos] (18.6,13.3)
		to[nos] (20.6,13.3) coordinate (leg4);
		\draw
		(16.6,10.3)
		to[nos] (18.6,10.3)
		to[nos] (20.6,10.3) coordinate (leg6);
		\draw 
		(leg6) -- ++(0,5.7)
		(16.6,10.3) -- ++(0,5.7)
		;
		\draw
		(20.6,15.3)
		to[C,invert, l=$C_{\textrm{DC}}$, v>=$V_{\textrm{DC}}$, i=$i_{\textrm{C,M}}$, current/distance=-0.8]  (16.6,15.3)
		;
		\draw
		(16.6,16.7) node [ocirc]{} node [above] {{-M}}  -- (16.6,16)
		(20.6,16.7) node [ocirc]{} node [above] {{+M}}  -- (20.6,16);
		\coordinate (V4) at (18.1,13.3);
	    \coordinate (V6) at (19.3,10.3);
		\draw (V6) node [circ] {}  to[short]  (T1A1);
				\draw (T1A2)to[crossing]++(0,5.7) to
		(V4) node[circ]{};
		\coordinate (V4S) at (13.1,10.3);
		\coordinate (V6S) at (11.9,13.3);
		\draw (V4S) node[circ]{}  to[short]  (T1B1);
		\draw (T1B2)to[crossing]++(0,5.7) to
		(V6S) node[circ]{};
		\draw
		(11.15,10.7) node(igbt1a){$Q_{2S}$}
		(13.65,10.7) node(igbt1b){$Q_{1S}$}
		(11.15,13.7) node(igbt2a){$Q_{4S}$}
		(13.65,13.7) node(igbt2b){$Q_{3S}$}; 
		\draw
		(10.4,13.3)
		to[nos] (12.4,13.3)
		to[nos] (14.4,13.3) coordinate (leg2)
		(10.4,10.3)
		to[nos] (12.4,10.3)
		to[nos] (14.4,10.3) coordinate (leg1)
		-- ++(0,5.7)
		(10.4,10.3) -- ++(0,5.7)
		;
		\draw
		(14.4,15.3)
		to[C,invert, l=$C_{\textrm{DC}}$, v>=$V_{\textrm{DC}}$, i=$i_{\textrm{C,S}}$, current/distance=-0.8]  (10.4,15.3);
		\draw
		(14.4,16.7) node [ocirc]{} node [above] {{+S}}  -- (14.4,16)
		(10.4,16.7) node [ocirc]{} node [above] {{-S}}  -- (10.4,16);
		\coordinate (V4T) at (7.9,10.3);
		\coordinate (V6T) at (6.7,13.3);
		\draw (V4T) node[circ]{} to[short] (T1C1);
		\draw (T1C2)to[crossing]++(0,5.7) to
		(V6T) node[circ]{};
		\draw
		(8.45,10.7) node(igbt3a){$Q_{1P}$}
		(5.95,10.7) node(igbt3b){$Q_{2P}$}
		(8.45,13.7) node(igbt5a){$Q_{3P}$}
		(5.95,13.7) node(igbt5b){$Q_{4P}$}; 
		\draw
		(5.2,13.3) coordinate (leg5)
		to[nos] (7.2,13.3)
		to[nos] (9.2,13.3);
		\draw
		(5.2,10.3) coordinate (leg3)
		to[nos] (7.2,10.3)
		to[nos] (9.2,10.3);
		\draw
		(leg3)
		-- ++(0,5.7)
		(9.2,10.3) -- ++(0,5.7)
		;
		\draw
		(9.2,15.3)
		to[C,invert, l=$C_{\textrm{DC}}$, v>=$V_{\textrm{S}}$, i=$i_{\textrm{C,P}}$, current/distance=-0.8]  (5.2,15.3);
		\draw
		(9.2,16.7) node [ocirc]{} node [above] {{+P}}  -- (9.2,16)
		(5.2,16.7) node [ocirc]{} node [above] {{GND}}  -- (5.2,16);
        \node at (15.6,6.90) {$\mathbf{\hdots}$};
        \node at (15.6,10.30) {$\mathbf{\hdots}$};
        \node at (15.6,12) {$\mathbf{\hdots}$};
        \node at (15.6,13.30) {$\mathbf{\hdots}$};
        \node at (15.6,15.30) {$\mathbf{\hdots}$};
        \node at (15.6,17.00) {$\mathbf{\hdots}$};
		\end{tikzpicture}
	}
\caption{Circuit diagram of the multi-active bridge converter with ideal switches. 
This DC/AC topology consists of $M$ active bridges that are connected through a multi-winding transformer in the AC domain. 
If all the windings of the transformer share a single core, the DC-voltages of the circuit will be balanced. The conversion from DC to AC is achieved through the switching pattern of $Q_1$, $Q_2$, $Q_3$, and $Q_4$.}
\label{fig:GD2}
\end{figure}
 \newline
     A promising multi-winding transformer-based DC/DC topology is the multi-active bridge (MAB) \cite{MABMIMO, MABDCDC, MAB, MABC, MAB2, MABBB, MABHRE, MABMAB, MABER}. 
    The circuit diagram of a MAB converter is shown in Fig. \ref{fig:GD2}.
    In \cite{MABDCDC}, a more detailed introduction to MAB-based DC/DC converters is given.
    The MAB topology is a combination of many active bridges connected magnetically through a multi-winding transformer \cite{MAB}. 
    Its advantage over regular cascaded H-bridge-based multilevel converters is the strong self-balancing which can be achieved by using a multi-winding transformer with a single core. 
    In this way, all windings will be exposed to the same magnetic flux. 
    If all windings also possess the same number of turns, the AC voltage in each active bridge will be similar which results in a balanced state.
    This solves the problem of the requirement of large DC/DC converters which significantly increases the power density of the converter.
    Over the past few years, several applications for MABs have been identified such as wireless charging \cite{MABC}, grid connections \cite{MAB2}, and battery balancing \cite{MABBB}. 
    Furthermore, the usage of the MAB has been proposed as a versatile solution for hybrid renewable energy systems \cite{MABHRE}, multilevel inverters \cite{MABMAB} and energy routers \cite{MABER}. 
     \newline 
     Since the MAB is an AC-link-based topology, it requires constant switching at high frequencies to prevent its magnetics from going into saturation.
     For this reason, the reduction of switching losses in MAB-based topologies is of great importance. 
     Over the past few decades, several approaches to quick and lossless gate-driving of MOSFET switches have been introduced \cite{ GD_old_NISO2, GD_old_NISO4, MMCGD2,GD_old_NISO1, GD_old_NISO3, GD_old_ISO1}. 
     Common state-of-the-art approaches include charge-pump \cite{GD_old_NISO2} and bootstrap \cite{GD_old_NISO4, MMCGD2}  solutions.
     To reduce losses, \cite{GD_old_NISO1} suggested saving the energy stored in the input capacitance of the MOSFET in an external resonant circuit. 
     While those gate drivers rely on a half-bridge to generate the switching signal, an approach based on a class $\Phi_2$ inverter consisting only of a single switch has been proposed in \cite{GD_old_NISO3}.
     In addition to that, \cite{GD_old_ISO1} proposed to isolate the switching signal from the power circuit by using a transformer to transfer the information about the desired switching state as shown in Fig. \ref{fig:GD1} (a). 
     While those solutions have been proven to work well for driving a single MOSFET which is sufficient for simple circuits, more complex topologies might consist of more than one switch. 
     A key question, therefore, is how these concepts can be generalized to drive more than one switch with a minimal increase in complexity of the driving circuit. 
    \newline
    Isolating the driving circuits from each other is important not only because it offers protection of the gate driver from faulty switches but also allows to safely connect one power supply to many switches which are placed at different voltage levels to a common reference voltage \cite{GD_old_ISO2,GD1, GDT2, MMCGD1, MABT}.
    Therefore, based on the approach presented in \cite{GD_old_ISO1}, many solutions for isolated gate drivers have been introduced,  \cite{GD_old_ISO2,GD1, GDT2, MMCGD1, MABT}. 
    For certain applications, such as half-bridges, some of the switches always will be opened in a mutually exclusive manner. 
    In this case, all switches can be attached to the transformer isolation as a tertiary winding and the mutual exclusive switching can be implemented by different turn directions of the windings \cite{GD_old_ISO2}. 
    Furthermore, the concept has been extended to high-power switches, adding overcurrent protection and thermal protection in \cite{GD1}.
    In the case of a multilevel converter, the gate driver concepts have to be further extended to drive not only a few but an arbitrarily large amount of switches. 
    While the basic concept of gate driving remains similar, the transformer isolation now is realized with a multi-winding transformer such as the cascaded half-bridge \cite{GDT2}. 
    In \cite{MMCGD1} several techniques to reduce the EMI for this configuration have been examined.
    Furthermore, an isolated gate driver for a different multilevel inverter concept, the solid-state transformer, has been introduced in \cite{MABT}. 
    \newline
     While several approaches for the topology of power supplies of MABs have been proposed, the design of a transformer-isolation-based power supply for gate drivers comes with several challenges \cite{ GD_old_ISOP6,GD_old_ISOP1, GD_old_ISOP2, GD_old_ISOP4, GD_old_ISOP3}. 
    One of these challenges is designing the transformer which starts with choosing the correct basic transformer geometry. 
    While regular transformers, such as the toroidal transformer \cite{GD_old_ISOP6} have small parasitic capacitances, they are bulky and reduce the power density of the supply  \cite{GD_old_ISOP6}.
    Planar transformers, on the other hand side, provide a high power density but larger parasitic capacitances \cite{GD_old_ISOP1}. 
    Once the basic geometry has been chosen, its parameters can be optimized to obtain the best transformer tailored to the desired application \cite{GD_old_ISOP1,GD_old_ISOP2} and reduce the effects of parasitic elements. 
    To reduce the parasitic capacitance, several techniques have been examined such as geometric optimization with the help of finite element analysis \cite{GD_old_ISOP1}. 
    Alternatively, formulas to estimate the parasitic capacitance from the parameters of this geometry can be used as shown in \cite{GD_old_ISOP2}. 
    It is moreover possible to supply other low-power components using the same multi-winding transformer as the gate driver \cite{GD_old_ISOP4}. 
    This can be achieved by attaching more windings to the transformer. 
    If different voltage levels are required, the number of turns can be varied for those cases \cite{GD_old_ISOP4}.
    In addition to the optimization of the transformer, system theoretic aspects such as the stability of the gate driver can be taken into account when designing the circuit to ensure safe and reliable operation \cite{GD_old_ISOP3}.
    \newline
     Several solutions for zero voltage switching in MABs have been proposed in \cite{MABSPECIAL1, MABSPECIAL2, MABSPECIAL3}. 
    A resonant tank can be used to reduce the voltage oscillations within the circuit as shown in \cite{MABSPECIAL1}. 
    Furthermore, the resonant tank can be used to increase the balancing speed of the DC-voltages of the MAB which was presented for battery balancing in \cite{MABSPECIAL2} .
    In \cite{MABSPECIAL3}, the authors derived soft-switching conditions for MABs and gave guidelines for the selection of a suitable operating point.
\newline
    Although a variety of interesting concepts for power supplies for modular multilevel converters as well as resonant configurations for MABs has been presented in the literature, a combination of the two approaches, i.e. a power supply for MABs that allows lossless switching, is still subject to research. 
    \newline
In this paper, we propose an advanced isolated gate driver for MABs with a unique energy recovery technique that allows lossless switching of all transistors of the circuit.
To store the energy of the parasitic capacitances of the switching in the MAB and the gate driver, two multi-winding transformers will be employed, one being part of the gate driver and the other being the regular MAB transformer that connects the modules. 
Contrary to the regular transformer isolation shown in Fig. \ref{fig:GD1} (a), the control part of the gate driver is entirely on the primary side of the multi-winding transformer which results in a centralized control approach, shown in Fig. \ref{fig:GD1} (b).
In addition, the proposed gate driver is capable of ensuring synchronized switching of all modules. 
\newline
The remainder of the paper is structured as follows:
Section \rom{2} provides an overview of the proposed topology and its basic operating principle. 
In Section \rom{3}, a state-space model and analysis of the proposed model are presented.
Experiment results are given in Section \rom{4} and the conclusion follows in Section \rom{5}.

\begin{figure}[!htbp]
    \centering
    	\ctikzset{bipoles/thickness=1}
	\ctikzset{bipoles/length=0.8cm}
	\ctikzset{tripoles/thyristor/height=.8}
	\ctikzset{tripoles/thyristor/width=1}
	\ctikzset{bipoles/diode/height=.375}
	\ctikzset{bipoles/diode/width=.3}
	\tikzstyle{block} = [draw,fill=white, rectangle, minimum height=1cm, minimum width=6em]
	\tikzstyle{sum} = [draw, fill=white, circle, node distance=1cm]
	\tikzstyle{pinstyle} = [pin edge={to-,thin,black}]
	\resizebox{0.5\textwidth}{!}{
		\begin{circuitikz}[longL/.style = {european inductor, inductors/scale=0.75,
 inductors/width=1.6, inductors/coils=9}]
		\draw
		(-17.9,5.3)  node[nmos,bodydiode](igbt6a){};
		\draw
		(-17.9,8.8)  node[nmos,bodydiode](igbt6b){};
		\draw
		(-19.4,5.3) node[nmos,bodydiode](igbt4a){};
		\draw
		(-19.4,8.8) node[nmos,bodydiode](igbt4b){}; 
		\draw (igbt6a)++(-0.4,0.4) node{$q_2$};
		\draw (igbt6b)++(-0.4,0.4) node{$q_1$};
		\draw (igbt4a)++(-0.4,0.4) node{$q_4$};
		\draw (igbt4b)++(-0.4,0.4) node{$q_3$};
		\draw
		(-19.4,9.5) node[circ]{}
		to (igbt4b.C);
		\draw
		(igbt4b.E) 
		to (igbt4a.C);
		\draw
		(igbt4a.E) to (-19.4,4.5) coordinate (leg4)node[circ]{};
		\draw
		(-17.9,9.5)
		to (igbt6b.C);
		\draw
		(igbt6b.E) 
		to (igbt6a.C);
		\draw
		(igbt6a.E) to (-17.9,4.5) coordinate (leg6);
		\draw 
		(leg6) -- (-21.6,4.5)
		(-21.6,9.5) --(-17.9,9.5)
		;
		\draw
		(-20.9,9.5) node[circ]{}
		to[C=$C_{\textrm{dc}}$]  (-20.9,4.5) node[circ]{}
		;
		\draw
		(-22.3,9.5) node [ocirc]{} node [left] {{$+V_{\textrm{gd}}$}}  -- (-21.6,9.5)
		(-22.3,4.5) node [ocirc]{} node [left] {{gnd}}  -- (-21.6,4.5);
		\coordinate (V4) at (-17.6,8.1);
	    \coordinate (V6) at (-17.6,6.2);
	    \draw
		(V6)  to (leg6 |- V6) node [circ] {};
		\draw
		(V4)   to[crossing]++(-0.6,0)to (leg4 |- V4) node [circ] {};
		\draw (-17,7.6)  node[nmos,bodydiode,yscale=-1](igbtCa){};
		\node[circ] at (-17,8.1){};
		\draw (-17,6.7)  node[nmos,bodydiode](igbtCb){};
		\node[circ] at (-17,6.2){};
		\draw (-17,8.1)--(igbtCa.E);
		\draw (igbtCa.C)--(igbtCb.C);
		\draw (-17,6.2)--(igbtCb.E);
		\draw (igbtCa) ++ (-0.5,0.3) node{$q_5$};
		\draw (igbtCb) ++ (-0.5,0.3) node{$q_6$};
        \node [circ] at (-16.5,8.1){};
        \node [circ] at (-16.5,6.2){};
        \draw (-16.5,8.1) to[longL=$L_m$] (-16.5,6.2);
        \node at (-16,7.15){$L_m$};
        \node at (-15.5,8.2) {$n$};
		\node at (-14.35,5.8) (I2AE2){};
        \node at (-14.35,8.55) (I1AB2){};
        \node[left of = I2AE2] (GDPMaux){};
        \node[left of = I1AB2] (GDPPaux){};
        \node[above of = GDPMaux,node distance=0.4cm] (GDPM){};
        \node[below of = GDPPaux,node distance=0.6cm] (GDPP){};
        \draw (GDPP) to[longL] (GDPM);
        \node[circ] at (-15.25,7.75){};
		\draw (V6) -- ++(2.24,0);
		\draw (V4) -- ++(2.24,0);
		\draw (GDPP)++(0,-0.2) |- (V4);
		\draw (GDPM)++(0,0.2) |- (V6);
		\end{circuitikz}
	}
    \caption{Circuit diagram of a midpoint-clamped active bridge with MOSFET switches and primary winding. 
    This topology represents an implementation of the DC/AC black box in Fig. \ref{fig:GD1} and will be used as basic gate driver topology in the following. 
    The switches $q_1$, ..., $q_6$ are low-power and controlled directly by the FPGA.}
    \label{fig:GD3}
\end{figure}
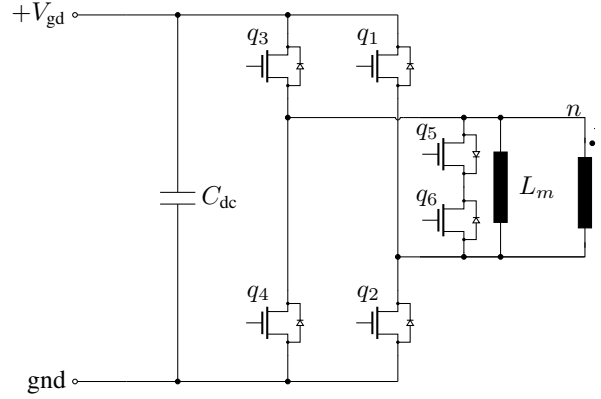

\begin{figure}[!htbp]
    \centering
    \includegraphics[scale=0.65]{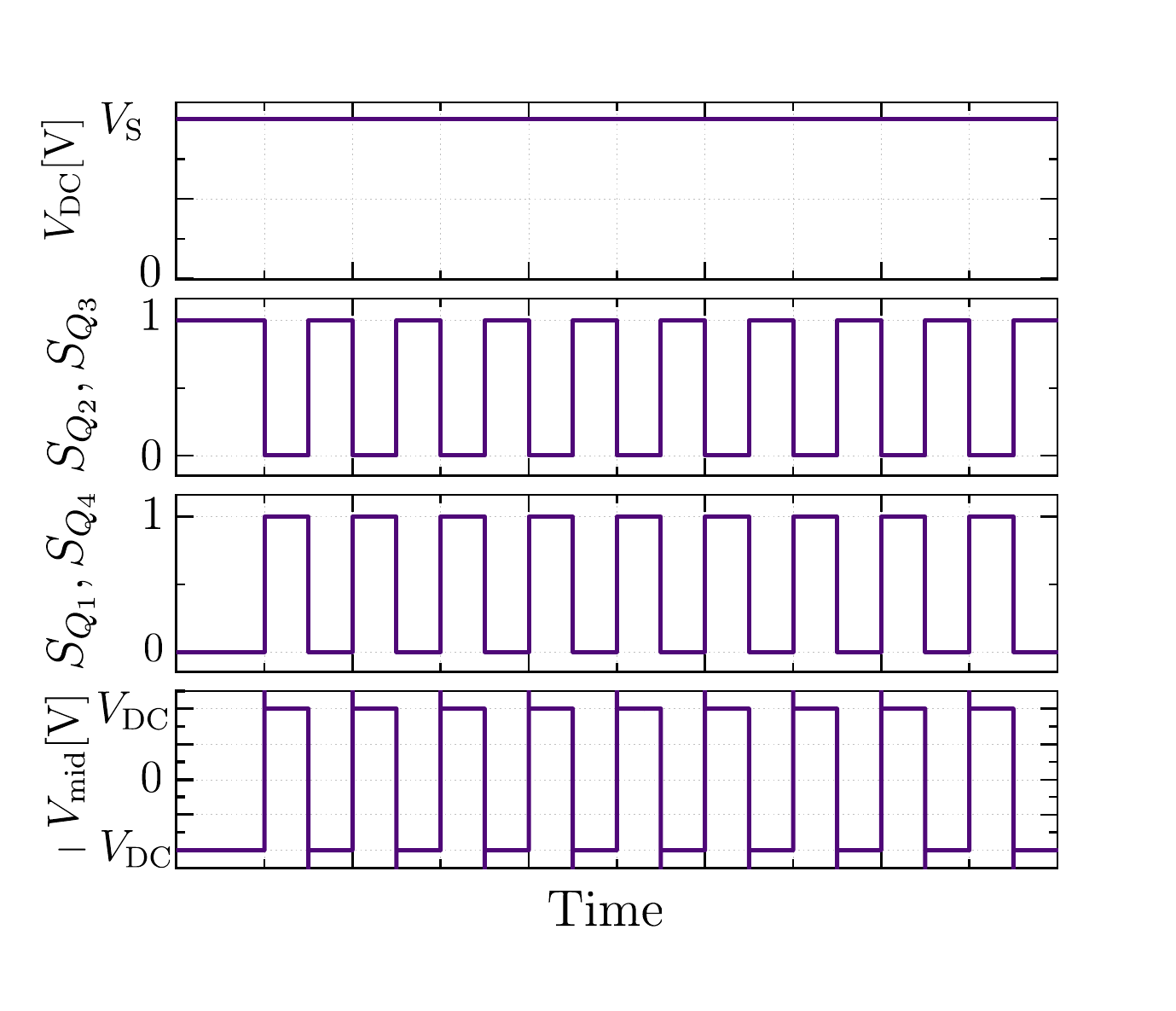}
    \caption{The waveform of the AC-voltage of a MAB. 
    To provide a stable DC-voltage shown in the top subfigure, the switches $Q_1$ and $Q_4$ must be switched alternating with $Q_2$ and $Q_3$ as depicted in the two middle subfigures which is the central requirement to the proposed gate driver. 
    This results in the square-wave AC-voltage shown in the bottom subfigure under the assumption of an ideal high-voltage transformer. 
    Phase shift control is not considered in this paper, thus every bridge uses the above switching pattern.}
    \label{fig:GD5}
\end{figure}

\section{Topology Overview and Operating Principle}
An overview of the overall design is shown in Fig. \ref{fig:GD1} (b). 
The switches in the MAB change their switching state when the gate-source capacitance is charged or discharged, for which an auxiliary power supply is used. 
A primary DC/AC converter topology is fed by the auxiliary power supply and determines whether the gate-source capacitors are charged or discharged.
To synchronously charge or discharge all gate-source capacitances and ensure isolation of the switches from each other, a multi-winding transformer is used that transmits the power from the primary DC/AC converter circuit to all switches in the MAB. 
The goal is to minimize the power consumption of this circuit which is provided by the auxiliary supply.
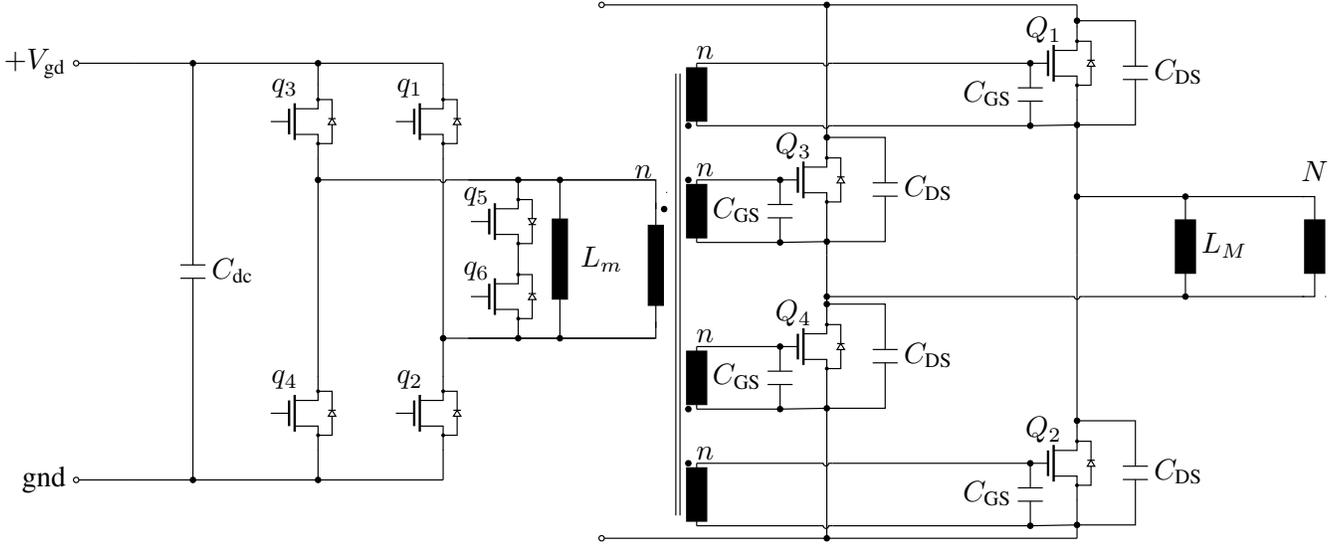
\begin{figure*}[!htbp]
    \centering
    	\ctikzset{bipoles/thickness=1}
	\ctikzset{bipoles/length=0.8cm}
	\ctikzset{tripoles/thyristor/height=.8}
	\ctikzset{tripoles/thyristor/width=1}
	\ctikzset{bipoles/diode/height=.375}
	\ctikzset{bipoles/diode/width=.3}
	\ctikzset{bipoles/capacitor/height=.375}
	\tikzstyle{block} = [draw,fill=white, rectangle, minimum height=1cm, minimum width=6em]
	\tikzstyle{sum} = [draw, fill=white, circle, node distance=1cm]
	\tikzstyle{pinstyle} = [pin edge={to-,thin,black}]
	\resizebox{\textwidth}{!}{
		\begin{circuitikz}[longL/.style = {european inductor, inductors/scale=0.75,
 inductors/width=1.6, inductors/coils=9}]
        \draw (-7.6,7.9)--(-7.45,7.9) node (T1C1) {} to[L] ++ (0,-1.2) node (T1C2) {}  --++ (-0.15,0);
		\node[above of = T1C1, node distance=0.3cm] {$N$};
		\node[circ] at (-9.0,7.9){};
		\node[circ] at (-9.0,6.7){};
		\draw (-9.0,7.9) to[L=$L_M$] (-9.0,6.7);
		\coordinate (V4T) at (-10.3,7.9);
		\coordinate (V6T) at (-13.3,6.7);
		\draw (V4T) node[circ]{} to[short] (T1C1);
		\draw (T1C2)to[crossing]++(-5.7,0) to
		(V6T) node[circ]{};
		\draw(-13.3,6.1) node[nmos,bodydiode](Q4){};
		\draw(-13.3,8.1) node[nmos,bodydiode](Q3){};
		\draw(-13.3,3.8) coordinate (leg5) node[circ]{} to(Q4.E);
		\draw (Q4.C) to (-13.3,7.2);
		\draw (-13.3,7.2) to (Q3.E);
		\draw (Q3.C)
		to(-13.3,10.2) node[circ]{};
		\draw(-10.3,4.7) node[nmos,bodydiode](Q2){};
		\draw(-10.3,9.5) node[nmos,bodydiode](Q1){};
		\draw(-10.3,3.8) coordinate (leg3) to (Q2.E);
		\draw (Q2.C) to (-10.3,7.2);
		\draw (-10.3,7.2) to (Q1.E);
		\draw (Q1.C) to (-10.3,10.2);
        \draw (Q1)++(-0.4,0.4)  node(igbt3a)  {$Q_{1}$};
        \draw (Q2)++(-0.4,0.4) node(igbt3b) {$Q_{2}$};
        \draw (Q3)++(-0.4,0.4)  node(igbt5a)  {$Q_{3}$};
        \draw (Q4)++(-0.4,0.4) node(igbt5b) {$Q_{4}$};
        \draw (Q1.E)++(0,-0.3) node[circ](Q1Aux1) {};
        \draw(Q1.B)++(0,-0.75) coordinate (Q1Aux2){} to[C=$C_{\textrm{GS}}$] ++ (0,0.75);
        \draw(Q1Aux1) to (Q1Aux2){};
        \draw (Q2.E)++(0,-0.3) node[circ](Q2Aux1) {};
        \draw(Q2.B)++(0,-0.75) coordinate (Q2Aux2){} to[C=$C_{\textrm{GS}}$] ++ (0,0.75);
        \draw(Q2Aux1) to (Q2Aux2){};
        \draw (Q3.E)++(0,-0.3) node[circ](Q3Aux1) {};
        \draw(Q3.B)++(0,-0.75) coordinate (Q3Aux2){} to[C=$C_{\textrm{GS}}$] ++ (0,0.75);
        \draw(Q3Aux1) to (Q3Aux2){};
        \draw (Q4.E)++(0,-0.3) node[circ](Q4Aux1) {};
        \draw(Q4.B)++(0,-0.75) coordinate (Q4Aux2){} to[C=$C_{\textrm{GS}}$] ++ (0,0.75);
        \draw(Q4Aux1) to (Q4Aux2){};
        \draw (Q1Aux1)++(0,1.25)node[circ]{} -- ++ (0.7,0) to[C=$C_{\textrm{DS}}$] ++ (0,-1.25)  -- ++ (-0.7,0);
        \draw (Q2Aux1)++(0,1.25)node[circ]{} -- ++ (0.7,0) to[C=$C_{\textrm{DS}}$] ++ (0,-1.25)  -- ++ (-0.7,0);
        \draw (Q3Aux1)++(0,1.25)node[circ]{} -- ++ (0.7,0) to[C=$C_{\textrm{DS}}$] ++ (0,-1.25)  -- ++ (-0.7,0);
        \draw (Q4Aux1)++(0,1.25)node[circ]{} -- ++ (0.7,0) to[C=$C_{\textrm{DS}}$] ++ (0,-1.25)  -- ++ (-0.7,0);
        \node[circ] at (Q1.B){};
        \node[circ] at (Q2.B){};
        \node[circ] at (Q3.B){};
        \node[circ] at (Q4.B){};
        \node[circ] at (Q1Aux2){};
        \node[circ] at (Q2Aux2){};
        \node[circ] at (Q3Aux2){};
        \node[circ] at (Q4Aux2){};
        \draw(Q1Aux2) --++ (-0.9,0) to[crossing] ++ (-3.1,0) coordinate(Wi1Aux1){};
        \draw(Q2Aux2) --++ (-0.9,0) to[crossing] ++ (-3.1,0) coordinate(Wi2Aux1){};
        \draw(Q3Aux2) --++ (-1,0) coordinate(Wi3Aux1){};
        \draw(Q4Aux2) --++ (-1,0) coordinate(Wi4Aux1){};
        \draw(Q1.B) --++ (-0.9,0) to[crossing] ++ (-3.1,0) coordinate(Wi1Aux2){};
        \draw(Q2.B) --++ (-0.9,0) to[crossing] ++ (-3.1,0) coordinate(Wi2Aux2){};
        \draw(Q3.B) --++ (-1,0) coordinate(Wi3Aux2){};
        \draw(Q4.B) --++ (-1,0) coordinate(Wi4Aux2){};
        \draw (Wi1Aux1) to[L] (Wi1Aux2);
        \draw (Wi2Aux1) to[L] (Wi2Aux2);
        \draw (Wi3Aux1) to[L] (Wi3Aux2);
        \draw (Wi4Aux1) to[L] (Wi4Aux2);
        \draw (Wi1Aux1)++(-0.1,0) node[circ] {};
        \draw (Wi3Aux2)++(-0.1,0) node[circ] {};
        \draw (Wi4Aux1)++(-0.1,0) node[circ] {};
        \draw (Wi2Aux2)++(-0.1,0) node[circ] {};
        \draw (Wi1Aux2)++(0.1,0.125) node {$n$};
        \draw (Wi2Aux2)++(0.1,0.125) node {$n$};
        \draw (Wi3Aux2)++(0.1,0.125) node {$n$};
        \draw (Wi4Aux2)++(0.1,0.125) node {$n$};
        \draw  (Wi1Aux2)++(-0.2,0) node (CAUX1U){};
        \draw  (Wi1Aux2)++(-0.25,0) node (CAUX2U){};
        \draw  (Wi2Aux1)++(-0.2,0) node (CAUX1D){};
        \draw  (Wi2Aux1)++(-0.25,0) node (CAUX2D){};
        \draw (CAUX1U) -- (CAUX1D);
        \draw (CAUX2U) -- (CAUX2D);
		\draw
		(leg3)	-- ++(-5.7,0) node[ocirc]{};
		\draw
		(-10.3,10.2) -- ++(-5.7,0) node[ocirc]{};
		\draw
		(-17.9,5.3)  node[nmos,bodydiode](igbt6a){};
		\draw
		(-17.9,8.8)  node[nmos,bodydiode](igbt6b){};
		\draw
		(-19.4,5.3) node[nmos,bodydiode](igbt4a){};
		\draw
		(-19.4,8.8) node[nmos,bodydiode](igbt4b){}; 
		\draw (igbt6a)++(-0.4,0.4) node{$q_2$};
		\draw (igbt6b)++(-0.4,0.4) node{$q_1$};
		\draw (igbt4a)++(-0.4,0.4) node{$q_4$};
		\draw (igbt4b)++(-0.4,0.4) node{$q_3$};
		\draw
		(-19.4,9.5) node[circ]{}
		to (igbt4b.C);
		\draw
		(igbt4b.E) 
		to (igbt4a.C);
		\draw
		(igbt4a.E) to (-19.4,4.5) coordinate (leg4)node[circ]{};
		\draw
		(-17.9,9.5)
		to (igbt6b.C);
		\draw
		(igbt6b.E) 
		to (igbt6a.C);
		\draw
		(igbt6a.E) to (-17.9,4.5) coordinate (leg6);
		\draw 
		(leg6) -- (-21.6,4.5)
		(-21.6,9.5) --(-17.9,9.5)
		;
		\draw
		(-20.9,9.5) node[circ]{}
		to[C=$C_{\textrm{dc}}$]  (-20.9,4.5) node[circ]{}
		;
		\draw
		(-22.3,9.5) node [ocirc]{} node [left] {{$+V_{\textrm{gd}}$}}  -- (-21.6,9.5)
		(-22.3,4.5) node [ocirc]{} node [left] {{gnd}}  -- (-21.6,4.5);
		\coordinate (V4) at (-17.6,8.1);
	    \coordinate (V6) at (-17.6,6.2);
	    \draw
		(V6)  to (leg6 |- V6) node [circ] {};
		\draw
		(V4)   to[crossing]++(-0.6,0)to (leg4 |- V4) node [circ] {};
		\draw (-17,7.6)  node[nmos,bodydiode,yscale=-1](igbtCa){};
		\node[circ] at (-17,8.1){};
		\draw (-17,6.7)  node[nmos,bodydiode](igbtCb){};
		\node[circ] at (-17,6.2){};
		\draw (-17,8.1)--(igbtCa.E);
		\draw (igbtCa.C)--(igbtCb.C);
		\draw (-17,6.2)--(igbtCb.E);
		\draw (igbtCa) ++ (-0.5,0.3) node{$q_5$};
		\draw (igbtCb) ++ (-0.5,0.3) node{$q_6$};
        \node [circ] at (-16.5,8.1){};
        \node [circ] at (-16.5,6.2){};
        \draw (-16.5,8.1) to[longL=$L_m$] (-16.5,6.2);
        \node at (-16,7.15){$L_m$};
        \node at (-15.5,8.2) {$n$};
		\node at (-14.35,5.8) (I2AE2){};
        \node at (-14.35,8.55) (I1AB2){};
        \node[left of = I2AE2] (GDPMaux){};
        \node[left of = I1AB2] (GDPPaux){};
        \node[above of = GDPMaux,node distance=0.4cm] (GDPM){};
        \node[below of = GDPPaux,node distance=0.6cm] (GDPP){};
        \draw (GDPP) to[longL] (GDPM);
        \node[circ] at (-15.25,7.75){};
		\draw (V6) -- ++(2.24,0);
		\draw (V4) -- ++(2.24,0);
		\draw (GDPP)++(0,-0.2) |- (V4);
		\draw (GDPM)++(0,0.2) |- (V6);
		\end{circuitikz}
	}
    \caption{Connection of an active bridge with MOSFETs as switches to the proposed gate driver. The dot above the windings indicates winding directions. For all windings with no dot, the direction does not matter. $L_M$ and $L_m$ denote magnetizing inductances for the MAB and the gate driver transformer respectively. Gate drain capacitances of the MOSFETs are not shown. Small case indices imply that the quantity belongs to the gate driver while capitalized indices represent components of the main circuit. $N$ and $n$ represent the turns number of the MAB and gate driver transformers respectively.}
    \label{fig:GD4}
\end{figure*}
\subsection{Proposed topology} 
To properly drive the MOSFETs of an active bridge, the gate driver topology must be able to produce the following switching states:
\begin{enumerate}
    \item a positive output voltage which indicates that the switches which are to be driven are closed
    \item a negative output voltage which shows that the switches to be driven by the gate driver are open. In case there are other switches connected to the gate driver transformer with the opposite turns direction, those switches will be closed when the negative voltage is applied.
    \item A third switching state that gives the possibility to save the energy which is stored in the gate-source capacitance of the MOSFETs.
    \item A switching state that sets the midpoint voltage to zero and allows energy exchange between the high power transformer and the drain-source capacitances of the MOSFETs to be switched to reduce switching losses.
\end{enumerate} 
Quantities that refer to the low-power gate driver will use small case letters while quantities that refer to the high-power MAB will be capitalized for the remainder of this paper. 
An example of a topology that satisfies those requirements is the clamped active bridge, shown in 
Fig. \ref{fig:GD3}. 
Using the auxiliary supply $V_{\textrm{gd}}$ as well as the switches $q_i$, $i \in \{1,2,3,4,5,6\}$, the following midpoint voltages can be created:
If the switches $q_2$ and $q_3$ are closed the midpoint voltage will be positive. 
In case the switches $q_1$ and $q_4$ are closed, the midpoint voltage will be negative.
Closing the switches $q_5$ and $q_6$ yields a zero midpoint voltage.
Opening all switches gives a floating midpoint. 
An overview of the switching states can be seen in table \ref{tab:GD}.
\begin{table}[!htbp]
\caption{Overview over the switching states of the gate driver and its implications to the midpoint voltage}
\label{tab:GD}
\centering
\begin{tabular}{l|lllllll}
         & $q_1$ & $q_2$ & $q_3$ & $q_4$ & $q_5$ & $q_6$ & $V_{\textrm{mid}}$   \\ \hline
High     & 0     & 1     & 1     & 0     & 0     & 0     & $+V_{\textrm{gd}}$ \\ \hline
Low      & 1     & 0     & 0     & 1     & 0     & 0     & $-V_{\textrm{gd}}$ \\ \hline
Zero     & 0     & 0     & 0     & 0     & 1     & 1     & 0                  \\ \hline
Floating & 0     & 0     & 0     & 0     & 0     & 0     & Varying      
\end{tabular}
\end{table}

\begin{figure*}[!htbp]
    	\ctikzset{bipoles/thickness=1}
	\ctikzset{bipoles/length=0.8cm}
	\ctikzset{tripoles/thyristor/height=.8}
	\ctikzset{tripoles/thyristor/width=1}
	\ctikzset{bipoles/diode/height=.375}
	\ctikzset{bipoles/diode/width=.3}
	\ctikzset{bipoles/capacitor/height=.375}
	\tikzstyle{block} = [draw,fill=white, rectangle, minimum height=1cm, minimum width=6em]
	\tikzstyle{sum} = [draw, fill=white, circle, node distance=1cm]
	\tikzstyle{pinstyle} = [pin edge={to-,thin,black}]
    \centering
	\subfloat[Operation with $Q_1$ and $Q_4$ closed.]{
	\resizebox{0.475\linewidth}{!}{
			\begin{circuitikz}[longL/.style = {european inductor, inductors/scale=0.75,
 inductors/width=1.6, inductors/coils=9}]
        \draw (-7.6,7.9)--(-7.45,7.9) node (T1C1) {} to[L] ++ (0,-1.2) node (T1C2) {}  --++ (-0.15,0);
		\node[above of = T1C1, node distance=0.3cm] {$N$};
		\node[circ] at (-9.0,7.9){};
		\node[circ] at (-9.0,6.7){};
		\draw[gray] (-9.0,7.9) to[L=$L_M$,color=gray] (-9.0,6.7);
		\coordinate (V4T) at (-10.3,7.9);
		\coordinate (V6T) at (-13.3,6.7);
		\draw (V4T) node[circ]{} to[short] (T1C1);
		\draw (T1C2)to[crossing]++(-5.7,0) to
		(V6T) node[circ]{};
		\draw(-13.3,6.1) node[nmos,bodydiode](Q4){};
		\draw(-13.3,8.1) node[nmos,bodydiode,color=gray](Q3){};
		\draw(-13.3,3.8) coordinate (leg5) node[circ]{} to(Q4.E);
		\draw (Q4.C) to (-13.3,6.7);
		\draw[gray] (-13.3,6.7) to (Q3.E);
		\draw[gray] (Q3.C)
		to(-13.3,10.2) node[circ]{};
		\draw(-10.3,4.7) node[nmos,bodydiode,color=gray](Q2){};
		\draw(-10.3,9.5) node[nmos,bodydiode](Q1){};
		\draw[gray](-10.3,3.8) coordinate (leg3) to (Q2.E);
		\draw[gray] (Q2.C) to (-10.3,7.9);
		\draw (-10.3,7.9) to (Q1.E);
		\draw (Q1.C) to (-10.3,10.2);
        \draw (Q1)++(-0.4,0.4)  node(igbt3a)  {$Q_{1}$};
        \draw (Q2)++(-0.4,0.4) node(igbt3b) {$Q_{2}$};
        \draw (Q3)++(-0.4,0.4)  node(igbt5a)  {$Q_{3}$};
        \draw (Q4)++(-0.4,0.4) node(igbt5b) {$Q_{4}$};
        \draw (Q1.E)++(0,-0.3) node[circ](Q1Aux1) {};
        \draw(Q1.B)++(0,-0.75) coordinate (Q1Aux2){} to[C=$C_{\textrm{GS}}$] ++ (0,0.75);
        \draw(Q1Aux1) to (Q1Aux2){};
        \draw (Q2.E)++(0,-0.3) node[circ,color=gray](Q2Aux1) {};
        \draw[gray](Q2.B)++(0,-0.75) coordinate (Q2Aux2){} to[C=$C_{\textrm{GS}}$,color=gray] ++ (0,0.75);
        \draw[gray](Q2Aux1) to (Q2Aux2){};
        \draw (Q3.E)++(0,-0.3) node[circ,color=gray](Q3Aux1) {};
        \draw[gray](Q3.B)++(0,-0.75) coordinate (Q3Aux2){} to[C=$C_{\textrm{GS}}$,color=gray] ++ (0,0.75);
        \draw[gray](Q3Aux1) to (Q3Aux2){};
        \draw (Q4.E)++(0,-0.3) node[circ](Q4Aux1) {};
        \draw(Q4.B)++(0,-0.75) coordinate (Q4Aux2){} to[C=$C_{\textrm{GS}}$] ++ (0,0.75);
        \draw(Q4Aux1) to (Q4Aux2){};
        \draw (Q1Aux1)++(0,1.25)node[circ]{} -- ++ (0.7,0) to[C=$C_{\textrm{DS}}$] ++ (0,-1.25)  -- ++ (-0.7,0);
        \draw[gray] (Q2Aux1)++(0,1.25)node[circ,color=gray]{} -- ++ (0.7,0) to[C=$C_{\textrm{DS}}$,color=gray] ++ (0,-1.25)  -- ++ (-0.7,0);
        \draw[gray] (Q3Aux1)++(0,1.25)node[circ,color=gray]{} -- ++ (0.7,0) to[C=$C_{\textrm{DS}}$,color=gray] ++ (0,-1.25)  -- ++ (-0.7,0);
        \draw (Q4Aux1)++(0,1.25)node[circ]{} -- ++ (0.7,0) to[C=$C_{\textrm{DS}}$] ++ (0,-1.25)  -- ++ (-0.7,0);
        \node[circ] at (Q1.B){};
        \node[circ,color=gray] at (Q2.B){};
        \node[circ,color=gray] at (Q3.B){};
        \node[circ] at (Q4.B){};
        \node[circ] at (Q1Aux2){};
        \node[circ,color=gray] at (Q2Aux2){};
        \node[circ,color=gray] at (Q3Aux2){};
        \node[circ] at (Q4Aux2){};
        \draw(Q1Aux2) --++ (-0.9,0) to[crossing] ++ (-3.1,0) coordinate(Wi1Aux1){};
        \draw[gray](Q2Aux2) --++ (-0.9,0) to[crossing,color=gray] ++ (-3.1,0) coordinate(Wi2Aux1){};
        \draw[gray](Q3Aux2) --++ (-1,0) coordinate(Wi3Aux1){};
        \draw(Q4Aux2) --++ (-1,0) coordinate(Wi4Aux1){};
        \draw(Q1.B) --++ (-0.9,0) to[crossing] ++ (-3.1,0) coordinate(Wi1Aux2){};
        \draw[gray](Q2.B) --++ (-0.9,0) to[crossing,color=gray] ++ (-3.1,0) coordinate(Wi2Aux2){};
        \draw[gray](Q3.B) --++ (-1,0) coordinate(Wi3Aux2){};
        \draw(Q4.B) --++ (-1,0) coordinate(Wi4Aux2){};
        \draw (Wi1Aux1) to[L] (Wi1Aux2);
        \draw[gray] (Wi2Aux1) to[L,color=gray] (Wi2Aux2);
        \draw[gray] (Wi3Aux1) to[L,color=gray] (Wi3Aux2);
        \draw (Wi4Aux1) to[L] (Wi4Aux2);
        \draw (Wi1Aux1)++(-0.1,0) node[circ] {};
        \draw (Wi2Aux2)++(-0.1,0) node[circ,color=gray] {};
        \draw (Wi3Aux2)++(-0.1,0) node[circ,color=gray] {};
        \draw (Wi4Aux1)++(-0.1,0) node[circ] {};
        \draw (Wi1Aux2)++(0.1,0.125) node {$n$};
        \draw (Wi2Aux2)++(0.1,0.125) node {$n$};
        \draw (Wi3Aux2)++(0.1,0.125) node {$n$};
        \draw (Wi4Aux2)++(0.1,0.125) node {$n$};
        \draw  (Wi1Aux2)++(-0.2,0) node (CAUX1U){};
        \draw  (Wi1Aux2)++(-0.25,0) node (CAUX2U){};
        \draw  (Wi2Aux1)++(-0.2,0) node (CAUX1D){};
        \draw  (Wi2Aux1)++(-0.25,0) node (CAUX2D){};
        \draw (CAUX1U) -- (CAUX1D);
        \draw (CAUX2U) -- (CAUX2D);
		\draw[gray]
		(leg3)	-- (leg5);
		\draw
		 (leg5) --++(-2.7,0)node[ocirc]{};
		\draw
		(-10.3,10.2) -- ++(-5.7,0) node[ocirc]{};
		\draw
		(-17.9,5.3)  node[color=gray,nmos,bodydiode](igbt6a){};
		\draw
		(-17.9,8.8)  node[nmos,bodydiode](igbt6b){};
		\draw
		(-19.4,5.3) node[nmos,bodydiode](igbt4a){};
		\draw
		(-19.4,8.8) node[color=gray,nmos,bodydiode](igbt4b){}; 
		\draw (igbt6a)++(-0.4,0.4) node{$q_2$};
		\draw (igbt6b)++(-0.4,0.4) node{$q_1$};
		\draw (igbt4a)++(-0.4,0.4) node{$q_4$};
		\draw (igbt4b)++(-0.4,0.4) node{$q_3$};
		\draw[gray]
		(-19.4,9.5) node[circ]{}
		to (igbt4b.C);
		\draw[gray]
		(igbt4b.E) to (-19.4,8.1);
		\draw (-19.4,8.1)
		to (igbt4a.C);
		\draw
		(igbt4a.E) to (-19.4,4.5) coordinate (leg4)node[circ]{};
		\draw
		(-17.9,9.5)
		to (igbt6b.C);
		\draw
		(igbt6b.E) to (-17.9,6.2);
		\draw[gray] (-17.9,6.2)
		to (igbt6a.C);
		\draw[gray]
		(igbt6a.E) to (-17.9,4.5) coordinate (leg6);
		\draw[gray] 
		(leg6) -- (-19.4,4.5); 
		\draw
		 (-19.4,4.5)--(-21.6,4.5);
		\draw
		(-21.6,9.5) --(-17.9,9.5)
		;
		\draw
		(-20.9,9.5) node[circ]{}
		to[C=$C_{\textrm{dc}}$]  (-20.9,4.5) node[circ]{}
		;
		\draw
		(-22.3,9.5) node [ocirc]{} node [left] {{$+V_{\textrm{gd}}$}}  -- (-21.6,9.5);
		\draw
		(-22.3,4.5) node [ocirc]{} node [left] {{gnd}}  -- (-21.6,4.5);
		\coordinate (V4) at (-17.6,8.1);
	    \coordinate (V6) at (-17.6,6.2);
	    \draw
		(V6)  to (leg6 |- V6) node [circ] {};
		\draw
		(V4)   to[crossing]++(-0.6,0)to (leg4 |- V4) node [circ] {};
		\draw (-17,7.6)  node[color=gray,nmos,bodydiode,yscale=-1](igbtCa){};
		\node[circ] at (-17,8.1){};
		\draw (-17,6.7)  node[color=gray,nmos,bodydiode](igbtCb){};
		\node[circ] at (-17,6.2){};
		\draw[color=gray] (-17,8.1)--(igbtCa.E);
		\draw[color=gray] (igbtCa.C)--(igbtCb.C);
		\draw[color=gray] (-17,6.2)--(igbtCb.E);
		\draw (igbtCa) ++ (-0.5,0.3) node{$q_5$};
		\draw (igbtCb) ++ (-0.5,0.3) node{$q_6$};
        \node [circ] at (-16.5,8.1){};
        \node [circ] at (-16.5,6.2){};
        \draw[gray] (-16.5,8.1) to[color=gray,longL=$L_m$] (-16.5,6.2);
        \node at (-16,7.15){$L_m$};
        \node at (-15.5,8.2) {$n$};
		\node at (-14.35,5.8) (I2AE2){};
        \node at (-14.35,8.55) (I1AB2){};
        \node[left of = I2AE2] (GDPMaux){};
        \node[left of = I1AB2] (GDPPaux){};
        \node[above of = GDPMaux,node distance=0.4cm] (GDPM){};
        \node[below of = GDPPaux,node distance=0.6cm] (GDPP){};
        \draw (GDPP) to[longL] (GDPM);
        \node[circ] at (-15.25,7.75){};
		\draw (V6) -- ++(2.24,0);
		\draw (V4) -- ++(2.24,0);
		\draw (GDPP)++(0,-0.2) |- (V4);
		\draw (GDPM)++(0,0.2) |- (V6);
		\end{circuitikz}
	}
	}
	\subfloat[Discharge Gate-source capacitors using the gate driver transformer.]{
	\resizebox{0.475\linewidth}{!}{
			\begin{circuitikz}[longL/.style = {european inductor, inductors/scale=0.75,
 inductors/width=1.6, inductors/coils=9}]
        \draw (-7.6,7.9)--(-7.45,7.9) node (T1C1) {} to[L] ++ (0,-1.2) node (T1C2) {}  --++ (-0.15,0);
		\node[above of = T1C1, node distance=0.3cm] {$N$};
		\node[circ] at (-9.0,7.9){};
		\node[circ] at (-9.0,6.7){};
		\draw[gray] (-9.0,7.9) to[L=$L_M$,color=gray] (-9.0,6.7);
		\coordinate (V4T) at (-10.3,7.9);
		\coordinate (V6T) at (-13.3,6.7);
		\draw (V4T) node[circ]{} to[short] (T1C1);
		\draw (T1C2)to[crossing]++(-5.7,0) to
		(V6T) node[circ]{};
		\draw(-13.3,6.1) node[nmos,bodydiode](Q4){};
		\draw(-13.3,8.1) node[nmos,bodydiode,color=gray](Q3){};
		\draw(-13.3,3.8) coordinate (leg5) node[circ]{} to(Q4.E);
		\draw (Q4.C) to (-13.3,6.7);
		\draw[gray] (-13.3,6.7) to (Q3.E);
		\draw[gray] (Q3.C)
		to(-13.3,10.2) node[circ]{};
		\draw(-10.3,4.7) node[nmos,bodydiode,color=gray](Q2){};
		\draw(-10.3,9.5) node[nmos,bodydiode](Q1){};
		\draw[gray](-10.3,3.8) coordinate (leg3) to (Q2.E);
		\draw[gray] (Q2.C) to (-10.3,7.9);
		\draw (-10.3,7.9) to (Q1.E);
		\draw (Q1.C) to (-10.3,10.2);
        \draw (Q1)++(-0.4,0.4)  node(igbt3a)  {$Q_{1}$};
        \draw (Q2)++(-0.4,0.4) node(igbt3b) {$Q_{2}$};
        \draw (Q3)++(-0.4,0.4)  node(igbt5a)  {$Q_{3}$};
        \draw (Q4)++(-0.4,0.4) node(igbt5b) {$Q_{4}$};
        \draw (Q1.E)++(0,-0.3) node[circ](Q1Aux1) {};
        \draw(Q1.B)++(0,-0.75) coordinate (Q1Aux2){} to[C=$C_{\textrm{GS}}$] ++ (0,0.75);
        \draw(Q1Aux1) to (Q1Aux2){};
        \draw (Q2.E)++(0,-0.3) node[circ,color=gray](Q2Aux1) {};
        \draw[gray](Q2.B)++(0,-0.75) coordinate (Q2Aux2){} to[C=$C_{\textrm{GS}}$,color=gray] ++ (0,0.75);
        \draw[gray](Q2Aux1) to (Q2Aux2){};
        \draw (Q3.E)++(0,-0.3) node[circ,color=gray](Q3Aux1) {};
        \draw[gray](Q3.B)++(0,-0.75) coordinate (Q3Aux2){} to[C=$C_{\textrm{GS}}$,color=gray] ++ (0,0.75);
        \draw[gray](Q3Aux1) to (Q3Aux2){};
        \draw (Q4.E)++(0,-0.3) node[circ](Q4Aux1) {};
        \draw(Q4.B)++(0,-0.75) coordinate (Q4Aux2){} to[C=$C_{\textrm{GS}}$] ++ (0,0.75);
        \draw(Q4Aux1) to (Q4Aux2){};
        \draw (Q1Aux1)++(0,1.25)node[circ]{} -- ++ (0.7,0) to[C=$C_{\textrm{DS}}$] ++ (0,-1.25)  -- ++ (-0.7,0);
        \draw[gray] (Q2Aux1)++(0,1.25)node[circ,color=gray]{} -- ++ (0.7,0) to[C=$C_{\textrm{DS}}$,color=gray] ++ (0,-1.25)  -- ++ (-0.7,0);
        \draw[gray] (Q3Aux1)++(0,1.25)node[circ,color=gray]{} -- ++ (0.7,0) to[C=$C_{\textrm{DS}}$,color=gray] ++ (0,-1.25)  -- ++ (-0.7,0);
        \draw (Q4Aux1)++(0,1.25)node[circ]{} -- ++ (0.7,0) to[C=$C_{\textrm{DS}}$] ++ (0,-1.25)  -- ++ (-0.7,0);
        \node[circ] at (Q1.B){};
        \node[circ,color=gray] at (Q2.B){};
        \node[circ,color=gray] at (Q3.B){};
        \node[circ] at (Q4.B){};
        \node[circ] at (Q1Aux2){};
        \node[circ,color=gray] at (Q2Aux2){};
        \node[circ,color=gray] at (Q3Aux2){};
        \node[circ] at (Q4Aux2){};
        \draw(Q1Aux2) --++ (-0.9,0) to[crossing] ++ (-3.1,0) coordinate(Wi1Aux1){};
        \draw[gray](Q2Aux2) --++ (-0.9,0) to[crossing,color=gray] ++ (-3.1,0) coordinate(Wi2Aux1){};
        \draw[gray](Q3Aux2) --++ (-1,0) coordinate(Wi3Aux1){};
        \draw(Q4Aux2) --++ (-1,0) coordinate(Wi4Aux1){};
        \draw(Q1.B) --++ (-0.9,0) to[crossing] ++ (-3.1,0) coordinate(Wi1Aux2){};
        \draw[gray](Q2.B) --++ (-0.9,0) to[crossing,color=gray] ++ (-3.1,0) coordinate(Wi2Aux2){};
        \draw[gray](Q3.B) --++ (-1,0) coordinate(Wi3Aux2){};
        \draw(Q4.B) --++ (-1,0) coordinate(Wi4Aux2){};
        \draw (Wi1Aux1) to[L] (Wi1Aux2);
        \draw[gray] (Wi2Aux1) to[L,color=gray] (Wi2Aux2);
        \draw[gray] (Wi3Aux1) to[L,color=gray] (Wi3Aux2);
        \draw (Wi4Aux1) to[L] (Wi4Aux2);
        \draw (Wi1Aux1)++(-0.1,0) node[circ] {};
        \draw (Wi2Aux2)++(-0.1,0) node[circ,color=gray] {};
        \draw (Wi3Aux2)++(-0.1,0) node[circ,color=gray] {};
        \draw (Wi4Aux1)++(-0.1,0) node[circ] {};
        \draw (Wi1Aux2)++(0.1,0.125) node {$n$};
        \draw (Wi2Aux2)++(0.1,0.125) node {$n$};
        \draw (Wi3Aux2)++(0.1,0.125) node {$n$};
        \draw (Wi4Aux2)++(0.1,0.125) node {$n$};
        \draw  (Wi1Aux2)++(-0.2,0) node (CAUX1U){};
        \draw  (Wi1Aux2)++(-0.25,0) node (CAUX2U){};
        \draw  (Wi2Aux1)++(-0.2,0) node (CAUX1D){};
        \draw  (Wi2Aux1)++(-0.25,0) node (CAUX2D){};
        \draw (CAUX1U) -- (CAUX1D);
        \draw (CAUX2U) -- (CAUX2D);
		\draw[gray]
		(leg3)	-- (leg5);
		\draw
		 (leg5) --++(-2.7,0)node[ocirc]{};
		\draw
		(-10.3,10.2) -- ++(-5.7,0) node[ocirc]{};
		\draw
		(-17.9,5.3)  node[color=gray,nmos,bodydiode](igbt6a){};
		\draw
		(-17.9,8.8)  node[color=gray,nmos,bodydiode](igbt6b){};
		\draw
		(-19.4,5.3) node[color=gray,nmos,bodydiode](igbt4a){};
		\draw
		(-19.4,8.8) node[color=gray,nmos,bodydiode](igbt4b){}; 
		\draw (igbt6a)++(-0.4,0.4) node{$q_2$};
		\draw (igbt6b)++(-0.4,0.4) node{$q_1$};
		\draw (igbt4a)++(-0.4,0.4) node{$q_4$};
		\draw (igbt4b)++(-0.4,0.4) node{$q_3$};
		\draw[gray]
		(-19.4,9.5) node[circ, color=gray]{}
		to (igbt4b.C);
		\draw[gray]
		(igbt4b.E) to (-19.4,8.1);
		\draw[gray] (-19.4,8.1)
		to (igbt4a.C);
		\draw[gray]
		(igbt4a.E) to (-19.4,4.5) coordinate (leg4)node[circ,color=gray]{};
		\draw[gray]
		(-17.9,9.5)
		to (igbt6b.C);
		\draw[gray]
		(igbt6b.E) to (-17.9,6.2);
		\draw[gray] (-17.9,6.2)
		to (igbt6a.C);
		\draw[gray]
		(igbt6a.E) to (-17.9,4.5) coordinate (leg6);
		\draw[gray] 
		(leg6) -- (-19.4,4.5); 
		\draw[gray] 
		 (-19.4,4.5)--(-21.6,4.5);
		\draw[gray] 
		(-21.6,9.5) --(-17.9,9.5)
		;
		\draw[gray] 
		(-20.9,9.5) node[circ,color=gray]{}
		to[C=$C_{\textrm{dc}}$,color=gray]  (-20.9,4.5) node[circ,color=gray]{}
		;
		\draw[gray] 
		(-22.3,9.5) node [ocirc,color=gray]{} node [left] {{$+V_{\textrm{gd}}$}}  -- (-21.6,9.5);
		\draw[gray] 
		(-22.3,4.5) node [ocirc,color=gray]{} node [left] {{gnd}}  -- (-21.6,4.5);
		\coordinate (V4) at (-17.6,8.1);
	    \coordinate (V6) at (-17.6,6.2);
	    \draw[gray] 
		(V6)  to (leg6 |- V6) node [circ,color=gray] {};
		\draw[gray] 
		(V4)   to[crossing]++(-0.6,0)to (leg4 |- V4) node [circ,color=gray] {};
		\draw (-17,7.6)  node[color=gray,nmos,bodydiode,yscale=-1](igbtCa){};
		\node[circ,color=gray] at (-17,8.1){};
		\draw (-17,6.7)  node[color=gray,nmos,bodydiode](igbtCb){};
		\node[circ,color=gray] at (-17,6.2){};
		\draw[color=gray] (-17,8.1)--(igbtCa.E);
		\draw[color=gray] (igbtCa.C)--(igbtCb.C);
		\draw[color=gray] (-17,6.2)--(igbtCb.E);
		\draw (igbtCa) ++ (-0.5,0.3) node{$q_5$};
		\draw (igbtCb) ++ (-0.5,0.3) node{$q_6$};
        \node [circ] at (-16.5,8.1){};
        \node [circ] at (-16.5,6.2){};
        \draw (-16.5,8.1) to[longL=$L_m$] (-16.5,6.2);
        \node at (-16,7.15){$L_m$};
        \node at (-15.5,8.2) {$n$};
		\node at (-14.35,5.8) (I2AE2){};
        \node at (-14.35,8.55) (I1AB2){};
        \node[left of = I2AE2] (GDPMaux){};
        \node[left of = I1AB2] (GDPPaux){};
        \node[above of = GDPMaux,node distance=0.4cm] (GDPM){};
        \node[below of = GDPPaux,node distance=0.6cm] (GDPP){};
        \draw (GDPP) to[longL] (GDPM);
        \node[circ] at (-15.25,7.75){};
		\draw[gray]  (V6) -- ++(2.24,0);
		\draw[gray]  (V4) -- ++(2.24,0);
		\draw  (GDPP)++(0,-0.2) |- (-16.5,8.1);
		\draw  (GDPM)++(0,0.2) |- (-16.5,6.2);
		\end{circuitikz}
	}
	}
	\\
	\subfloat[Discharge Drain-source capacitors via MAB transformer.]{
	\resizebox{0.475\linewidth}{!}{
			\begin{circuitikz}[longL/.style = {european inductor, inductors/scale=0.75,
 inductors/width=1.6, inductors/coils=9}]
        \draw (-7.6,7.9)--(-7.45,7.9) node (T1C1) {} to[L] ++ (0,-1.2) node (T1C2) {}  --++ (-0.15,0);
		\node[above of = T1C1, node distance=0.3cm] {$N$};
		\node[circ] at (-9.0,7.9){};
		\node[circ] at (-9.0,6.7){};
		\draw (-9.0,7.9) to[L=$L_M$] (-9.0,6.7);
		\coordinate (V4T) at (-10.3,7.9);
		\coordinate (V6T) at (-13.3,6.7);
		\draw (V4T) node[circ]{} to[short] (T1C1);
		\draw (T1C2)to[crossing]++(-5.7,0) to
		(V6T) node[circ]{};
		\draw(-13.3,6.1) node[nmos,bodydiode](Q4){};
		\draw(-13.3,8.1) node[nmos,bodydiode,color=gray](Q3){};
		\draw(-13.3,3.8) coordinate (leg5) node[circ]{} to(Q4.E);
		\draw (Q4.C) to (-13.3,6.7);
		\draw[gray] (-13.3,6.7) to (Q3.E);
		\draw[gray] (Q3.C)
		to(-13.3,10.2) node[circ]{};
		\draw(-10.3,4.7) node[nmos,bodydiode,color=gray](Q2){};
		\draw(-10.3,9.5) node[nmos,bodydiode](Q1){};
		\draw[gray](-10.3,3.8) coordinate (leg3) to (Q2.E);
		\draw[gray] (Q2.C) to (-10.3,7.9);
		\draw (-10.3,7.9) to (Q1.E);
		\draw (Q1.C) to (-10.3,10.2);
        \draw (Q1)++(-0.4,0.4)  node(igbt3a)  {$Q_{1}$};
        \draw (Q2)++(-0.4,0.4) node(igbt3b) {$Q_{2}$};
        \draw (Q3)++(-0.4,0.4)  node(igbt5a)  {$Q_{3}$};
        \draw (Q4)++(-0.4,0.4) node(igbt5b) {$Q_{4}$};
        \draw (Q1.E)++(0,-0.3) node[circ](Q1Aux1) {};
        \draw[gray](Q1.B)++(0,-0.75) coordinate (Q1Aux2){} to[C=$C_{\textrm{GS}}$,color=gray] ++ (0,0.75);
        \draw[gray](Q1Aux1) to (Q1Aux2){};
        \draw (Q2.E)++(0,-0.3) node[circ,color=gray](Q2Aux1) {};
        \draw[gray](Q2.B)++(0,-0.75) coordinate (Q2Aux2){} to[C=$C_{\textrm{GS}}$,color=gray] ++ (0,0.75);
        \draw[gray](Q2Aux1) to (Q2Aux2){};
        \draw (Q3.E)++(0,-0.3) node[circ,color=gray](Q3Aux1) {};
        \draw[gray](Q3.B)++(0,-0.75) coordinate (Q3Aux2){} to[C=$C_{\textrm{GS}}$,color=gray] ++ (0,0.75);
        \draw[gray](Q3Aux1) to (Q3Aux2){};
        \draw (Q4.E)++(0,-0.3) node[circ](Q4Aux1) {};
        \draw[gray](Q4.B)++(0,-0.75) coordinate (Q4Aux2){} to[C=$C_{\textrm{GS}}$,color=gray] ++ (0,0.75);
        \draw[gray](Q4Aux1) to (Q4Aux2){};
        \draw (Q1Aux1)++(0,1.25)node[circ]{} -- ++ (0.7,0) to[C=$C_{\textrm{DS}}$] ++ (0,-1.25)  -- ++ (-0.7,0);
        \draw[gray] (Q2Aux1)++(0,1.25)node[circ,color=gray]{} -- ++ (0.7,0) to[C=$C_{\textrm{DS}}$,color=gray] ++ (0,-1.25)  -- ++ (-0.7,0);
        \draw[gray] (Q3Aux1)++(0,1.25)node[circ,color=gray]{} -- ++ (0.7,0) to[C=$C_{\textrm{DS}}$,color=gray] ++ (0,-1.25)  -- ++ (-0.7,0);
        \draw (Q4Aux1)++(0,1.25)node[circ]{} -- ++ (0.7,0) to[C=$C_{\textrm{DS}}$] ++ (0,-1.25)  -- ++ (-0.7,0);
        \node[circ,color=gray] at (Q1.B){};
        \node[circ,color=gray] at (Q2.B){};
        \node[circ,color=gray] at (Q3.B){};
        \node[circ,color=gray] at (Q4.B){};
        \node[circ,color=gray] at (Q1Aux2){};
        \node[circ,color=gray] at (Q2Aux2){};
        \node[circ,color=gray] at (Q3Aux2){};
        \node[circ,color=gray] at (Q4Aux2){};
        \draw[gray](Q1Aux2) --++ (-0.9,0) to[crossing,color=gray] ++ (-3.1,0) coordinate(Wi1Aux1){};
        \draw[gray](Q2Aux2) --++ (-0.9,0) to[crossing,color=gray] ++ (-3.1,0) coordinate(Wi2Aux1){};
        \draw[gray](Q3Aux2) --++ (-1,0) coordinate(Wi3Aux1){};
        \draw[gray](Q4Aux2) --++ (-1,0) coordinate(Wi4Aux1){};
        \draw[gray](Q1.B) --++ (-0.9,0) to[crossing,color=gray] ++ (-3.1,0) coordinate(Wi1Aux2){};
        \draw[gray](Q2.B) --++ (-0.9,0) to[crossing,color=gray] ++ (-3.1,0) coordinate(Wi2Aux2){};
        \draw[gray](Q3.B) --++ (-1,0) coordinate(Wi3Aux2){};
        \draw[gray](Q4.B) --++ (-1,0) coordinate(Wi4Aux2){};
        \draw[gray] (Wi1Aux1) to[L,color=gray] (Wi1Aux2);
        \draw[gray] (Wi2Aux1) to[L,color=gray] (Wi2Aux2);
        \draw[gray] (Wi3Aux1) to[L,color=gray] (Wi3Aux2);
        \draw[gray] (Wi4Aux1) to[L,color=gray] (Wi4Aux2);
        \draw (Wi1Aux1)++(-0.1,0) node[circ,color=gray] {};
        \draw (Wi2Aux2)++(-0.1,0) node[circ,color=gray] {};
        \draw (Wi3Aux2)++(-0.1,0) node[circ,color=gray] {};
        \draw (Wi4Aux1)++(-0.1,0) node[circ,color=gray] {};
        \draw (Wi1Aux2)++(0.1,0.125) node {$n$};
        \draw (Wi2Aux2)++(0.1,0.125) node {$n$};
        \draw (Wi3Aux2)++(0.1,0.125) node {$n$};
        \draw (Wi4Aux2)++(0.1,0.125) node {$n$};
        \draw  (Wi1Aux2)++(-0.2,0) node (CAUX1U){};
        \draw  (Wi1Aux2)++(-0.25,0) node (CAUX2U){};
        \draw  (Wi2Aux1)++(-0.2,0) node (CAUX1D){};
        \draw  (Wi2Aux1)++(-0.25,0) node (CAUX2D){};
        \draw (CAUX1U) -- (CAUX1D);
        \draw (CAUX2U) -- (CAUX2D);
		\draw[gray]
		(leg3)	-- (leg5);
		\draw
		 (leg5) --++(-2.7,0)node[ocirc]{};
		\draw
		(-10.3,10.2) -- ++(-5.7,0) node[ocirc]{};
		\draw
		(-17.9,5.3)  node[color=gray,nmos,bodydiode](igbt6a){};
		\draw
		(-17.9,8.8)  node[color=gray,nmos,bodydiode](igbt6b){};
		\draw
		(-19.4,5.3) node[color=gray,nmos,bodydiode](igbt4a){};
		\draw
		(-19.4,8.8) node[color=gray,nmos,bodydiode](igbt4b){}; 
		\draw (igbt6a)++(-0.4,0.4) node{$q_2$};
		\draw (igbt6b)++(-0.4,0.4) node{$q_1$};
		\draw (igbt4a)++(-0.4,0.4) node{$q_4$};
		\draw (igbt4b)++(-0.4,0.4) node{$q_3$};
		\draw[gray]
		(-19.4,9.5) node[circ, color=gray]{}
		to (igbt4b.C);
		\draw[gray]
		(igbt4b.E) to (-19.4,8.1);
		\draw[gray] (-19.4,8.1)
		to (igbt4a.C);
		\draw[gray]
		(igbt4a.E) to (-19.4,4.5) coordinate (leg4)node[circ,color=gray]{};
		\draw[gray]
		(-17.9,9.5)
		to (igbt6b.C);
		\draw[gray]
		(igbt6b.E) to (-17.9,6.2);
		\draw[gray] (-17.9,6.2)
		to (igbt6a.C);
		\draw[gray]
		(igbt6a.E) to (-17.9,4.5) coordinate (leg6);
		\draw[gray] 
		(leg6) -- (-19.4,4.5); 
		\draw[gray] 
		 (-19.4,4.5)--(-21.6,4.5);
		\draw[gray] 
		(-21.6,9.5) --(-17.9,9.5)
		;
		\draw[gray] 
		(-20.9,9.5) node[circ,color=gray]{}
		to[C=$C_{\textrm{dc}}$,color=gray]  (-20.9,4.5) node[circ,color=gray]{}
		;
		\draw[gray] 
		(-22.3,9.5) node [ocirc,color=gray]{} node [left] {{$+V_{\textrm{gd}}$}}  -- (-21.6,9.5);
		\draw[gray] 
		(-22.3,4.5) node [ocirc,color=gray]{} node [left] {{gnd}}  -- (-21.6,4.5);
		\coordinate (V4) at (-17.6,8.1);
	    \coordinate (V6) at (-17.6,6.2);
	    \draw[gray] 
		(V6)  to (leg6 |- V6) node [circ,color=gray] {};
		\draw[gray] 
		(V4)   to[crossing]++(-0.6,0)to (leg4 |- V4) node [circ,color=gray] {};
		\draw (-17,7.6)  node[nmos,bodydiode,yscale=-1](igbtCa){};
		\node[circ] at (-17,8.1){};
		\draw (-17,6.7)  node[nmos,bodydiode](igbtCb){};
		\node[circ] at (-17,6.2){};
		\draw (-17,8.1)--(igbtCa.E);
		\draw (igbtCa.C)--(igbtCb.C);
		\draw (-17,6.2)--(igbtCb.E);
		\draw (igbtCa) ++ (-0.5,0.3) node{$q_5$};
		\draw (igbtCb) ++ (-0.5,0.3) node{$q_6$};
        \node [circ] at (-16.5,8.1){};
        \node [circ] at (-16.5,6.2){};
        \draw (-16.5,8.1) to[longL=$L_m$] (-16.5,6.2);
        \node at (-16,7.15){$L_m$};
        \node at (-15.5,8.2) {$n$};
		\node at (-14.35,5.8) (I2AE2){};
        \node at (-14.35,8.55) (I1AB2){};
        \node[left of = I2AE2] (GDPMaux){};
        \node[left of = I1AB2] (GDPPaux){};
        \node[above of = GDPMaux,node distance=0.4cm] (GDPM){};
        \node[below of = GDPPaux,node distance=0.6cm] (GDPP){};
        \draw (GDPP) to[longL] (GDPM);
        \node[circ] at (-15.25,7.75){};
		\draw[gray]  (V6) -- ++(2.24,0);
		\draw[gray]  (V4) -- ++(2.24,0);
		\draw  (-16.5,8.1) -- (-17,8.1);
		\draw  (-16.5,6.2)  -- (-17,6.2);
		\draw  (GDPP)++(0,-0.2) |- (-16.5,8.1);
		\draw  (GDPM)++(0,0.2) |- (-16.5,6.2);
		\end{circuitikz}
	}
	}
	\subfloat[Charge Gate-source capacitors using the MAB transformer.]{
	\resizebox{0.475\linewidth}{!}{
			\begin{circuitikz}[longL/.style = {european inductor, inductors/scale=0.75,
 inductors/width=1.6, inductors/coils=9}]
        \draw (-7.6,7.9)--(-7.45,7.9) node (T1C1) {} to[L] ++ (0,-1.2) node (T1C2) {}  --++ (-0.15,0);
		\node[above of = T1C1, node distance=0.3cm] {$N$};
		\node[circ] at (-9.0,7.9){};
		\node[circ] at (-9.0,6.7){};
		\draw (-9.0,7.9) to[L=$L_M$] (-9.0,6.7);
		\coordinate (V4T) at (-10.3,7.9);
		\coordinate (V6T) at (-13.3,6.7);
		\draw (V4T) node[circ]{} to[short] (T1C1);
		\draw (T1C2)to[crossing]++(-5.7,0) to
		(V6T) node[circ]{};
		\draw(-13.3,6.1) node[nmos,bodydiode,color=gray](Q4){};
		\draw(-13.3,8.1) node[nmos,bodydiode](Q3){};
		\draw[gray](-13.3,3.8) coordinate (leg5) node[circ]{} to(Q4.E);
		\draw[gray] (Q4.C) to (-13.3,6.7);
		\draw (-13.3,6.7) to (Q3.E);
		\draw (Q3.C)
		to(-13.3,10.2) node[circ]{};
		\draw(-10.3,4.7) node[nmos,bodydiode](Q2){};
		\draw(-10.3,9.5) node[nmos,bodydiode,color=gray](Q1){};
		\draw(-10.3,3.8) coordinate (leg3) to (Q2.E);
		\draw (Q2.C) to (-10.3,7.9);
		\draw[gray] (-10.3,7.9) to (Q1.E);
		\draw[gray] (Q1.C) to (-10.3,10.2);
        \draw (Q1)++(-0.4,0.4)  node(igbt3a)  {$Q_{1}$};
        \draw (Q2)++(-0.4,0.4) node(igbt3b) {$Q_{2}$};
        \draw (Q3)++(-0.4,0.4)  node(igbt5a)  {$Q_{3}$};
        \draw (Q4)++(-0.4,0.4) node(igbt5b) {$Q_{4}$};
        \draw (Q1.E)++(0,-0.3) node[circ,color=gray](Q1Aux1) {};
        \draw[gray](Q1.B)++(0,-0.75) coordinate (Q1Aux2){} to[C=$C_{\textrm{GS}}$,color=gray] ++ (0,0.75);
        \draw[gray](Q1Aux1) to (Q1Aux2){};
        \draw (Q2.E)++(0,-0.3) node[circ](Q2Aux1) {};
        \draw[gray](Q2.B)++(0,-0.75) coordinate (Q2Aux2){} to[C=$C_{\textrm{GS}}$,color=gray] ++ (0,0.75);
        \draw[gray](Q2Aux1) to (Q2Aux2){};
        \draw (Q3.E)++(0,-0.3) node[circ](Q3Aux1) {};
        \draw[gray](Q3.B)++(0,-0.75) coordinate (Q3Aux2){} to[C=$C_{\textrm{GS}}$,color=gray] ++ (0,0.75);
        \draw[gray](Q3Aux1) to (Q3Aux2){};
        \draw (Q4.E)++(0,-0.3) node[circ,color=gray](Q4Aux1) {};
        \draw[gray](Q4.B)++(0,-0.75) coordinate (Q4Aux2){} to[C=$C_{\textrm{GS}}$,color=gray] ++ (0,0.75);
        \draw[gray](Q4Aux1) to (Q4Aux2){};
        \draw[gray] (Q1Aux1)++(0,1.25)node[circ,color=gray]{} -- ++ (0.7,0) to[C=$C_{\textrm{DS}}$,color=gray] ++ (0,-1.25)  -- ++ (-0.7,0);
        \draw (Q2Aux1)++(0,1.25)node[circ]{} -- ++ (0.7,0) to[C=$C_{\textrm{DS}}$] ++ (0,-1.25)  -- ++ (-0.7,0);
        \draw (Q3Aux1)++(0,1.25)node[circ]{} -- ++ (0.7,0) to[C=$C_{\textrm{DS}}$] ++ (0,-1.25)  -- ++ (-0.7,0);
        \draw[gray] (Q4Aux1)++(0,1.25)node[circ,color=gray]{} -- ++ (0.7,0) to[C=$C_{\textrm{DS}}$,color=gray] ++ (0,-1.25)  -- ++ (-0.7,0);
        \node[circ,color=gray] at (Q1.B){};
        \node[circ,color=gray] at (Q2.B){};
        \node[circ,color=gray] at (Q3.B){};
        \node[circ,color=gray] at (Q4.B){};
        \node[circ,color=gray] at (Q1Aux2){};
        \node[circ,color=gray] at (Q2Aux2){};
        \node[circ,color=gray] at (Q3Aux2){};
        \node[circ,color=gray] at (Q4Aux2){};
        \draw[gray](Q1Aux2) --++ (-0.9,0) to[crossing,color=gray] ++ (-3.1,0) coordinate(Wi1Aux1){};
        \draw[gray](Q2Aux2) --++ (-0.9,0) to[crossing,color=gray] ++ (-3.1,0) coordinate(Wi2Aux1){};
        \draw[gray](Q3Aux2) --++ (-1,0) coordinate(Wi3Aux1){};
        \draw[gray](Q4Aux2) --++ (-1,0) coordinate(Wi4Aux1){};
        \draw[gray](Q1.B) --++ (-0.9,0) to[crossing,color=gray] ++ (-3.1,0) coordinate(Wi1Aux2){};
        \draw[gray](Q2.B) --++ (-0.9,0) to[crossing,color=gray] ++ (-3.1,0) coordinate(Wi2Aux2){};
        \draw[gray](Q3.B) --++ (-1,0) coordinate(Wi3Aux2){};
        \draw[gray](Q4.B) --++ (-1,0) coordinate(Wi4Aux2){};
        \draw[gray] (Wi1Aux1) to[L,color=gray] (Wi1Aux2);
        \draw[gray] (Wi2Aux1) to[L,color=gray] (Wi2Aux2);
        \draw[gray] (Wi3Aux1) to[L,color=gray] (Wi3Aux2);
        \draw[gray] (Wi4Aux1) to[L,color=gray] (Wi4Aux2);
        \draw (Wi1Aux1)++(-0.1,0) node[circ,color=gray] {};
        \draw (Wi2Aux2)++(-0.1,0) node[circ,color=gray] {};
        \draw (Wi3Aux2)++(-0.1,0) node[circ,color=gray] {};
        \draw (Wi4Aux1)++(-0.1,0) node[circ,color=gray] {};
        \draw (Wi1Aux2)++(0.1,0.125) node {$n$};
        \draw (Wi2Aux2)++(0.1,0.125) node {$n$};
        \draw (Wi3Aux2)++(0.1,0.125) node {$n$};
        \draw (Wi4Aux2)++(0.1,0.125) node {$n$};
        \draw  (Wi1Aux2)++(-0.2,0) node (CAUX1U){};
        \draw  (Wi1Aux2)++(-0.25,0) node (CAUX2U){};
        \draw  (Wi2Aux1)++(-0.2,0) node (CAUX1D){};
        \draw  (Wi2Aux1)++(-0.25,0) node (CAUX2D){};
        \draw (CAUX1U) -- (CAUX1D);
        \draw (CAUX2U) -- (CAUX2D);
		\draw
		(leg3)	-- (leg5);
		\draw
		 (leg5) --++(-2.7,0)node[ocirc]{};
		\draw[gray]
		(-10.3,10.2) -- (-13.3,10.2);
		\draw
		(-13.3,10.2) --++
		(-2.7,0) node[ocirc]{};
		\draw
		(-17.9,5.3)  node[color=gray,nmos,bodydiode](igbt6a){};
		\draw
		(-17.9,8.8)  node[color=gray,nmos,bodydiode](igbt6b){};
		\draw
		(-19.4,5.3) node[color=gray,nmos,bodydiode](igbt4a){};
		\draw
		(-19.4,8.8) node[color=gray,nmos,bodydiode](igbt4b){}; 
		\draw (igbt6a)++(-0.4,0.4) node{$q_2$};
		\draw (igbt6b)++(-0.4,0.4) node{$q_1$};
		\draw (igbt4a)++(-0.4,0.4) node{$q_4$};
		\draw (igbt4b)++(-0.4,0.4) node{$q_3$};
		\draw[gray]
		(-19.4,9.5) node[circ, color=gray]{}
		to (igbt4b.C);
		\draw[gray]
		(igbt4b.E) to (-19.4,8.1);
		\draw[gray] (-19.4,8.1)
		to (igbt4a.C);
		\draw[gray]
		(igbt4a.E) to (-19.4,4.5) coordinate (leg4)node[circ,color=gray]{};
		\draw[gray]
		(-17.9,9.5)
		to (igbt6b.C);
		\draw[gray]
		(igbt6b.E) to (-17.9,6.2);
		\draw[gray] (-17.9,6.2)
		to (igbt6a.C);
		\draw[gray]
		(igbt6a.E) to (-17.9,4.5) coordinate (leg6);
		\draw[gray] 
		(leg6) -- (-19.4,4.5); 
		\draw[gray] 
		 (-19.4,4.5)--(-21.6,4.5);
		\draw[gray] 
		(-21.6,9.5) --(-17.9,9.5)
		;
		\draw[gray] 
		(-20.9,9.5) node[circ,color=gray]{}
		to[C=$C_{\textrm{dc}}$,color=gray]  (-20.9,4.5) node[circ,color=gray]{}
		;
		\draw[gray] 
		(-22.3,9.5) node [ocirc,color=gray]{} node [left] {{$+V_{\textrm{gd}}$}}  -- (-21.6,9.5);
		\draw[gray] 
		(-22.3,4.5) node [ocirc,color=gray]{} node [left] {{gnd}}  -- (-21.6,4.5);
		\coordinate (V4) at (-17.6,8.1);
	    \coordinate (V6) at (-17.6,6.2);
	    \draw[gray] 
		(V6)  to (leg6 |- V6) node [circ,color=gray] {};
		\draw[gray] 
		(V4)   to[crossing]++(-0.6,0)to (leg4 |- V4) node [circ,color=gray] {};
		\draw (-17,7.6)  node[nmos,bodydiode,yscale=-1](igbtCa){};
		\node[circ] at (-17,8.1){};
		\draw (-17,6.7)  node[nmos,bodydiode](igbtCb){};
		\node[circ] at (-17,6.2){};
		\draw (-17,8.1)--(igbtCa.E);
		\draw (igbtCa.C)--(igbtCb.C);
		\draw (-17,6.2)--(igbtCb.E);
		\draw (igbtCa) ++ (-0.5,0.3) node{$q_5$};
		\draw (igbtCb) ++ (-0.5,0.3) node{$q_6$};
        \node [circ] at (-16.5,8.1){};
        \node [circ] at (-16.5,6.2){};
        \draw (-16.5,8.1) to[longL=$L_m$] (-16.5,6.2);
        \node at (-16,7.15){$L_m$};
        \node at (-15.5,8.2) {$n$};
		\node at (-14.35,5.8) (I2AE2){};
        \node at (-14.35,8.55) (I1AB2){};
        \node[left of = I2AE2] (GDPMaux){};
        \node[left of = I1AB2] (GDPPaux){};
        \node[above of = GDPMaux,node distance=0.4cm] (GDPM){};
        \node[below of = GDPPaux,node distance=0.6cm] (GDPP){};
        \draw (GDPP) to[longL] (GDPM);
        \node[circ] at (-15.25,7.75){};
		\draw[gray]  (V6) -- ++(2.24,0);
		\draw[gray]  (V4) -- ++(2.24,0);
		\draw  (-16.5,8.1) -- (-17,8.1);
		\draw  (-16.5,6.2)  -- (-17,6.2);
		\draw  (GDPP)++(0,-0.2) |- (-16.5,8.1);
		\draw  (GDPM)++(0,0.2) |- (-16.5,6.2);
		\end{circuitikz}
	}
	}
	\\
	\subfloat[Charge Gate-source capacitors using the gate driver transformer.]{
	\resizebox{0.475\linewidth}{!}{
		\begin{circuitikz}[longL/.style = {european inductor, inductors/scale=0.75,
 inductors/width=1.6, inductors/coils=9}]
        \draw (-7.6,7.9)--(-7.45,7.9) node (T1C1) {} to[L] ++ (0,-1.2) node (T1C2) {}  --++ (-0.15,0);
		\node[above of = T1C1, node distance=0.3cm] {$N$};
		\node[circ] at (-9.0,7.9){};
		\node[circ] at (-9.0,6.7){};
		\draw[gray] (-9.0,7.9) to[L=$L_M$,color=gray] (-9.0,6.7);
		\coordinate (V4T) at (-10.3,7.9);
		\coordinate (V6T) at (-13.3,6.7);
		\draw (V4T) node[circ]{} to[short] (T1C1);
		\draw (T1C2)to[crossing]++(-5.7,0) to
		(V6T) node[circ]{};
		\draw(-13.3,6.1) node[nmos,bodydiode,color=gray](Q4){};
		\draw(-13.3,8.1) node[nmos,bodydiode](Q3){};
		\draw[gray](-13.3,3.8) coordinate (leg5) node[circ]{} to(Q4.E);
		\draw[gray] (Q4.C) to (-13.3,6.7);
		\draw (-13.3,6.7) to (Q3.E);
		\draw (Q3.C)
		to(-13.3,10.2) node[circ]{};
		\draw(-10.3,4.7) node[nmos,bodydiode](Q2){};
		\draw(-10.3,9.5) node[nmos,bodydiode,color=gray](Q1){};
		\draw(-10.3,3.8) coordinate (leg3) to (Q2.E);
		\draw (Q2.C) to (-10.3,7.9);
		\draw[gray] (-10.3,7.9) to (Q1.E);
		\draw[gray] (Q1.C) to (-10.3,10.2);
        \draw (Q1)++(-0.4,0.4)  node(igbt3a)  {$Q_{1}$};
        \draw (Q2)++(-0.4,0.4) node(igbt3b) {$Q_{2}$};
        \draw (Q3)++(-0.4,0.4)  node(igbt5a)  {$Q_{3}$};
        \draw (Q4)++(-0.4,0.4) node(igbt5b) {$Q_{4}$};
        \draw (Q1.E)++(0,-0.3) node[circ,color=gray](Q1Aux1) {};
        \draw[gray](Q1.B)++(0,-0.75) coordinate (Q1Aux2){} to[C=$C_{\textrm{GS}}$,color=gray] ++ (0,0.75);
        \draw[gray](Q1Aux1) to (Q1Aux2){};
        \draw (Q2.E)++(0,-0.3) node[circ](Q2Aux1) {};
        \draw(Q2.B)++(0,-0.75) coordinate (Q2Aux2){} to[C=$C_{\textrm{GS}}$] ++ (0,0.75);
        \draw(Q2Aux1) to (Q2Aux2){};
        \draw (Q3.E)++(0,-0.3) node[circ](Q3Aux1) {};
        \draw(Q3.B)++(0,-0.75) coordinate (Q3Aux2){} to[C=$C_{\textrm{GS}}$] ++ (0,0.75);
        \draw(Q3Aux1) to (Q3Aux2){};
        \draw (Q4.E)++(0,-0.3) node[circ,color=gray](Q4Aux1) {};
        \draw[gray](Q4.B)++(0,-0.75) coordinate (Q4Aux2){} to[C=$C_{\textrm{GS}}$,color=gray] ++ (0,0.75);
        \draw[gray](Q4Aux1) to (Q4Aux2){};
        \draw[gray] (Q1Aux1)++(0,1.25)node[circ,color=gray]{} -- ++ (0.7,0) to[C=$C_{\textrm{DS}}$,color=gray] ++ (0,-1.25)  -- ++ (-0.7,0);
        \draw (Q2Aux1)++(0,1.25)node[circ]{} -- ++ (0.7,0) to[C=$C_{\textrm{DS}}$] ++ (0,-1.25)  -- ++ (-0.7,0);
        \draw (Q3Aux1)++(0,1.25)node[circ]{} -- ++ (0.7,0) to[C=$C_{\textrm{DS}}$] ++ (0,-1.25)  -- ++ (-0.7,0);
        \draw[gray] (Q4Aux1)++(0,1.25)node[circ,color=gray]{} -- ++ (0.7,0) to[C=$C_{\textrm{DS}}$,color=gray] ++ (0,-1.25)  -- ++ (-0.7,0);
        \node[circ,color=gray] at (Q1.B){};
        \node[circ] at (Q2.B){};
        \node[circ] at (Q3.B){};
        \node[circ,color=gray] at (Q4.B){};
        \node[circ,color=gray] at (Q1Aux2){};
        \node[circ] at (Q2Aux2){};
        \node[circ] at (Q3Aux2){};
        \node[circ,color=gray] at (Q4Aux2){};
        \draw[gray](Q1Aux2) --++ (-0.9,0) to[crossing,color=gray] ++ (-3.1,0) coordinate(Wi1Aux1){};
        \draw(Q2Aux2) --++ (-0.9,0) to[crossing] ++ (-3.1,0) coordinate(Wi2Aux1){};
        \draw(Q3Aux2) --++ (-1,0) coordinate(Wi3Aux1){};
        \draw[gray](Q4Aux2) --++ (-1,0) coordinate(Wi4Aux1){};
        \draw[gray](Q1.B) --++ (-0.9,0) to[crossing,color=gray] ++ (-3.1,0) coordinate(Wi1Aux2){};
        \draw(Q2.B) --++ (-0.9,0) to[crossing] ++ (-3.1,0) coordinate(Wi2Aux2){};
        \draw(Q3.B) --++ (-1,0) coordinate(Wi3Aux2){};
        \draw[gray](Q4.B) --++ (-1,0) coordinate(Wi4Aux2){};
        \draw[gray] (Wi1Aux1) to[L,color=gray] (Wi1Aux2);
        \draw (Wi2Aux1) to[L] (Wi2Aux2);
        \draw (Wi3Aux1) to[L] (Wi3Aux2);
        \draw[gray] (Wi4Aux1) to[L,color=gray] (Wi4Aux2);
        \draw (Wi1Aux1)++(-0.1,0) node[circ,color=gray] {};
        \draw (Wi2Aux2)++(-0.1,0) node[circ] {};
        \draw (Wi3Aux2)++(-0.1,0) node[circ] {};
        \draw (Wi4Aux1)++(-0.1,0) node[circ,color=gray] {};
        \draw (Wi1Aux2)++(0.1,0.125) node {$n$};
        \draw (Wi2Aux2)++(0.1,0.125) node {$n$};
        \draw (Wi3Aux2)++(0.1,0.125) node {$n$};
        \draw (Wi4Aux2)++(0.1,0.125) node {$n$};
        \draw  (Wi1Aux2)++(-0.2,0) node (CAUX1U){};
        \draw  (Wi1Aux2)++(-0.25,0) node (CAUX2U){};
        \draw  (Wi2Aux1)++(-0.2,0) node (CAUX1D){};
        \draw  (Wi2Aux1)++(-0.25,0) node (CAUX2D){};
        \draw (CAUX1U) -- (CAUX1D);
        \draw (CAUX2U) -- (CAUX2D);
		\draw
		(leg3)	-- (leg5);
		\draw
		 (leg5) --++(-2.7,0)node[ocirc]{};
		\draw[gray]
		(-10.3,10.2) -- (-13.3,10.2);
		\draw 
		(-13.3,10.2)--++(-2.7,0) node[ocirc]{};
		\draw
		(-17.9,5.3)  node[color=gray,nmos,bodydiode](igbt6a){};
		\draw
		(-17.9,8.8)  node[color=gray,nmos,bodydiode](igbt6b){};
		\draw
		(-19.4,5.3) node[color=gray,nmos,bodydiode](igbt4a){};
		\draw
		(-19.4,8.8) node[color=gray,nmos,bodydiode](igbt4b){}; 
		\draw (igbt6a)++(-0.4,0.4) node{$q_2$};
		\draw (igbt6b)++(-0.4,0.4) node{$q_1$};
		\draw (igbt4a)++(-0.4,0.4) node{$q_4$};
		\draw (igbt4b)++(-0.4,0.4) node{$q_3$};
		\draw[gray]
		(-19.4,9.5) node[circ, color=gray]{}
		to (igbt4b.C);
		\draw[gray]
		(igbt4b.E) to (-19.4,8.1);
		\draw[gray] (-19.4,8.1)
		to (igbt4a.C);
		\draw[gray]
		(igbt4a.E) to (-19.4,4.5) coordinate (leg4)node[circ,color=gray]{};
		\draw[gray]
		(-17.9,9.5)
		to (igbt6b.C);
		\draw[gray]
		(igbt6b.E) to (-17.9,6.2);
		\draw[gray] (-17.9,6.2)
		to (igbt6a.C);
		\draw[gray]
		(igbt6a.E) to (-17.9,4.5) coordinate (leg6);
		\draw[gray] 
		(leg6) -- (-19.4,4.5); 
		\draw[gray] 
		 (-19.4,4.5)--(-21.6,4.5);
		\draw[gray] 
		(-21.6,9.5) --(-17.9,9.5)
		;
		\draw[gray] 
		(-20.9,9.5) node[circ,color=gray]{}
		to[C=$C_{\textrm{dc}}$,color=gray]  (-20.9,4.5) node[circ,color=gray]{}
		;
		\draw[gray] 
		(-22.3,9.5) node [ocirc,color=gray]{} node [left] {{$+V_{\textrm{gd}}$}}  -- (-21.6,9.5);
		\draw[gray] 
		(-22.3,4.5) node [ocirc,color=gray]{} node [left] {{gnd}}  -- (-21.6,4.5);
		\coordinate (V4) at (-17.6,8.1);
	    \coordinate (V6) at (-17.6,6.2);
	    \draw[gray] 
		(V6)  to (leg6 |- V6) node [circ,color=gray] {};
		\draw[gray] 
		(V4)   to[crossing]++(-0.6,0)to (leg4 |- V4) node [circ,color=gray] {};
		\draw (-17,7.6)  node[color=gray,nmos,bodydiode,yscale=-1](igbtCa){};
		\node[circ,color=gray] at (-17,8.1){};
		\draw (-17,6.7)  node[color=gray,nmos,bodydiode](igbtCb){};
		\node[circ,color=gray] at (-17,6.2){};
		\draw[color=gray] (-17,8.1)--(igbtCa.E);
		\draw[color=gray] (igbtCa.C)--(igbtCb.C);
		\draw[color=gray] (-17,6.2)--(igbtCb.E);
		\draw (igbtCa) ++ (-0.5,0.3) node{$q_5$};
		\draw (igbtCb) ++ (-0.5,0.3) node{$q_6$};
        \node [circ] at (-16.5,8.1){};
        \node [circ] at (-16.5,6.2){};
        \draw (-16.5,8.1) to[longL=$L_m$] (-16.5,6.2);
        \node at (-16,7.15){$L_m$};
        \node at (-15.5,8.2) {$n$};
		\node at (-14.35,5.8) (I2AE2){};
        \node at (-14.35,8.55) (I1AB2){};
        \node[left of = I2AE2] (GDPMaux){};
        \node[left of = I1AB2] (GDPPaux){};
        \node[above of = GDPMaux,node distance=0.4cm] (GDPM){};
        \node[below of = GDPPaux,node distance=0.6cm] (GDPP){};
        \draw (GDPP) to[longL] (GDPM);
        \node[circ] at (-15.25,7.75){};
		\draw[gray]  (V6) -- ++(2.24,0);
		\draw[gray]  (V4) -- ++(2.24,0);
		\draw  (GDPP)++(0,-0.2) |- (-16.5,8.1);
		\draw  (GDPM)++(0,0.2) |- (-16.5,6.2);
		\end{circuitikz}
	}
	}
	\subfloat[Operation with $Q_2$ and $Q_3$ closed.]{
	\resizebox{0.475\linewidth}{!}{
	\begin{circuitikz}[longL/.style = {european inductor, inductors/scale=0.75,
 inductors/width=1.6, inductors/coils=9}]
        \draw (-7.6,7.9)--(-7.45,7.9) node (T1C1) {} to[L] ++ (0,-1.2) node (T1C2) {}  --++ (-0.15,0);
		\node[above of = T1C1, node distance=0.3cm] {$N$};
		\node[circ] at (-9.0,7.9){};
		\node[circ] at (-9.0,6.7){};
		\draw[gray] (-9.0,7.9) to[L=$L_M$,color=gray] (-9.0,6.7);
		\coordinate (V4T) at (-10.3,7.9);
		\coordinate (V6T) at (-13.3,6.7);
		\draw (V4T) node[circ]{} to[short] (T1C1);
		\draw (T1C2)to[crossing]++(-5.7,0) to
		(V6T) node[circ]{};
		\draw(-13.3,6.1) node[nmos,bodydiode,color=gray](Q4){};
		\draw(-13.3,8.1) node[nmos,bodydiode](Q3){};
		\draw[gray](-13.3,3.8) coordinate (leg5) node[circ]{} to(Q4.E);
		\draw[gray] (Q4.C) to (-13.3,6.7);
		\draw (-13.3,6.7) to (Q3.E);
		\draw (Q3.C)
		to(-13.3,10.2) node[circ]{};
		\draw(-10.3,4.7) node[nmos,bodydiode](Q2){};
		\draw(-10.3,9.5) node[nmos,bodydiode,color=gray](Q1){};
		\draw(-10.3,3.8) coordinate (leg3) to (Q2.E);
		\draw (Q2.C) to (-10.3,7.9);
		\draw[gray] (-10.3,7.9) to (Q1.E);
		\draw[gray] (Q1.C) to (-10.3,10.2);
        \draw (Q1)++(-0.4,0.4)  node(igbt3a)  {$Q_{1}$};
        \draw (Q2)++(-0.4,0.4) node(igbt3b) {$Q_{2}$};
        \draw (Q3)++(-0.4,0.4)  node(igbt5a)  {$Q_{3}$};
        \draw (Q4)++(-0.4,0.4) node(igbt5b) {$Q_{4}$};
        \draw (Q1.E)++(0,-0.3) node[circ,color=gray](Q1Aux1) {};
        \draw[gray](Q1.B)++(0,-0.75) coordinate (Q1Aux2){} to[C=$C_{\textrm{GS}}$,color=gray] ++ (0,0.75);
        \draw[gray](Q1Aux1) to (Q1Aux2){};
        \draw (Q2.E)++(0,-0.3) node[circ](Q2Aux1) {};
        \draw(Q2.B)++(0,-0.75) coordinate (Q2Aux2){} to[C=$C_{\textrm{GS}}$] ++ (0,0.75);
        \draw(Q2Aux1) to (Q2Aux2){};
        \draw (Q3.E)++(0,-0.3) node[circ](Q3Aux1) {};
        \draw(Q3.B)++(0,-0.75) coordinate (Q3Aux2){} to[C=$C_{\textrm{GS}}$] ++ (0,0.75);
        \draw(Q3Aux1) to (Q3Aux2){};
        \draw (Q4.E)++(0,-0.3) node[circ,color=gray](Q4Aux1) {};
        \draw[gray](Q4.B)++(0,-0.75) coordinate (Q4Aux2){} to[C=$C_{\textrm{GS}}$,color=gray] ++ (0,0.75);
        \draw[gray](Q4Aux1) to (Q4Aux2){};
        \draw[gray] (Q1Aux1)++(0,1.25)node[circ,color=gray]{} -- ++ (0.7,0) to[C=$C_{\textrm{DS}}$,color=gray] ++ (0,-1.25)  -- ++ (-0.7,0);
        \draw (Q2Aux1)++(0,1.25)node[circ]{} -- ++ (0.7,0) to[C=$C_{\textrm{DS}}$] ++ (0,-1.25)  -- ++ (-0.7,0);
        \draw (Q3Aux1)++(0,1.25)node[circ]{} -- ++ (0.7,0) to[C=$C_{\textrm{DS}}$] ++ (0,-1.25)  -- ++ (-0.7,0);
        \draw[gray] (Q4Aux1)++(0,1.25)node[circ,color=gray]{} -- ++ (0.7,0) to[C=$C_{\textrm{DS}}$,color=gray] ++ (0,-1.25)  -- ++ (-0.7,0);
        \node[circ,color=gray] at (Q1.B){};
        \node[circ] at (Q2.B){};
        \node[circ] at (Q3.B){};
        \node[circ,color=gray] at (Q4.B){};
        \node[circ,color=gray] at (Q1Aux2){};
        \node[circ] at (Q2Aux2){};
        \node[circ] at (Q3Aux2){};
        \node[circ,color=gray] at (Q4Aux2){};
        \draw[gray](Q1Aux2) --++ (-0.9,0) to[crossing,color=gray] ++ (-3.1,0) coordinate(Wi1Aux1){};
        \draw(Q2Aux2) --++ (-0.9,0) to[crossing] ++ (-3.1,0) coordinate(Wi2Aux1){};
        \draw(Q3Aux2) --++ (-1,0) coordinate(Wi3Aux1){};
        \draw[gray](Q4Aux2) --++ (-1,0) coordinate(Wi4Aux1){};
        \draw[gray](Q1.B) --++ (-0.9,0) to[crossing,color=gray] ++ (-3.1,0) coordinate(Wi1Aux2){};
        \draw(Q2.B) --++ (-0.9,0) to[crossing] ++ (-3.1,0) coordinate(Wi2Aux2){};
        \draw(Q3.B) --++ (-1,0) coordinate(Wi3Aux2){};
        \draw[gray](Q4.B) --++ (-1,0) coordinate(Wi4Aux2){};
        \draw[gray] (Wi1Aux1) to[L,color=gray] (Wi1Aux2);
        \draw (Wi2Aux1) to[L] (Wi2Aux2);
        \draw (Wi3Aux1) to[L] (Wi3Aux2);
        \draw[gray] (Wi4Aux1) to[L,color=gray] (Wi4Aux2);
        \draw (Wi1Aux1)++(-0.1,0) node[circ,color=gray] {};
        \draw (Wi2Aux2)++(-0.1,0) node[circ] {};
        \draw (Wi3Aux2)++(-0.1,0) node[circ] {};
        \draw (Wi4Aux1)++(-0.1,0) node[circ,color=gray] {};
        \draw (Wi1Aux2)++(0.1,0.125) node {$n$};
        \draw (Wi2Aux2)++(0.1,0.125) node {$n$};
        \draw (Wi3Aux2)++(0.1,0.125) node {$n$};
        \draw (Wi4Aux2)++(0.1,0.125) node {$n$};
        \draw  (Wi1Aux2)++(-0.2,0) node (CAUX1U){};
        \draw  (Wi1Aux2)++(-0.25,0) node (CAUX2U){};
        \draw  (Wi2Aux1)++(-0.2,0) node (CAUX1D){};
        \draw  (Wi2Aux1)++(-0.25,0) node (CAUX2D){};
        \draw (CAUX1U) -- (CAUX1D);
        \draw (CAUX2U) -- (CAUX2D);
		\draw
		(leg3)	-- (leg5);
		\draw
		 (leg5) --++(-2.7,0)node[ocirc]{};
		\draw[gray]
		(-10.3,10.2) -- (-13.3,10.2);
		\draw 
		(-13.3,10.2)--++(-2.7,0) node[ocirc]{};
		\draw
		(-17.9,5.3)  node[nmos,bodydiode](igbt6a){};
		\draw
		(-17.9,8.8)  node[color=gray,nmos,bodydiode](igbt6b){};
		\draw
		(-19.4,5.3) node[color=gray,nmos,bodydiode](igbt4a){};
		\draw
		(-19.4,8.8) node[nmos,bodydiode](igbt4b){}; 
		\draw (igbt6a)++(-0.4,0.4) node{$q_2$};
		\draw (igbt6b)++(-0.4,0.4) node{$q_1$};
		\draw (igbt4a)++(-0.4,0.4) node{$q_4$};
		\draw (igbt4b)++(-0.4,0.4) node{$q_3$};
		\draw
		(-19.4,9.5) node[circ]{}
		to (igbt4b.C);
		\draw
		(igbt4b.E) to (-19.4,8.1);
		\draw[gray] (-19.4,8.1)
		to (igbt4a.C);
		\draw[gray]
		(igbt4a.E) to (-19.4,4.5) coordinate (leg4)node[circ]{};
		\draw[gray]
		(-17.9,9.5)
		to (igbt6b.C);
		\draw[gray]
		(igbt6b.E) to (-17.9,6.2);
		\draw (-17.9,6.2)
		to (igbt6a.C);
		\draw
		(igbt6a.E) to (-17.9,4.5) coordinate (leg6);
		\draw
		(leg6) -- (-19.4,4.5); 
		\draw
		 (-19.4,4.5)--(-21.6,4.5);
		\draw
		(-21.6,9.5) -- (-19.4,9.5);
		\draw[gray]
	    (-19.4,9.5)--	(-17.9,9.5)
		;
		\draw
		(-20.9,9.5) node[circ]{}
		to[C=$C_{\textrm{dc}}$]  (-20.9,4.5) node[circ]{}
		;
		\draw
		(-22.3,9.5) node [ocirc]{} node [left] {{$+V_{\textrm{gd}}$}}  -- (-21.6,9.5);
		\draw
		(-22.3,4.5) node [ocirc]{} node [left] {{gnd}}  -- (-21.6,4.5);
		\coordinate (V4) at (-17.6,8.1);
	    \coordinate (V6) at (-17.6,6.2);
	    \draw
		(V6)  to (leg6 |- V6) node [circ] {};
		\draw
		(V4)   to[crossing]++(-0.6,0)to (leg4 |- V4) node [circ] {};
		\draw (-17,7.6)  node[color=gray,nmos,bodydiode,yscale=-1](igbtCa){};
		\node[circ] at (-17,8.1){};
		\draw (-17,6.7)  node[color=gray,nmos,bodydiode](igbtCb){};
		\node[circ] at (-17,6.2){};
		\draw[color=gray] (-17,8.1)--(igbtCa.E);
		\draw[color=gray] (igbtCa.C)--(igbtCb.C);
		\draw[color=gray] (-17,6.2)--(igbtCb.E);
		\draw (igbtCa) ++ (-0.5,0.3) node{$q_5$};
		\draw (igbtCb) ++ (-0.5,0.3) node{$q_6$};
        \node [circ] at (-16.5,8.1){};
        \node [circ] at (-16.5,6.2){};
        \draw[gray] (-16.5,8.1) to[longL=$L_m$,color=gray] (-16.5,6.2);
        \node at (-16,7.15){$L_m$};
        \node at (-15.5,8.2) {$n$};
		\node at (-14.35,5.8) (I2AE2){};
        \node at (-14.35,8.55) (I1AB2){};
        \node[left of = I2AE2] (GDPMaux){};
        \node[left of = I1AB2] (GDPPaux){};
        \node[above of = GDPMaux,node distance=0.4cm] (GDPM){};
        \node[below of = GDPPaux,node distance=0.6cm] (GDPP){};
        \draw (GDPP) to[longL] (GDPM);
        \node[circ] at (-15.25,7.75){};
		\draw  (V6) -- ++(2.24,0);
		\draw  (V4) -- ++(2.24,0);
		\draw  (GDPP)++(0,-0.2) |- (-16.5,8.1);
		\draw  (GDPM)++(0,0.2) |- (-16.5,6.2);
		\end{circuitikz}
	}
	}
    \caption{The operating principle of the proposed gate driver connected to an active bridge converter. Black: ON-path, Gray: OFF-path.
    To simplify the figure, only one module of the MAB is shown. 
    Starting from a regular operation with $Q_1$ and $Q_4$ closed shown in subfigure (a), the gate driver discharges all gate-source capacitances shown in subfigure (b) and drain-source capacitances (c) of the MAB by exchanging energy with the multi-winding transformer magnetizing inductances $L_m$ and $L_M$. 
    The energy exchange is triggered by changing the switching state of the gate driver.
    After that, first, the drain-source capacitors of $Q_2$ and $Q_3$ are charged as shown in subfigure (d) and then the gate-source capacitance (e). 
    This again is achieved by changing the switching state of the gate driver.
    As soon as $Q_2$ and $Q_3$ are closed, normal operation continues as shown in subfigure (f). Switching back to $Q_1$ and $Q_4$ follows the same principle, in reverse order.}
    \label{fig:GD6}
\end{figure*}
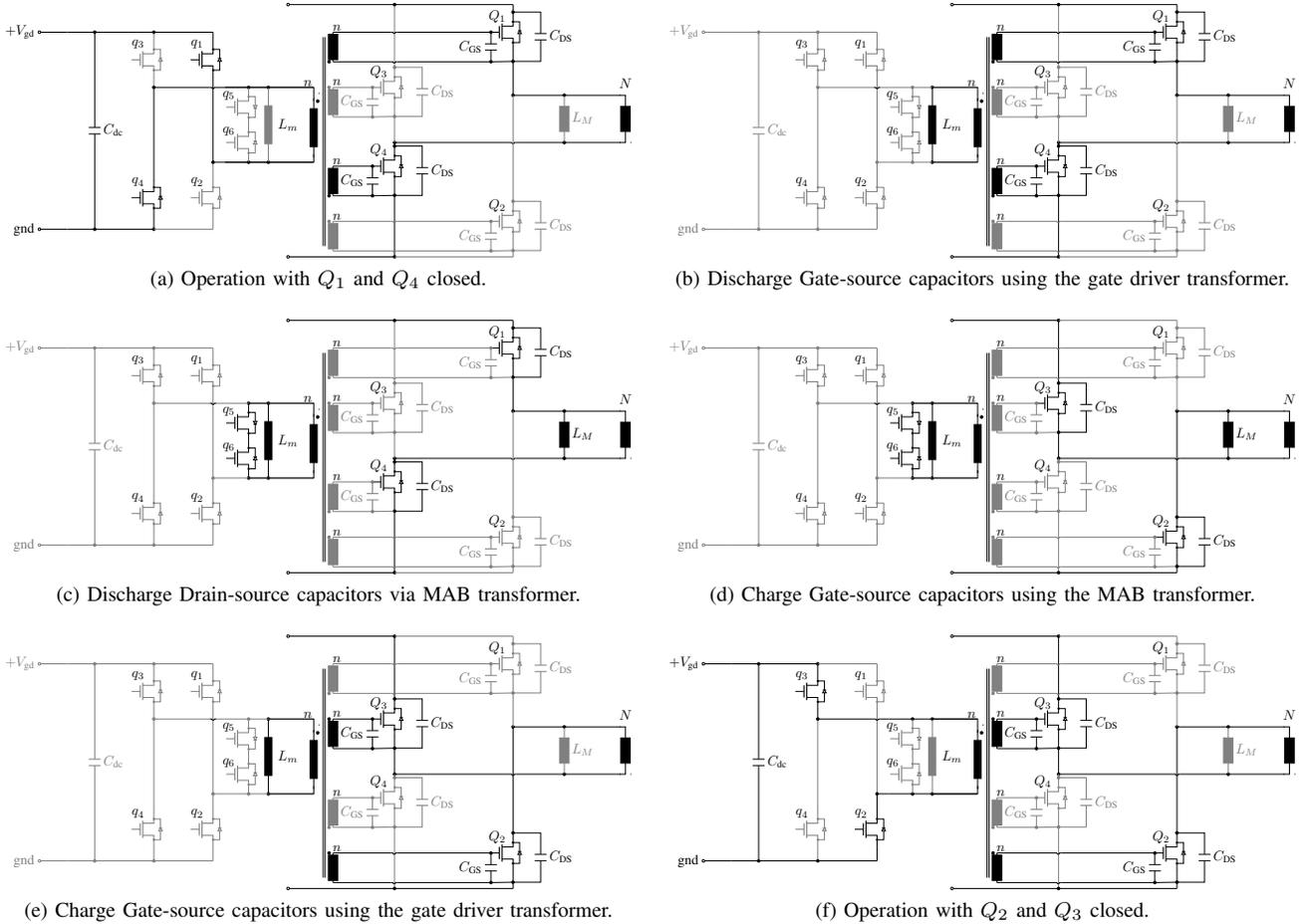

\begin{figure}[!htbp]
			\begin{center}
 			\subfloat[Step 1 and Step 6]{
		    \resizebox{!}{0.07\textheight}{
		\begin{circuitikz}
		\draw
		(0,-0.375) 
		to[L=$L_{\textrm{m}}$,name=Lm, i_<=$i_{\textrm{Lm}}$,v<=$V_{\textrm{Lm}}$]
		++(0,2.75) coordinate(LmR)
		;
		\node[transformer core,circuitikz/inductors/coils=9,circuitikz/inductors/width=1.3,  name=trafo, yscale=1.3,anchor=A1] at(2.375,2.375) {};
		\draw
		(0,-0.375)--++(1.2,0)  node[ocirc](Ap) {};
		\draw(Ap) --(trafo.A2);
		\draw((0,2.375) to[short, i=$i_{\textrm{P}}$]++(1.2,0)  node[ocirc](Bp) {}   --(trafo.A1)
		;
		\draw 
		(trafo.B1)|- ++ (0.375,0.0) node[ocirc](As) {} to[short, i=$i_{\textrm{S}}$]++(1,0) to[L=$L_{\textrm{s}}$, i=$i_{\textrm{Ls}}$,v=$V_{\textrm{Ls}}$] ++ (3,0) coordinate (Ls1)
		(trafo.B2)  -- ++ (0.375,0)  node[ocirc](Bs) {}
		--++(4,0) coordinate (Ls2)
		(Ls2) to [C=$C_{\textrm{gs}}$, i<=$i_{\textrm{gs}}$,v<=$V_{\textrm{gs}}$] (Ls1)
		;
		\node[above of =trafo, node distance = 1.7cm]{$n:n$};
		\draw
		(Ap) to[open,v<=$V_p$] (Bp)
		(As) to[open,v=$V_s$] (Bs);
		\draw 
		(-3,2.375) to[short]
		++ (3,0);
		\draw 
		(-3,2.375)   to [R=$R_{\textrm{eq}}$, i=$i_{\textrm{R}}$,v=$V_{\textrm{R}}$]++(-3,0);
		\draw 
		(-6,-0.375)
		to[battery2, l=$\pm V_{\textrm{gd}}$]++ (0,2.75);
		\draw 
		(-6,-0.375) --++ (6,0);
		\end{circuitikz}
 		}
 		}
 		\end{center}
    \begin{center}
		\subfloat[Step 2 and Step 5]{
		    \resizebox{!}{0.07\textheight}{
			\begin{circuitikz}
			\draw
			(0,-0.375) 
			to[L=$L_{\textrm{m}}$,name=Lm, i_<=$i_{\textrm{Lm}}$,v<=$V_{\textrm{Lm}}$]
			++(0,2.75) coordinate(LmR);
			\node[transformer core,circuitikz/inductors/coils=9,circuitikz/inductors/width=1.3,  name=trafo, yscale=1.3,anchor=A1] at(2.375,2.375) {};
			\draw
			(0,-0.375)--++(1.2,0)  node[ocirc](Ap) {} -|(trafo.A2)
			(0,2.375)to[short, i=$i_{\textrm{P}}$]++(1.2,0)  node[ocirc](Bp) {}   -|(trafo.A1)
			;
			\draw 
			(trafo.B1)-- ++ (0.375,0.0) node[ocirc](As) {} to[short, i=$i_{\textrm{S}}$]++(1,0) to[L=$L_{\textrm{s}}$, i=$i_{\textrm{Ls}}$,v=$V_{\textrm{Ls}}$] ++ (3,0) coordinate (Ls1)
			(trafo.B2)  -- ++ (0.375,0)  node[ocirc](Bs) {}
			--++(4,0) coordinate (Ls2)
			(Ls2) to [C=$C_{\textrm{gs}}$, i<=$i_{\textrm{gs}}$,v<=$V_{\textrm{gs}}$] (Ls1)
			;
			\node[above of =trafo, node distance = 1.7cm]{$n:n$};
			\draw
			(Ap) to[open,v<=$V_p$] (Bp)
			(As) to[open,v=$V_s$] (Bs);
			\end{circuitikz}
			}
			}
			\end{center}
			\begin{center}
			\subfloat[Step 3 and Step 4]{
		    \resizebox{!}{0.07\textheight}{
			\begin{circuitikz}
			\draw
			(0,-0.375) 
			to[L=$L_{\textrm{m}}$,name=Lm, i_<=$i_{\textrm{Lm}}$,v<=$V_{\textrm{Lm}}$]
			++(0,2.75) coordinate(LmR)
			;
			\node[transformer core,circuitikz/inductors/coils=9,circuitikz/inductors/width=1.3,  name=trafo, yscale=1.3,anchor=A1] at(2.375,2.375) {};
			\draw
			(0,-0.375)--++(1.2,0)  node[ocirc](Ap) {} -|(trafo.A2)
			(0,2.375) to[short, i=$i_{\textrm{P}}$]++(1.2,0)  node[ocirc](Bp) {}   -|(trafo.A1)
			;
			\draw 
			(trafo.B1)-- ++ (0.375,0) node[ocirc](As) {} to[short, i=$i_{\textrm{S}}$]++(1,0) to[L=$L_{\textrm{s}}$, i=$i_{\textrm{Ls}}$,v=$V_{\textrm{Ls}}$] ++ (3,0) coordinate (Ls1)
			(trafo.B2)  -- ++ (0.375,0)  node[ocirc](Bs) {}
			--++(4,0) coordinate (Ls2)
			(Ls2) to [C=$C_{\textrm{gs}}$, i<=$i_{\textrm{gs}}$,v<=$V_{\textrm{gs}}$] (Ls1)
			;
			\node[above of =trafo, node distance = 1.7cm]{$n:n$};
			\draw
			(Ap) to[open,v<=$V_p$] (Bp)
			(As) to[open,v=$V_s$] (Bs);
			\draw 
			(-3,2.375) to[short]
			++ (3,0);
			\draw 
			(-3,2.375) to [R=$R_{\textrm{eq}}$, i=$i_{\textrm{R}}$,v=$V_{\textrm{R}}$] ++ (0,-2.75) 
			--++ (3,0);
			\end{circuitikz}
			}
			}
			\end{center}
    \caption{Equivalent circuits of the proposed gate driver during one switching period.
    The transformer is modeled with a primary winding and a single secondary winding representing all switches under the assumption that the windings are identical and perfectly synchronized. 
    Each winding has an equivalent stray inductance $L_s$ and the primary side furthermore has a magnetizing inductance $L_m$ that models the behavior of the core.
    The equivalent resistance $R_{\textrm{eq}}$ represents the ON-state resistance of the switches of the and the winding resistance of the transformer. 
    In the first and last Step, shown in (a), the auxiliary supply of the gate driver is used for the model as well. 
    The difference in the topology in these Steps is implemented as a different sign for the voltage source.
    For Step 2 and Step 5 this voltage does not influence the behavior of the midpoint and is therefore removed as shown in (b). 
    In (c) the supply is disconnected as well, however, the ON-state resistance on the primary side is taken into account. 
    This represents the behavior during Step 3 and Step 4.}
    \label{fig:GD7}
\end{figure}
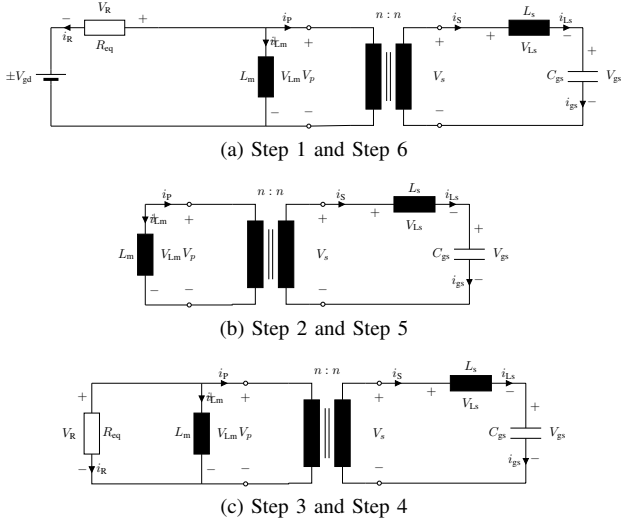

\begin{figure}[!htbp]
    \hspace{-0.5cm}
    \includegraphics[scale=0.475]{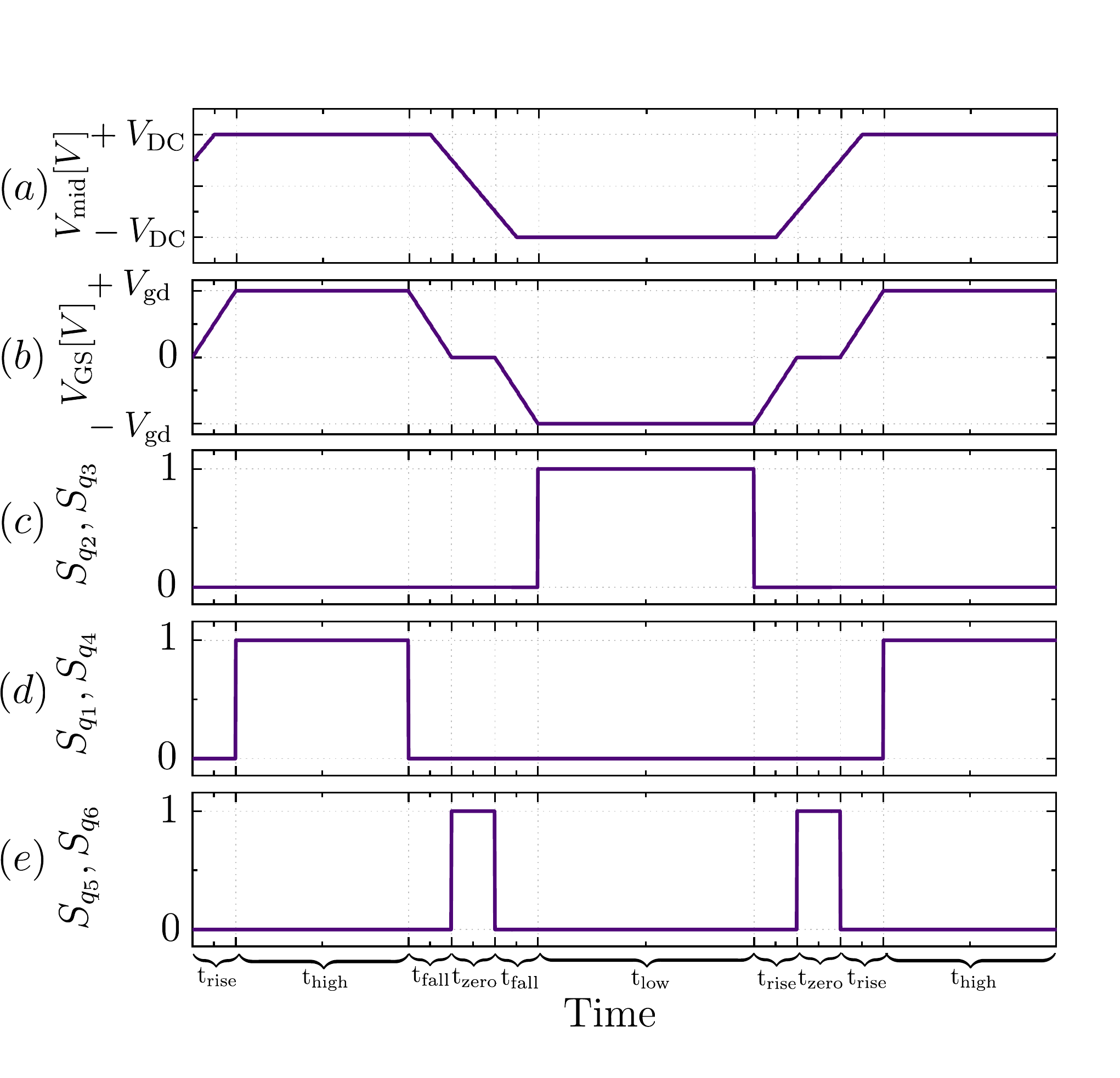}
    \caption{The waveform of the gate-source voltage of the proposed gate driver together with the AC-voltage of the MAB. 
    The resulting AC-voltage of the MAB is shown in (a). 
    Below, in (b), is the midpoint voltage of the gate driver measured at the primary winding of the multi-winding transformer. 
    The bottom three subfigures (c), (d), and (e) show the transients of the switching signals that are given to the gate driver switches $q_1$, $q_2$, $q_3$, $q_4$, $q_5$ and $q_6$ shown in Fig. \ref{fig:GD3}. 
    The timings are assumed to be ideal which results in a voltage waveform without discontinuous jumps as seen in the experiment results. 
    }
    \label{fig:GD8}
\end{figure}

\begin{figure}[!htbp]
    \centering
    \includegraphics[scale=0.725]{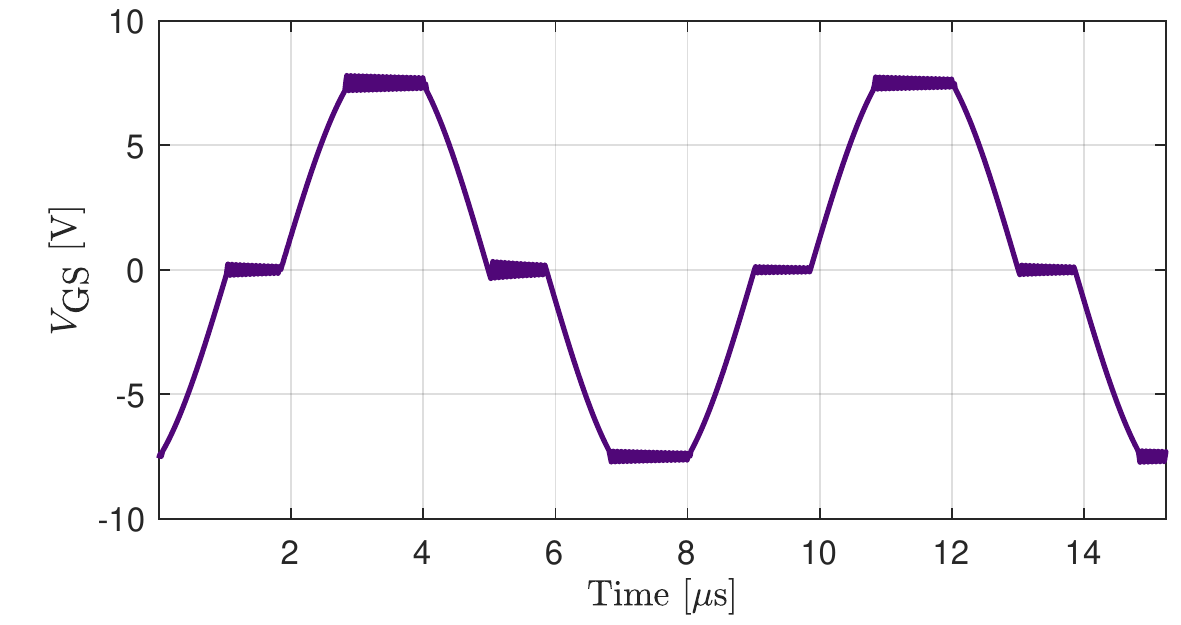}
    \caption{Simulation results of the gate driver midpoint waveform with correctly timed modulator. The timing parameters in this condition are $t_{\textrm{rise}} = 1000$ ns, $t_{\textrm{zero}} = 800$ ns and $f_{\textrm{s}} = 125$ kHz. The simulation is modeled after the prototype presented in section 5.}
    \label{fig:GDEPSIM}
\end{figure}
The goal is to use the gate driver topology to drive all switches of the MAB synchronously.
To successfully drive all switches in the MAB, the gate driver needs to be able to access all switches in the MAB converter. 
This is achieved by connecting it to each of the switches with a multi-winding transformer.
Each winding of the multi-winding transformer is connected to one MOSFET of the MAB. 
All switches of the MAB that are mutually exclusive (as seen in Fig. \ref{fig:GD5}) must furthermore have different turns directions in the secondary windings to ensure that they cannot be closed at the same time as shown in Fig. \ref{fig:GD4}. 
 To connect a MAB, other active bridges can be connected magnetically to the gate driver transformer in a similar way. 
Furthermore, the energy that is stored in the gate-source capacitors of the switches that are turned off can be transferred to the magnetization inductance $L_m$ of the gate-driver transformer during the floating state which achieves lossless operation of the gate driver. 
During the zero states, the energy stored in the drain-source capacitors will be stored in the magnetization inductance $L_M$ of the main circuit which results in the lossless operation of the MAB switches as well.
\subsection{Operating principle} 
An overview of the operating principle during one switching can be seen in
Figure  \ref{fig:GD6}. 
For this section, we assume that the switching frequency is high enough so that both transformers will not go into saturation. 
One operation cycle consists of $6$ steps which will be given in the following:
\begin{itemize}
\item 
\textbf{Step 1}:
In the beginning, $Q_1$ and $Q_4$ are closed, and their gate-source and drain-source capacitors are fully charged. 
The gate driver ensures this by keeping $q_1$ and $q_4$ closed which results in a voltage of $+V_{\textrm{gd}}$ at the transformer windings attached to $Q_1$ and $Q_4$ and $-V_{\textrm{gd}}$ at the transformer windings attached to $Q_2$ and $Q_3$. 
This configuration is shown in Fig. \ref{fig:GD6} (a).
\item
\textbf{Step 2}:
As soon as $Q_1$ and $Q_4$ should be opened, the gate driver opens $q_1$ and $q_4$ resulting in all gate driver switches being open and the midpoint voltage changing to floating. 
Since the voltage across the gate-driver magnetization inductance $L_m$ is now not fixed by the gate driver anymore, it will start discharging the gate-source capacitors of $Q_1$ and $Q_4$, storing the energy from those two capacitors. 
Fig. \ref{fig:GD6} (b) shows this configuration.
\item
\textbf{Step 3}:
When the gate-source capacitors are fully discharged, the gate-driver changes to its zero state by closing $q_5$ and $q_6$. 
This forces the midpoint voltage to zero and allows the drain-source capacitors of the gate driver to discharge through the main transformer's magnetization inductance $L_M$ in a similar manner as the gate-source discharge in the previous step. 
This setup can be seen in Fig. \ref{fig:GD6} (c). 
\item 
\textbf{Step 4}:
Once all capacitors of $Q_1$ and $Q_4$ are fully discharged, the drain-source capacitors of $Q_2$ and $Q_3$ are charged through the magnetization inductance of the MAB transformer. 
This step is shown in Fig. \ref{fig:GD6} (d). 
\item 
\textbf{Step 5}:
As soon as the drain-source capacitors of $Q_2$ and $Q_3$ are fully charged, the gate driver midpoint switches $q_5$ and $q_6$ are opened again leaving the midpoint floating. 
This allows energy exchange between the magnetization inductance $L_m$ and the gate-source capacitances of the MOSFETs to take place again.
This time, however, the capacitors belonging to $Q_1$ and $Q_4$ will be charged negatively while the capacitors belonging to $Q_2$ and $Q_3$ will be charged positively resulting in the switches $Q_2$ and $Q_3$ being closed.
Fig. \ref{fig:GD6} (e) shows this step. 
\item 
\textbf{Step 6}:
When $Q_2$ and $Q_3$ are closed, the corresponding gate-driver switches $q_2$ and $q_3$ will be closed as well to ensure that the midpoint voltage will stay constant and the switches will not open again.
The final step is shown in Fig. \ref{fig:GD6} (f). 
\end{itemize}
Switching $Q_2$ and $Q_3$ off and $Q_1$ and $Q_4$ on again follows a similar pattern, in reverse order. 
If the capacitors charge and discharge linearly, the switching losses of the MAB will be negligible if combined with the proposed gate driver. 
A hard-switching gate driver with the same auxiliary supply would yield losses of $\frac{1}{2} C_{\textrm{GD}} (2V_{\textrm{gd}})^2$ for each of the $4M$ switches per switching action.

\section{State-space Modelling and Analysis}
In this section, a system model of the gate driver is derived to get a more detailed insight into its behavior. 
The purpose of the state-space model is to analyze the behavior of the system and derive design criteria for a prototype.
The modeling is based on the following assumptions:
\begin{enumerate}
    \item Negligible parasitic capacitances within the transformer.
    \item All switches are synchronized. 
    \item All transformer windings are identical.
    \item Linear magnetization inductance, stray inductances, conduction losses, and Gate-source capacitances of the driven MOSFET. 
    \item $100 \%$ coupling of transformer windings.
    \item The behavior of all circuit elements is independent of the frequency of the signals given to the gate driver.
\end{enumerate}
A multi-active full-bridge converter consists of $4M$ high-power MOSFETs. 
Since all windings of the gate driver transformer share the same magnetic flux, the secondary AC voltages of each winding are identical.
The second assumption together with the third assumption allows replacing the $4M$ secondary windings attached to the gate-source capacitances $C_{\textrm{GS}}$ of the MAB with one single equivalent winding.
Furthermore, it is assumed that each element has a linear resistance resulting in one equivalent resistance $R_{\textrm{eq}}$ for each Step,
consisting of $R_{\textrm{ON}}$, the ON-state resistance of the gate-driver MOSFETs $q_1$, ..., $q_6$ and $R_c$ which is the conduction resistance of the multi-winding transformer. 
In addition to the equivalent capacitance, assuming negligible cross-coupling allows representing the multi-winding transformer as a single T-model shown in Fig. \ref{fig:GD7}. 
To fully describe the system, a three-dimensional state-space vector $\mathbf{x}(t)$ is necessary, consisting of the gate-source current $i_{\textrm{gs}}(t)$, the gate-source voltage $V_{\textrm{gs}}(t)$, as well as the magnetization current $i_{\textrm{Lm}}(t)$:
\begin{equation}
    \mathbf{x}(t) = \left[ \begin{array}{ccc} i_{\textrm{gs}}(t) & V_{\textrm{gs}}(t) & i_{\textrm{Lm}}(t) \end{array}\right]^T \in \mathbb{R}^3.
\end{equation} 
Furthermore, the input $u(t)$ of the gate driver shall be defined as the gate-driver supply voltage, which is changing for different different steps,
\begin{equation}
    u(t) = \begin{cases}
-V_{\textrm{gd}}, \quad \textrm{Step 1}\\
0, \quad \textrm{Step 2, Step 3, Step 4, Step 5}\\
+V_{\textrm{gd}}, \quad \textrm{Step 6}.
\end{cases}
\end{equation}
While the switches of the MAB are ON, the gate driver is represented by the equivalent circuit shown in Fig. \ref{fig:GD7} (a). 
Using Kirchhoff's equations, the dynamics of the system can be derived as
\begin{equation}
        \dot{\mathbf{x}}(t) = 
    \left[\begin{array}{ccc}
    -\frac{R_{\textrm{eq}}}{L_s} & -\frac{1}{L_s} & -\frac{R_{\textrm{eq}}}{L_s} \\
     \frac{1}{C_{\textrm{GS}}} & 0 & 0 \\
     -\frac{R_{\textrm{eq}}}{L_m} &  0 & -\frac{R_{\textrm{eq}}}{L_m}
    \end{array}\right]
     \mathbf{x}(t)
     +
     \left[\begin{array}{c}
     \frac{1}{L_s} \\
     0 \\
     \frac{1}{L_m}
     \end{array}\right] 
      u(t),
\end{equation}
for Step 1 and Step 6. 
For those cases, the system reaches steady state for 
\begin{equation}
    \mathbf{x}^{\textrm{ss}}_{1,6} = \left[  \begin{array}{ccc} 0 & 0 & \frac{u}{R_{\textrm{eq}}}  \end{array}\right]^T.
\end{equation}
This indicates that the transformer will go into saturation and the magnetization current will become $0$ after a while. 
In this situation, the gate driver loses its functionality and the supply will simply force a current going through the primary side. 
To avoid this, the circuit must be operated at a minimum operating frequency. 
\newline
During the floating phase, no external power supply is connected to the gate driver transformer. 
An equivalent circuit for this case is given in Fig. \ref{fig:GD7} (b).
Therefore, the only way to change the voltage across the magnetization inductance is through exchange with the gate-source capacitance. 
For this reason, the input $u(t)$ will not influence the system behavior in this case.
Since there is no resistor current $i_R$, the third row of the state-space matrix will be zero and the system matrix will be overdetermined.
 \begin{equation}
\dot{\mathbf{x}}(t) =\left[ \begin{array}{ccc} 0 &- \frac{1}{L_m+L_s} & 0 \\ \frac{1}{C_{\textrm{GS}}} & 0 & 0 \\
0 & 0 & 0 \end{array} \right] \mathbf{x}(t) + \left[\begin{array}{c} 0 \\ 0 \\ 0 \end{array}\right] \cdot u(t).
 \end{equation}
 In this situation, the entire system dynamics can be formulated as a second order differential equation,
 \begin{equation}
     \ddot{x}_1(t) = \frac{1}{C_{\textrm{GS}}(L_m-L_s)} x_1(t).
 \end{equation}
 This equation is a homogeneous linear differential equation of second order and thus has a sinusoidal solution in a steady-state condition.
 It can therefore be observed that there will be a constant fluctuation of energy between the magnetization inductance and the gate-source capacitance.
 \newline
 An equivalent circuit for the case when the switches at the midpoint clamping are closed is given in Fig. \ref{fig:GD7} (c). 
 As in the previous step, there is no external source supplying the primary side of the gate driver. 
 In this case, however, there is a current $i_R$ flowing through the midpoint clamping which can be characterized from the ON-state resistance of the closed switches $q_5$ and $q_6$. 
 The system equations for this situation are given as 
  \begin{equation}
    \dot{\mathbf{x}}(t) = 
    \left[\begin{array}{ccc}
    -\frac{R_{\textrm{eq}}}{L_s} & -\frac{1}{L_s} & -\frac{R_{\textrm{eq}}}{L_s} \\
     \frac{1}{C_{\textrm{GS}}} & 0 & 0 \\
     -\frac{R_{\textrm{eq}}}{L_m} &  0 & -\frac{R_{\textrm{eq}}}{L_m}
    \end{array}\right]
     \mathbf{x}(t) + \left[\begin{array}{c} 0 \\ 0 \\ 0 \end{array}\right] \cdot u(t).
 \end{equation}
 The steady state conditions for this case are given as 
 \begin{equation}
     \mathbf{x}^{\textrm{ss}}_{3,4} = \left[  \begin{array}{ccc} 0 & 0 & 0  \end{array}\right]^T.
 \end{equation}
 This indicates that the system will converge towards an idle state during this switching state, allowing charge exchange for the drain-source capacitances $C_{\textrm{DS}}$ with the main transformer magnetization inductance $L_M$. 
 It should be noted that the system matrix for this switching state is identical to the one for Steps 1 and 6. 
During all Steps, the system has a linear structure of 
 \begin{equation}
     \dot{\mathbf{x}}(t) = \mathbf{A}\mathbf{x}(t) + \mathbf{b} u(t).
 \end{equation}
 Having the system equations, it is possible to predict the behavior of the gate-source voltage of the MAB switches under ideal operating conditions:
 \newline
 Ideal operating conditions are defined as the switching frequency being high enough to have an almost constant gate-source voltage during Step 1 and Step 6 and a gate driver supply that ensures that the gate-source voltage is almost linear during the floating state. 
 Furthermore, the clamping switches shall be activated when the midpoint voltage reaches zero. 
 In this case, behavior such as the one depicted in Fig. \ref{fig:GD8} can be observed. 
 This behavior can also be seen in simulation results as shown in Fig. \ref{fig:GDEPSIM}.
\subsection{Design Guidelines}
\begin{figure}[!htbp]
     \centering
     \includegraphics[scale=0.6]{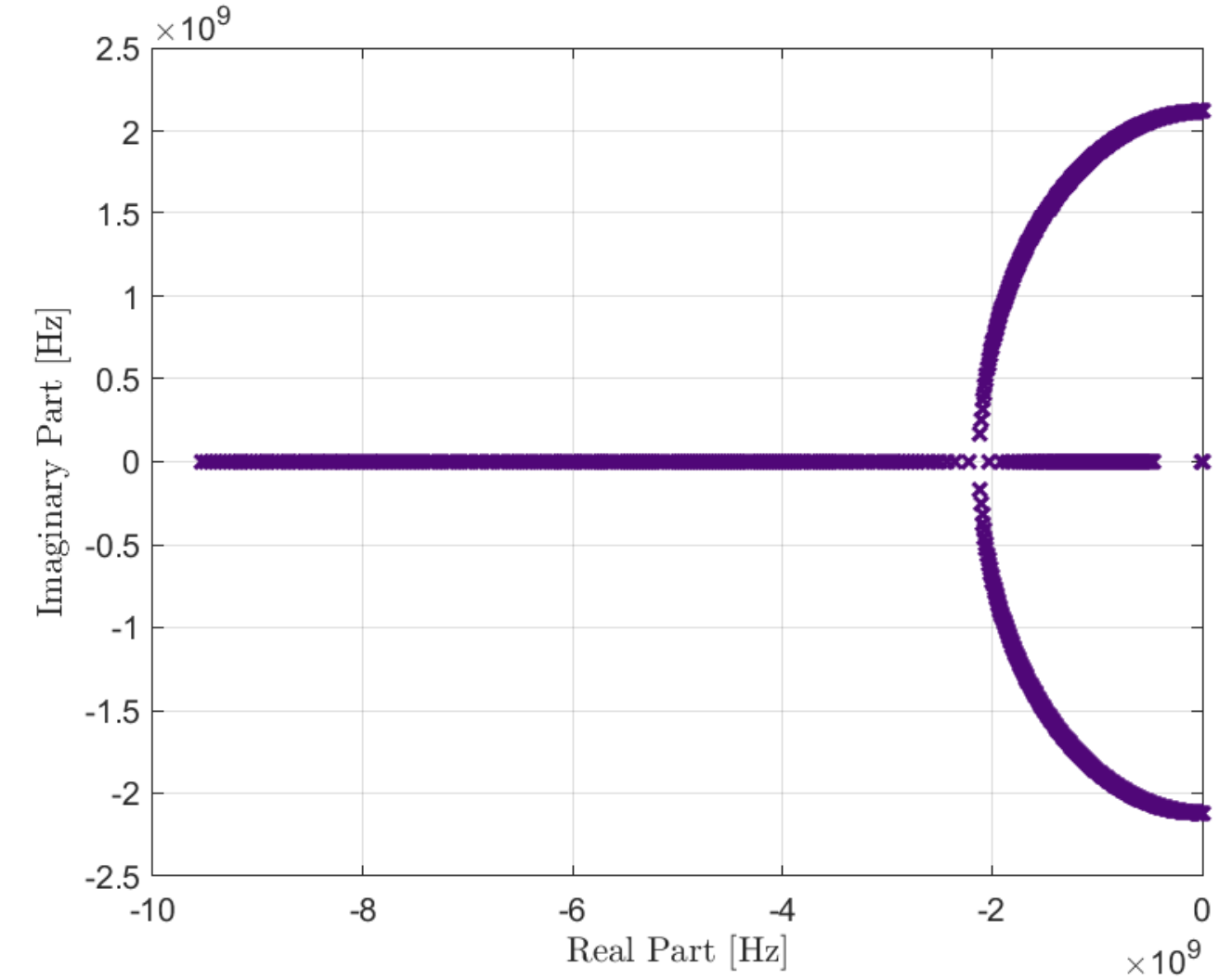}
     \caption{Root locus of the gate driver system matrix for changing $R_{\textrm{eq}}$. A magnetization inductance of $L_m = 4.1 \mu$H, a stray inductance of $L_s= 0.1$nH and a gate-source capacitance of $C_{\textrm{GS}}= 2.23$nF were selected for this scenario. Then, $500$  values with logarithmic spacing between $1 $n$\Omega$ and $1 \Omega$ were chosen for $R_{\textrm{eq}}$. The figure shows the change of the eigenvalues of the system matrix $\mathbf{A}$ for the different values of $R_{\textrm{eq}}$.}
     \label{fig:RLR}
 \end{figure}

\begin{figure}[!htbp]
     \centering
     \includegraphics[scale=0.6]{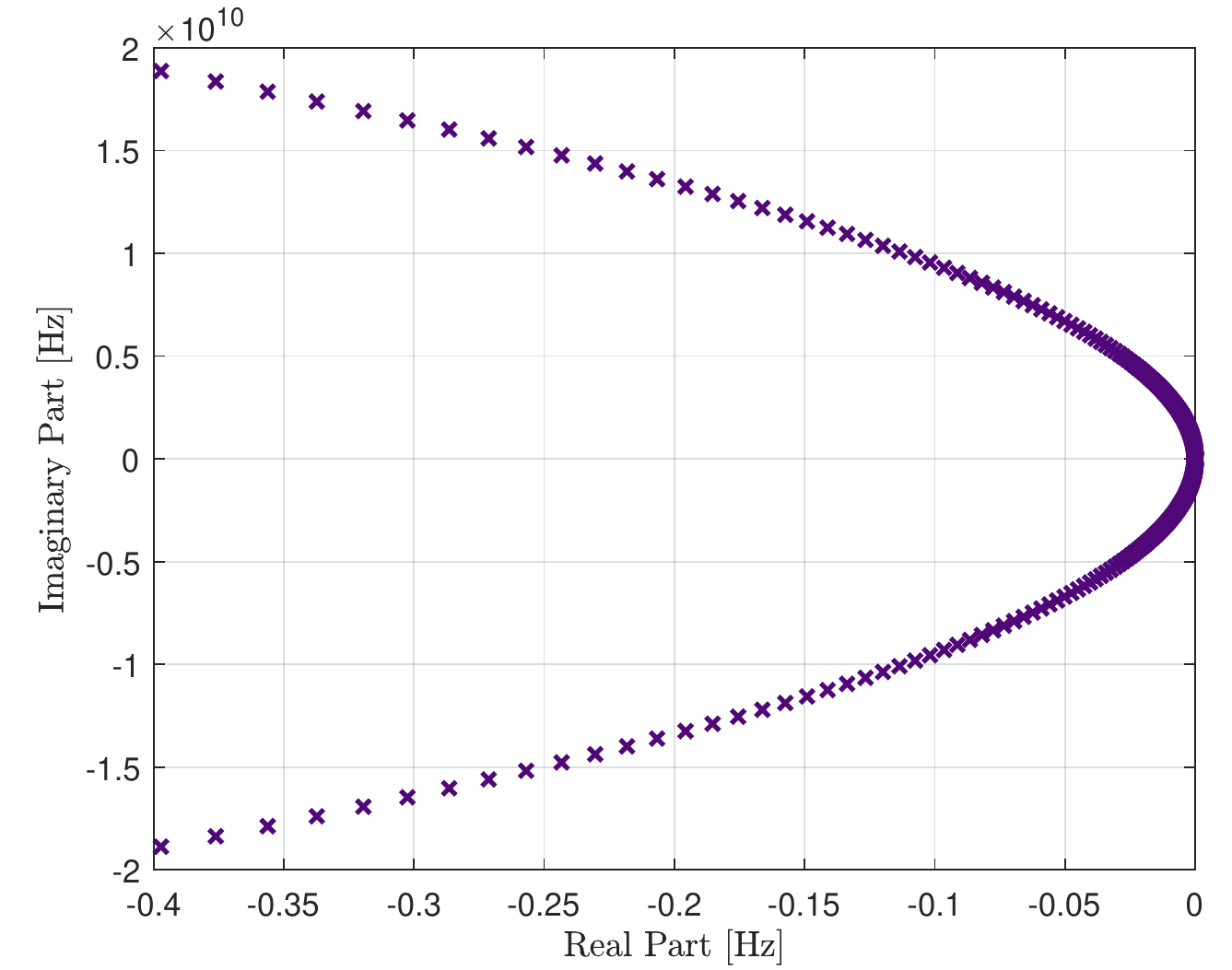}
     \caption{Root locus of the gate driver system matrix for different values of the stray inductance $L_s$. For this scenario, a magnetization inductance of $L_m = 4.1 \mu$H, an equivalent resistance of $R_{\textrm{eq}}=0.125 \Omega$ and a gate-source capacitance of $C_{\textrm{GS}} = 2.230$ nF were chosen. After that, the Eigenvalues of the system matrix $\mathbf{A}$ were evaluated for $500$ different values of the stray inductance which were logarithmically spaced between $L_s = 0.001$nH and $L_s = 1$H. Shown are the trajectories of the two complex conjugated Eigenvalues. The third Eigenvalue is independent of $L_s$ and does not move.}
     \label{fig:RLL}
 \end{figure}

This section discusses several design aspects of a prototype of the proposed gate driver topology. 
The design process in this paper is based on the assumption that the user has already selected a MAB with a known MOSFET gate-source capacitance $C_{\textrm{GS}}$. 
Then, the remaining steps are as follows: 
\begin{enumerate}
    \item Select a switching frequency and compute the desired magnetization inductance that can charge and discharge the MOSFET capacitors within this frequency.
    \item Compute values for $L_s$ and $R_{\textrm{eq}}$ that guarantee a stable system during $t_{\textrm{high}}$, $t_{\textrm{low}}$ and $t_{\textrm{zero}}$. 
    \item Tune the timing constants $t_{\textrm{zero}}$, $t_{\textrm{high}}$, $t_{\textrm{fall}}$, $t_{\textrm{rise}}$, $t_{\textrm{low}}$ to ensure operation with minimal switching losses.
\end{enumerate}
The following sections discuss these steps in more detail.
 Having selected the magnetization inductance, the question of how to choose the other parameters $L_s$ and $R_{\textrm{eq}}$ shall be selected. 
 Therefore, we examine the stability properties of the system matrix $\mathbf{A}$ derived in the previous section. 
 To examine the dynamic behavior of the system, the Eigenvalues $\lambda$ were computed for different scenarios:
 \begin{equation}
     \det \left( \mathbf{A} - \lambda \mathbf{I} \right) = 0,
 \end{equation}
 where $\mathbf{I}$ is a $3 \times 3$ identity matrix. 
 Since the system matrix is the same for the switching Steps 1,3,4,6 and the system is always boundary stable for Steps 2 and 5, only the system matrices for Steps 1,3,4,6 are discussed. 
 Although the presented system has a switching nature, and thus stability for all individual Steps does not necessitate stability for the system under switching conditions, it can be achieved in this case by changing to Step 3 or 4 and waiting until the dynamic components of the system discharge through the equivalent resistance. 
 \newline
Using the Eigenvalue analysis, it was found that $\mathbf{A}$ does not become unstable for all examined realizations of the system. 
 However, choosing the correct values for $L_s$ and $R_{\textrm{eq}}$ has an impact on the oscillations in the system and thus the power losses.
 \newline
 The system has two complex conjugated and one real eigenvalue as can be seen from Fig. \ref{fig:RLR}.
 In case the resistance $R_{\textrm{eq}}$ is zero, the real part of all Eigenvalues is $0$ and the system will continue to oscillate. 
 Increasing $R_{\textrm{eq}}$ will add a negative real part to all three Eigenvalues. This reduces the amplitude of the oscillations and introduces damping to the system. 
 A high resistance however also increases the conduction losses of the system and thus decreases the efficiency. 
 For this reason, a trade-off has to be reached for $R_{\textrm{eq}}$.
 \newline
 A variation in the stray inductance $L_s$ only influences the two complex conjugated eigenvalues as can be seen from Fig. \ref{fig:RLL}. 
 The smaller the value, the more negative the real part of the eigenvalue pair. 
In addition to that, small values of $L_s$ also increase the imaginary part of the eigenvalues which results in higher oscillation frequencies. 
Those oscillations would, however, decay over time due to the negative real part of the Eigenvalue. 
Furthermore, having no $L_s$ will result in an $LC$-resonant tank formed by the magnetization inductance $L_m$ and the MOSFET capacitance $C_{\textrm{DS}}$. 
For decreasing $L_s$ the behavior of the system will therefore converge towards the one of a conventional $RLC$ parallel circuit.
 \subsection{Operation Timing Selection}
 To successfully operate the gate driver, the timing constants have to be chosen correctly which shall be discussed in this paragraph. 
 To minimize losses of the gate driver, it is important to obtain a curve with continuous transitions at the midpoint as shown in Fig. \ref{fig:GD8}.
 In this case, the timings are selected in a way that the transition to the zero-state will take place when the midpoint voltage is zero and the transition to the high/low state takes place when the midpoint voltage is $\pm V_{\textrm{GD}}$.  
 Additionally, the zero-time must be chosen to be large enough to accommodate discharging and charging the drain-source capacitances of $Q_s$ and allowing a sign change in the midpoint voltage of the MAB, as seen in Fig. \ref{fig:GD8}.
 Furthermore, the overall operating frequency should be chosen to be small enough to allow the required energy transfers to take place. 
 For small transformers, where the number of windings is small enough to allow full identification of all parasitic inductances, an analytic formula using the system model can be used to tune the timing constants. 
In the case of larger transformers, where the number of required tests to identify all parameters becomes too numerous, the parameters can be tuned by inspection, using low-power tests as shown in the next section.

\section{Prototype Construction and Experimental Results}
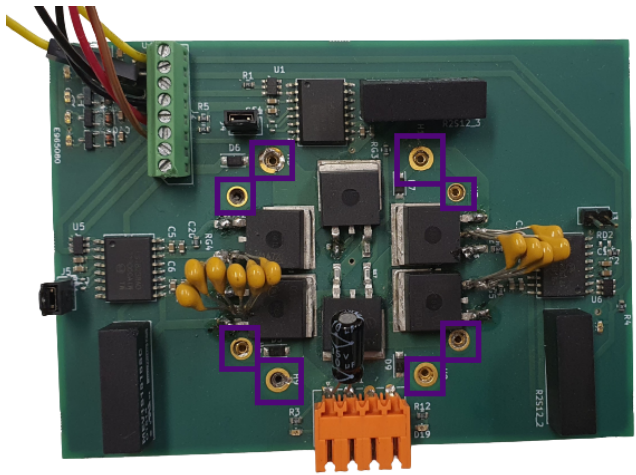
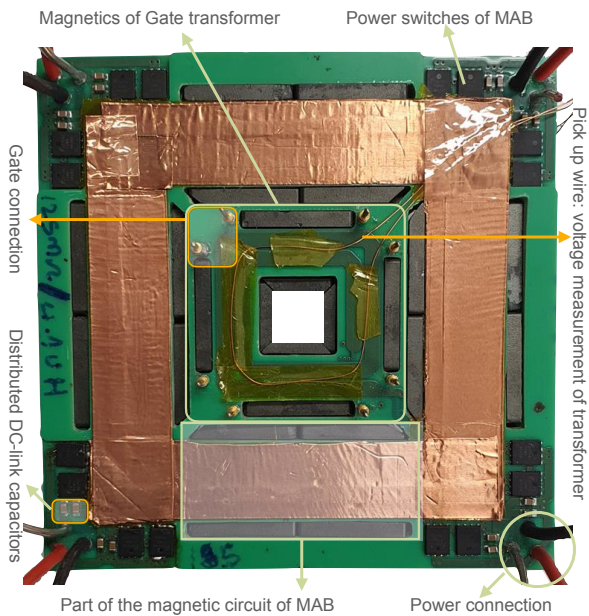
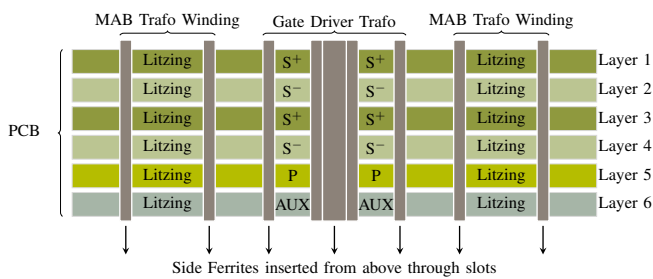
\begin{figure}[!htbp]
    \centering
    	\subfloat[Gate Driver Board]{
    \includegraphics[scale=0.56,angle=90,origin=c]{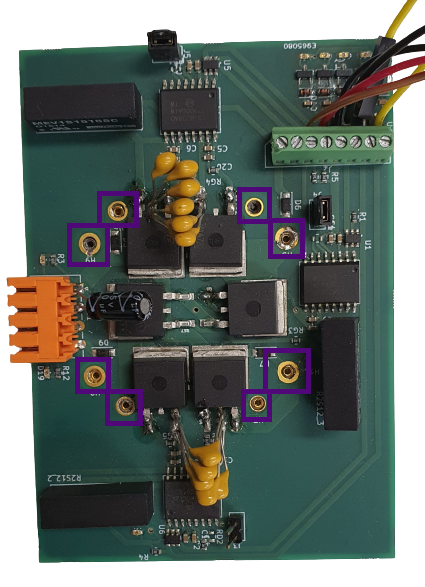}
    }
    \\
    	\subfloat[Gate Driver transformer and MAB]{
    \includegraphics[scale=0.28]{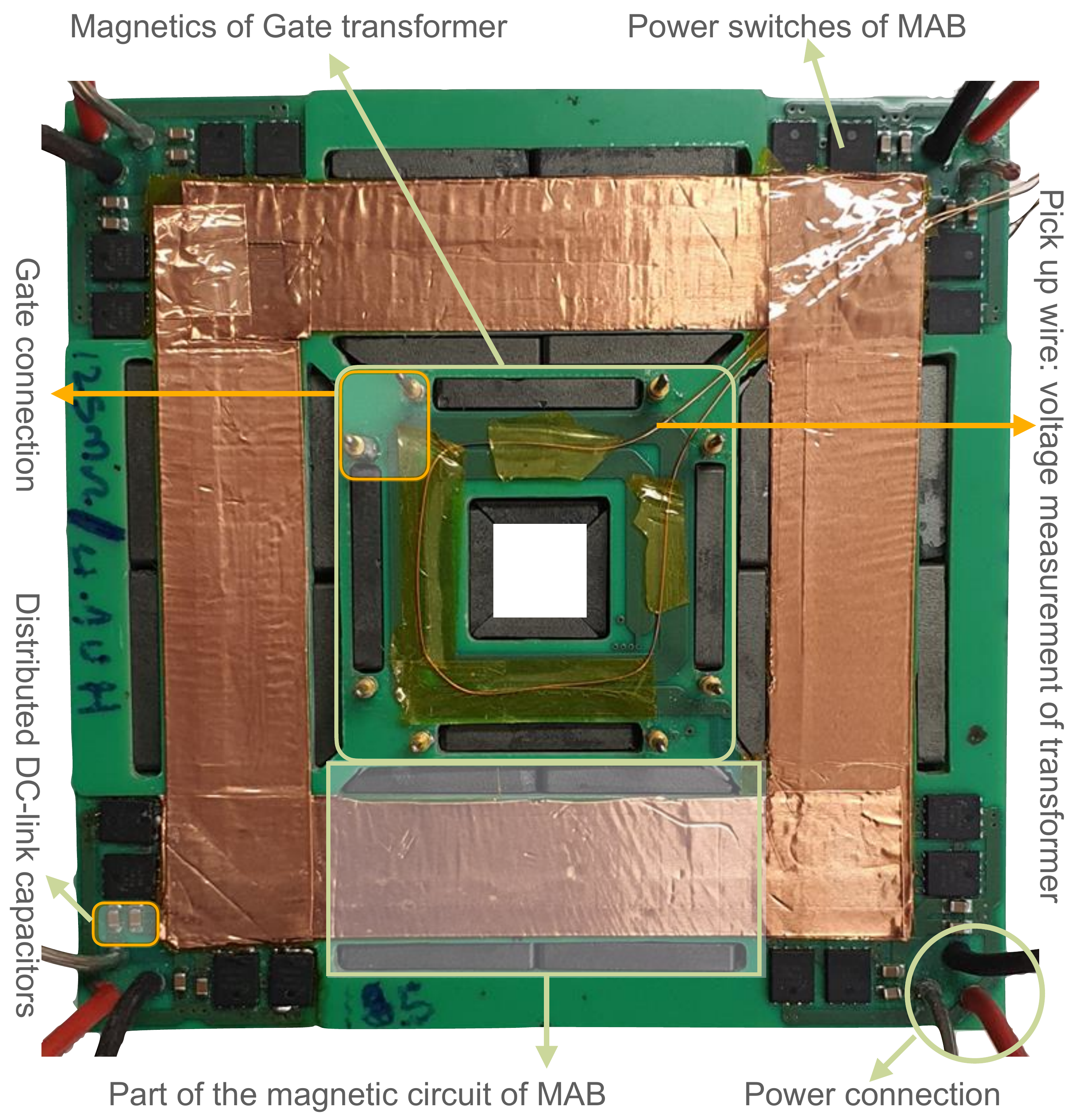}
    }
    \\
    \subfloat[Interleaving pattern of the Prototype]{
    \resizebox{\linewidth}{!}{
    \begin{tikzpicture}
    \draw[white,thick,fill=UCLGRAY] (0,0.1) rectangle (0.25,-3.7);
    \draw[white,thick,fill=UCLGRAY] (3,0.1) rectangle (3.25,-3.7);
    \draw[white,thick,fill=UCLGRAY] (7,0.1) rectangle (7.25,-3.7);
    \draw[white,thick,fill=UCLGRAY] (1.75,0.1) rectangle (2.0,-3.7);
    \draw[white,thick,fill=UCLGRAY] (5.75,0.1) rectangle (6.0,-3.7);
    \draw[white,thick,fill=UCLGRAY] (8.75,0.1) rectangle (9.0,-3.7);
    \draw[white,thick,fill=UCLGRAY] (4.0,0.1) rectangle (4.25,-3.7);
    \draw[white,thick,fill=UCLGRAY] (4.75,0.1) rectangle (5.0,-3.7);
    \draw[white,thick,fill=UCLGRAY] (4.25,0.1) rectangle (4.75,-3.7);
    \draw [-stealth,thick](0.125,-3.8) -- (0.125,-4.4);
    \draw [-stealth,thick](3.125,-3.8) -- (3.125,-4.4);
    \draw [-stealth,thick](7.125,-3.8) -- (7.125,-4.4);
    \draw [-stealth,thick](1.875,-3.8) -- (1.875,-4.4);
    \draw [-stealth,thick](5.875,-3.8) -- (5.875,-4.4);
    \draw [-stealth,thick](8.875,-3.8) -- (8.875,-4.4);
    \draw [-stealth,thick](4.5,-3.8) -- (4.5,-4.4);
    \node at(4.5,-4.7){Side Ferrites inserted from above through slots};
           \begin{scope}[on background layer]
          \draw[black!15,fill=UCLGreenM] (-1,-0.1)
                    rectangle (10,-0.6);
          \draw[black!15,fill=UCLGreenL] (-1,-0.7)
                    rectangle (10,-1.2);
          \draw[black!15,fill=UCLGreenM] (-1,-1.3)
                    rectangle (10,-1.8);
          \draw[black!15,fill=UCLGreenL] (-1,-1.9)
                    rectangle (10,-2.4);
          \draw[black!15,fill=UCLGreenB] (-1,-2.5)
                    rectangle (10,-3.0);
          \draw[black!15,fill=PCBgreen!35] (-1,-3.1)
                    rectangle (10,-3.6);
        \end{scope}
    \node at (10.6,-0.35){Layer 1};
    \node at (10.6,-0.95){Layer 2};
    \node at (10.6,-1.55){Layer 3};
    \node at (10.6,-2.15){Layer 4};
    \node at (10.6,-2.75){Layer 5};
    \node at (10.6,-3.35){Layer 6};
    \draw [decorate,
    decoration = {brace,mirror}] (-1.2,-0.1) --  (-1.2,-3.6);
    \node at (-2,-1.8){PCB};
    \node at (1,-0.35){Litzing};
    \node at (1,-0.95){Litzing};
    \node at (1,-1.55){Litzing};
    \node at (1,-2.15){Litzing};
    \node at (1,-2.75){Litzing};
    \node at (1,-3.35){Litzing};
    \node at (8,-0.35){Litzing};
    \node at (8,-0.95){Litzing};
    \node at (8,-1.55){Litzing};
    \node at (8,-2.15){Litzing};
    \node at (8,-2.75){Litzing};
    \node at (8,-3.35){Litzing};
    \node at (5.375,-0.35){S$^{+}$};
    \node at (5.375,-0.95){S$^{-}$};
    \node at (5.375,-1.55){S$^{+}$};
    \node at (5.375,-2.15){S$^{-}$};
    \node at (5.375,-2.75){P};
    \node at (5.375,-3.35){AUX};
    \node at (3.625,-0.35){S$^{+}$};
    \node at (3.625,-0.95){S$^{-}$};
    \node at (3.625,-1.55){S$^{+}$};
    \node at (3.625,-2.15){S$^{-}$};
    \node at (3.625,-2.75){P};
    \node at (3.625,-3.35){AUX};
        \draw [decorate,
    decoration = {brace,mirror}] (6.0,0.2) --  (3.0,0.2);
        \draw [decorate,
    decoration = {brace,mirror}] (2.0,0.2) --  (0.0,0.2);
        \draw [decorate,
    decoration = {brace,mirror}] (9.0,0.2) --  (7.0,0.2);
    \node at (4.5, 0.5){Gate Driver Trafo};
    \node at (1.0, 0.5){MAB Trafo Winding};
    \node at (8.0, 0.5){MAB Trafo Winding};
    \end{tikzpicture}
    }
    }
    \caption{The prototype realizes the proposed circuit. The prototype is split into two subsystems, the gate driver board shown in  (a) and the transformer board shown in (b). The transformer board contains both the gate driver transformer and MAB circuit. To drive the MOSFETs in the transformer board, the gate driver board is connected to the transformer board through the holes H5, H6, H7, H8, H9, H10, H11, H12 highlighted in purple. For the experiment results,  a Spartan 6 FPGA was used to generate the control signals and an auxiliary power supply was connected. Each PCB of the transformer board consists of one full bridge with gate driver windings. The interleaving pattern of each of the PCBs is shown in (c).}
    \label{fig:GDPROT}
\end{figure}
To verify the findings proposed in this paper, a gate driver prototype shown in Fig. \ref{fig:GDPROT} was built. 
The prototype consists of two assemblies, the primary gate driver board which can be seen in Fig. \ref{fig:GDPROT} (a) and the gate driver transformer together with a MAB shown in Fig. \ref{fig:GDPROT} (b). 
To verify the efficiency for large-scale multilevel converters, the gate driver is tested for a MAB with $4$ groups of modules, each of which is made up of $4$ active bridges that are connected in parallel.
The $400$ V MAB thus consists of $16$ modules with $64$ switches which were used to test the proposed gate driver. 
The $64$ switches each have an input capacitance of $2330$ pF and output capacitance of $220$ pF that need to be charged during the switching. 
To get insight into the efficiency of the proposed topology, the switching losses for hard-switching can be estimated by $E_{\textrm{sw}}= 0.5 C_{\textrm{GS}} V_{\textrm{GS}}^2 \approx 0.456 \mu  $J per switch.  
In the case of a frequency of $125$ kHz, the average power loss is equal to $57$ mW for each switch.
To change the switching state, it is furthermore necessary to maintain a threshold voltage of $2.9$ V across the gate-source terminals of the MOSFETs. 
A DC voltage of $7.0$ V was selected as the operating voltage of the gate driver to ensure the threshold voltage requirements are met. 
The gate driver prototype, therefore, needs a multi-winding transformer with $64$ secondary windings. 
To optimize the power packaging and increase the power density, the gate driver transformer is implemented as a planar transformer with $16$ primary windings (P) and $64$ secondary windings (S) as shown in Fig. \ref{fig:GDPROT} (c). 
The winding arrangement is S$^+$-S$^-$-S$^+$-S$^-$-P. 
All windings are identical and have $2$ turns where the notation S$^+$ and S$^-$ refer to the turn directions of each winding.
Since a full identification of this transformer requires $0.5 \cdot 80 \cdot 79 = 3160 $ measurements, a full characterization of the prototype will not be given in this paper and all tuning will be performed "by inspection" to obtain preliminary results for the verification of our findings. 
However, a qualitative assessement of the behaviour of the device
For the remainder of this section, symmetric operation is assumed, i.e. $t_{\textrm{rise}} = t_{\textrm{fall}}$ and $t_{\textrm{high}} = t_{\textrm{low}}$.
\newline
\begin{figure}[!htbp]
    \centering
    \includegraphics[scale=0.725]{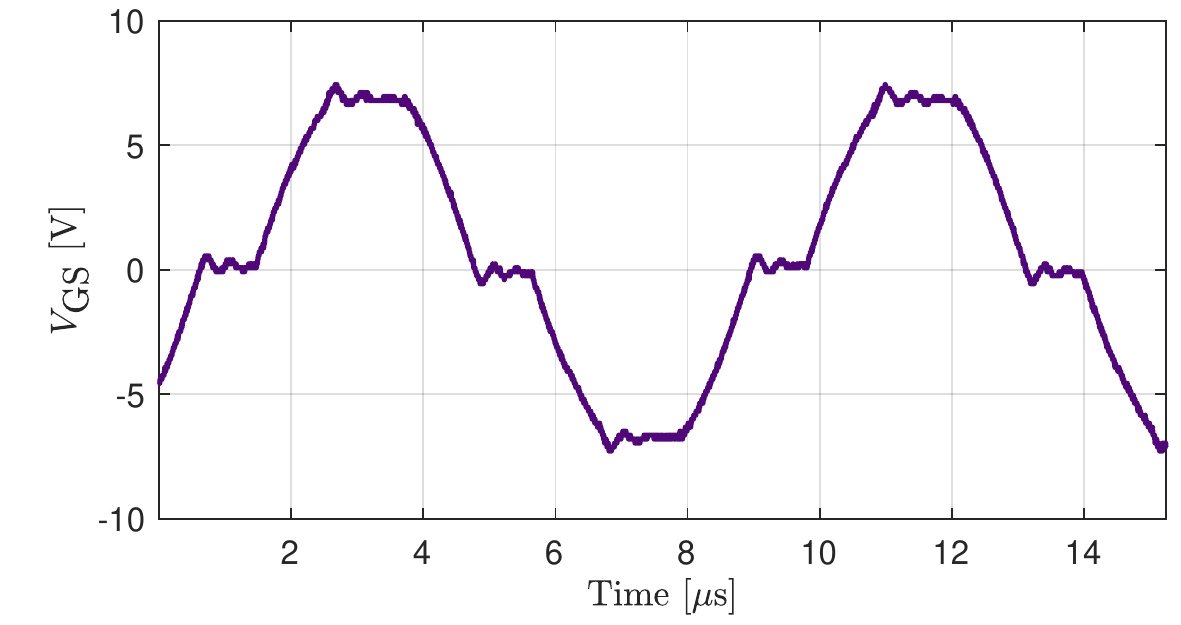}
    \caption{Experiment results of the gate driver midpoint waveform with correctly timed modulator. The timing parameters in this condition are $t_{\textrm{rise}} = 1000$ ns, $t_{\textrm{zero}} = 800$ ns and $f_{\textrm{s}} = 125$ kHz.}
    \label{fig:GDE0}
\end{figure}
Figure \ref{fig:GDE0} shows the midpoint waveform with the modulator being timed correctly. 
The curve verifies the theoretical analysis and modeling shown in Fig. \ref{fig:GD8}, which suggests a similar pattern in the second subfigure. 
The main difference is seen as a decaying sinusoidal noise during the high-, low- and zero states. 
This is due to the limited resolution of the FPGA which is controlling the modulator. 
All timings are selected as an integer multiple of the clock frequency of the FPGA. 
In case this value is not exactly equal to the optimal value, a small deviation will remain which is further amplified by the turn- ON and turn- OFF delays of the switches on the gate driver circuits. 
\newline
\begin{figure}[!htbp]
    \centering
    \includegraphics[scale=0.65]{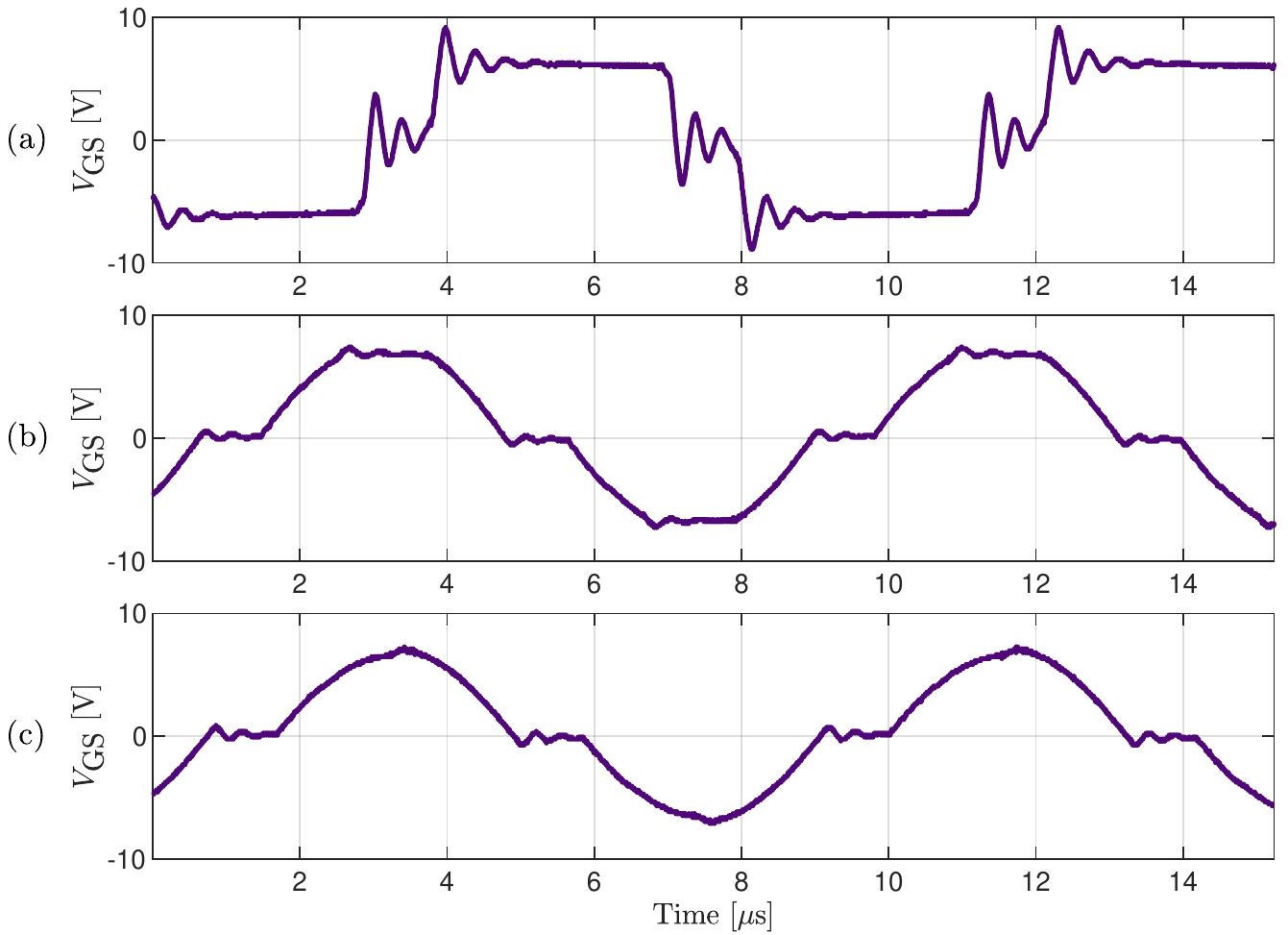}
    \caption{Experiment results of the effects of different timing selections on the gate driver waveform concerning $t_{\textrm{rise}}$. 
    The top subfigure (a) shows a waveform corresponding to a value of $t_{\textrm{rise}} = 100$ ns. The middle subfigure (b) depicts a waveform with $t_{\textrm{rise}} = 1000$ ns and the bottom subfigure (c) represents $t_{\textrm{rise}} = 1500$ ns. 
    In all cases, the zero-time is $t_{\textrm{zero}} = 800$ ns and the switching frequency is $f_{\textrm{s}} = 125$ kHz.
    }
    \label{fig:GDE1}
\end{figure}
\begin{figure}[!htbp]
    \centering
    \includegraphics[scale=0.6]{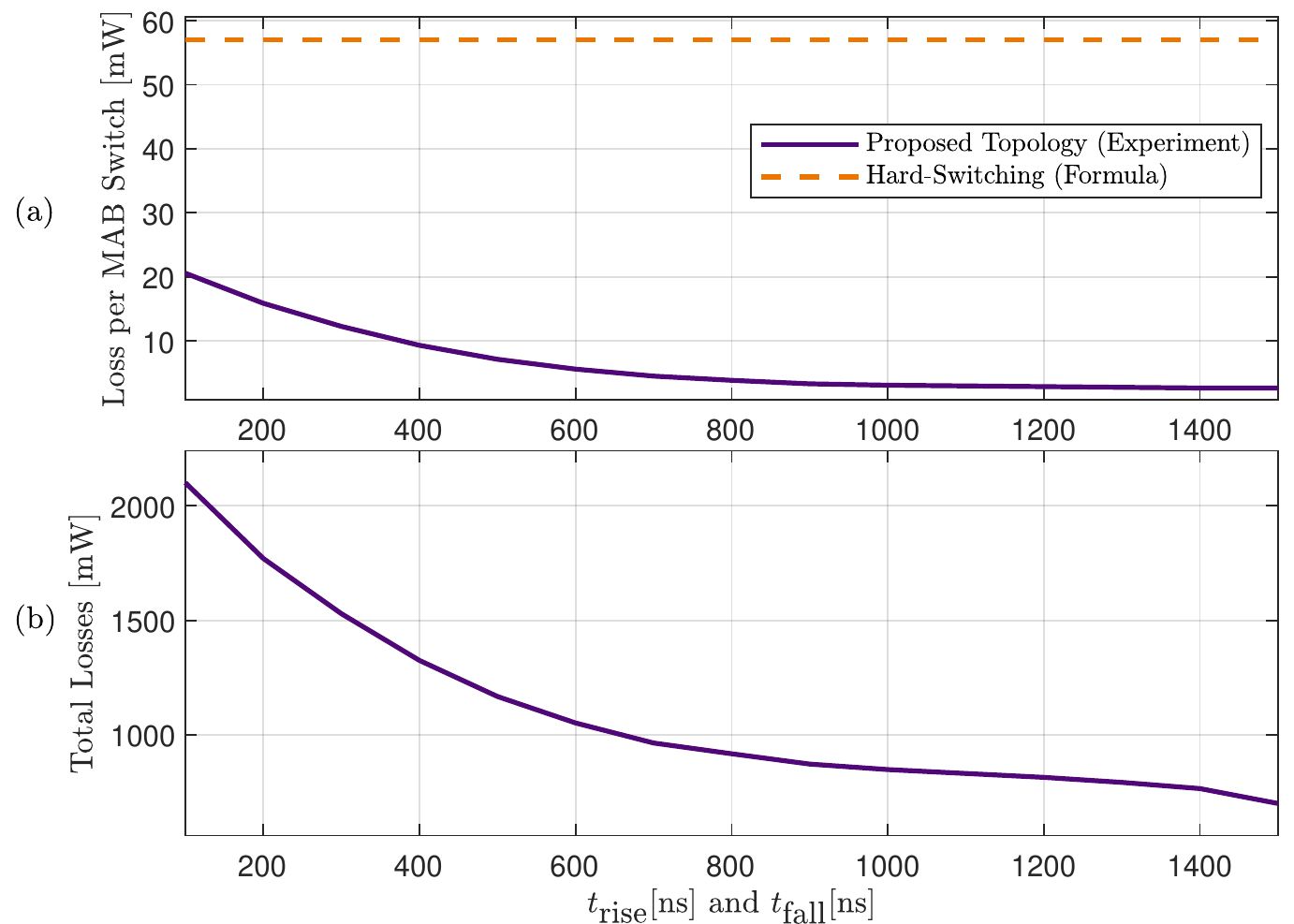}
    \caption{Experiment results of the power loss for different values of $t_{\textrm{rise}}$. The top figure (a) shows the energy loss of the main gate driver circuit per MAB MOSFET in mW. For reference, the power loss that would occur on a gate driver topology with hard switching is shown in orange. The lower subfigure (b) shows the overall losses of all $64$ MAB MOSFETs and the auxiliary power losses for DC/DC converters, diagnostics LEDs, ...
    For these experiments, the zero-time is $t_{\textrm{zero}} = 800$ ns and the switching frequency is $f_{\textrm{s}} = 125$ kHz.}
    \label{fig:GDE2}
\end{figure}
If the timings are not chosen correctly, the gate-source voltage waveform will be distorted. 
Choosing the rise/fall time too small will result in ripples as seen in Fig. \ref{fig:GDE1} which occur due to the gate-source voltage not yet reaching the desired value when the next switching state is applied. Choosing the timings too small will result in a curved waveform during the rise/ fall time. 
This is due to the sinusoidal nature of the voltage during that switching state which was derived in the previous section. 
Choosing a correct timing however, will result in only the approximately linear part of the waveform being present since the circuit switches to Step 1 or Step 6 before the peak of the sinusoidal is reached. 
Figure \ref{fig:GDE2} shows that the overall energy consumption of the gate driver circuit decreases until the optimal value is found and stays constant after that. 
\newline
\begin{figure}[!htbp]
    \centering
    \includegraphics[scale=0.65]{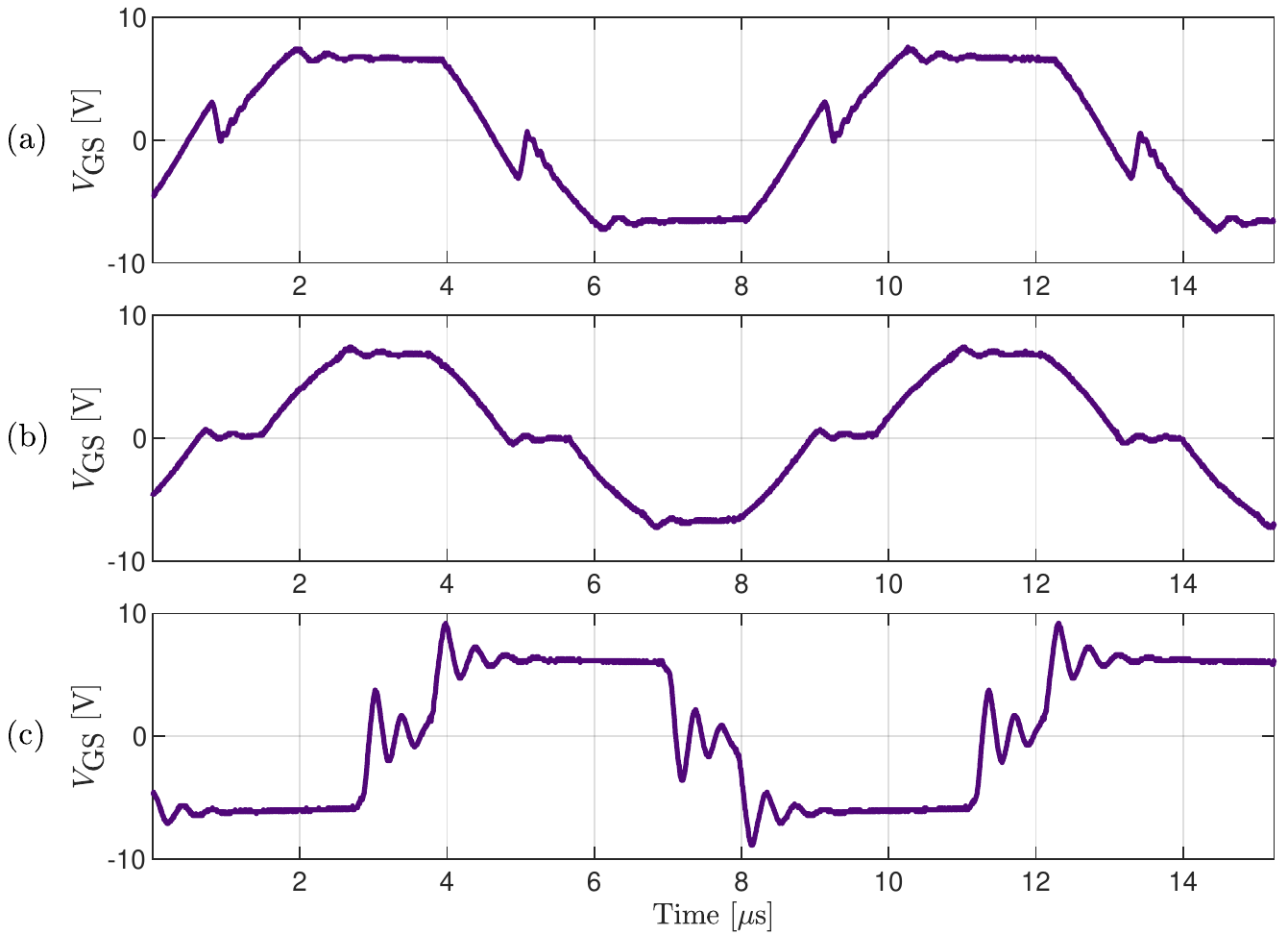}
    \caption{Experiment results of the effects of different timing selections on the gate driver waveform with respect to $t_{\textrm{zero}}$. 
    The top subfigure (a) shows the midpoint waveform for $t_{\textrm{zero}} = 100$ ns. A zero-time of $t_{\textrm{zero}} = 800$ ns can be seen in the middle subfigure (b) while the bottom subfigure (c) shows a waveform with $t_{\textrm{zero}} = 1500$ ns.
    A rise time of $t_{\textrm{rise}} = 1000$ ns and a switching frequency of  $f_{\textrm{s}} = 125$ kHz were selected during the experiments.}
    \label{fig:GDE3}
\end{figure}
\begin{figure}[!htbp]
    \centering
    \includegraphics[scale=0.6]{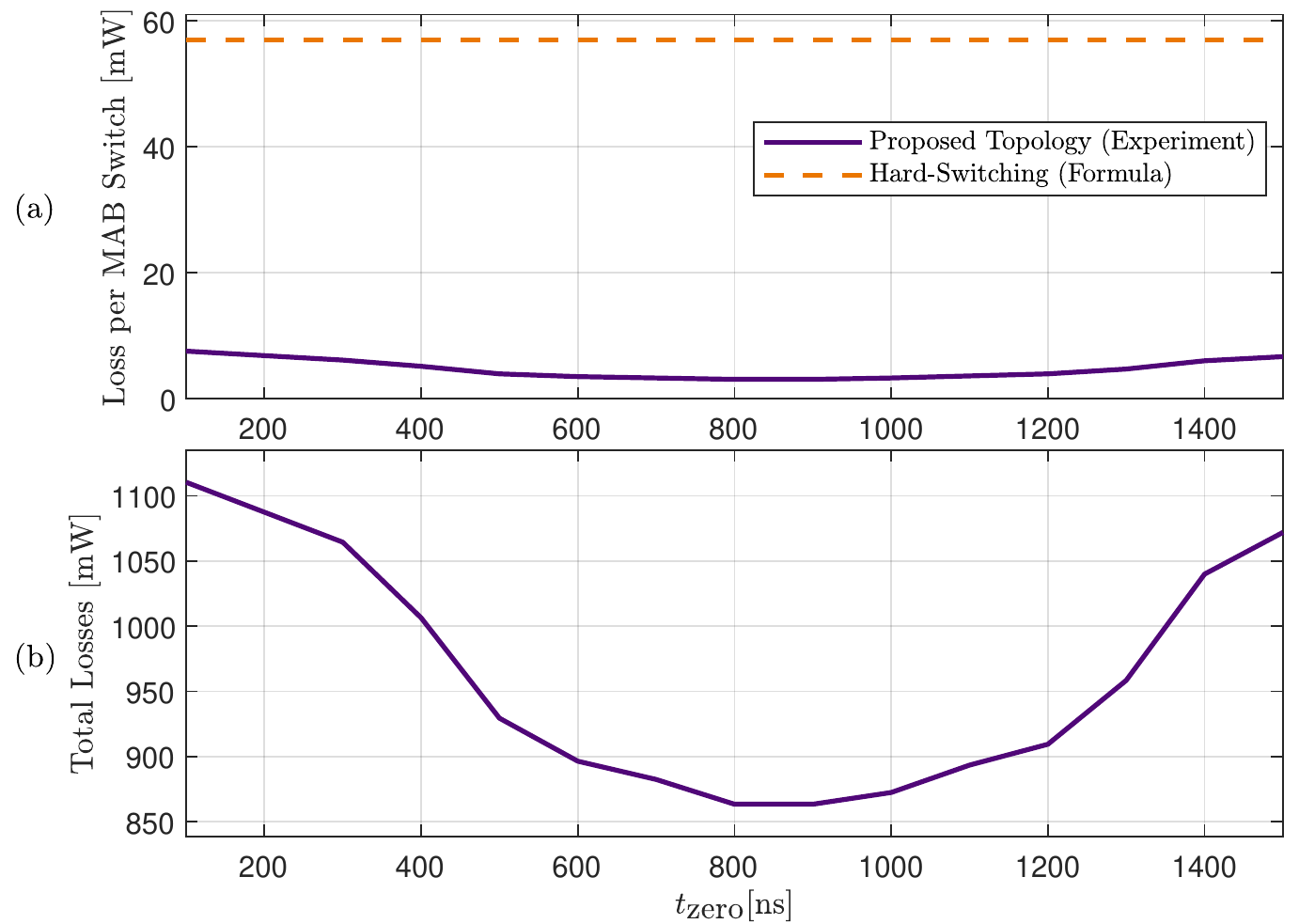}
    \caption{Experiment results of the power loss for different values of $t_{\textrm{zero}}$. The top figure (a) represents the power that is used by the primary active bridge to maintain operation, normalized for the total number of high-power MAB switches. 
    In addition, the power loss of a gate driver topology based on hard-switching is shown in orange.
    The bottom figure (b) depicts the total power losses, used by the MAB as well as the auxiliary supply for the gate driver circuit. 
    To obtain a fair comparison, the rise time of $t_{\textrm{rise}} = 1000$ ns and switching frequency of $f_{\textrm{s}} = 125$ kHz were used in all cases.
    }
    \label{fig:GDE4}
\end{figure}
A similar effect can be seen when the zero time is varied. 
Selecting a zero time too short will result in the drain-source capacitor not fully discharging which results in distorted waveforms (Fig. \ref{fig:GDE3}) and higher losses (Fig. \ref{fig:GDE4}). 
Similarly, choosing the zero time too large will result in an increase in losses as seen in Fig. \ref{fig:GDE4}, which can be explained by staying in the zero states while the energy is completely transferred and the energy might flow back to the drain-source capacitance. 
In between of those effects, an optimal value can be obtained as shown in Fig. \ref{fig:GDE4}.
\newline

\begin{figure}[!htbp]
    \centering
    \includegraphics[scale=0.6]{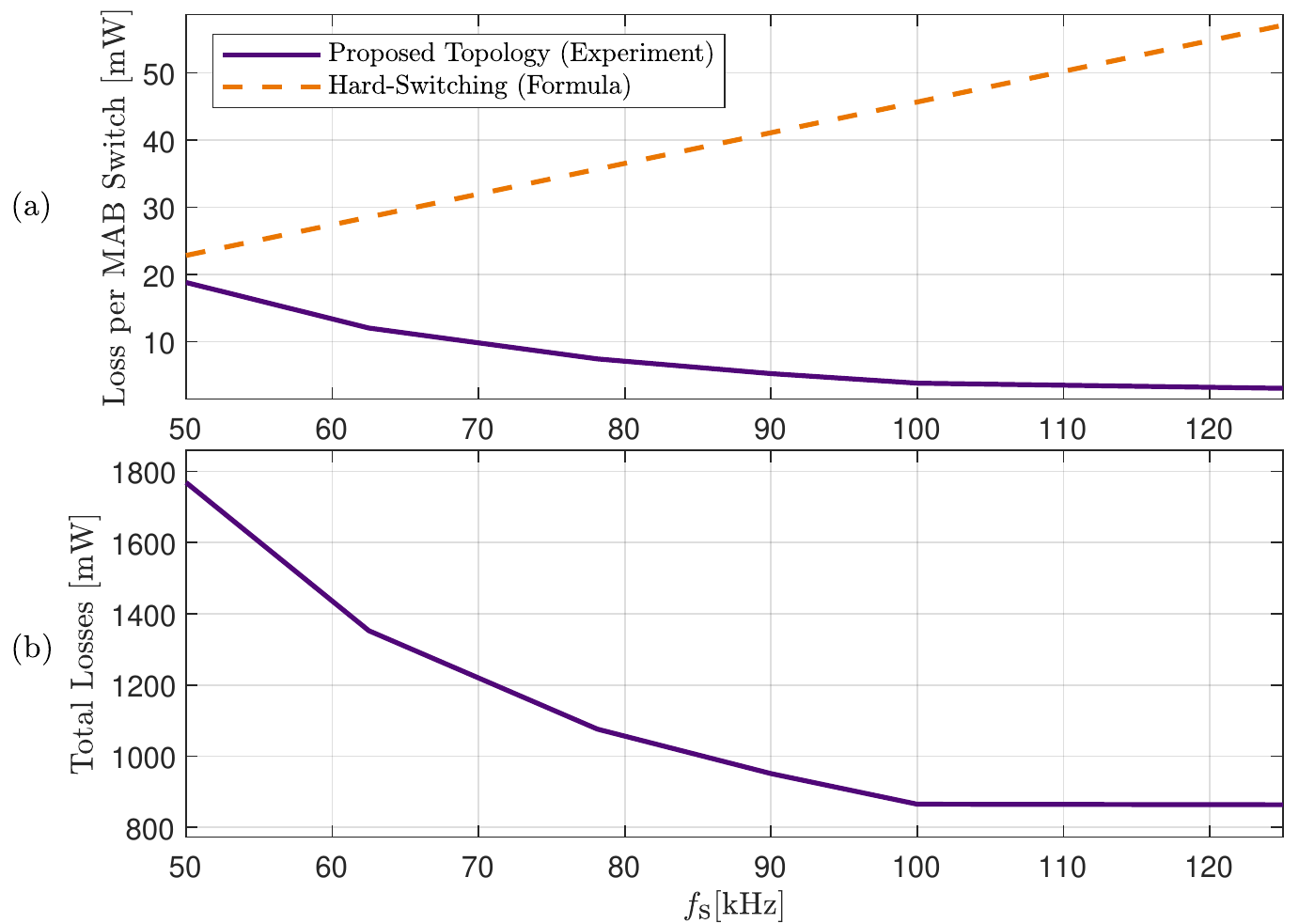}
    \caption{Experiment results of the power losses of the gate driver over different switching frequencies. At each frequency, the timings are tuned to optimal values. 
    An overview of the power losses of a hard-switching-based topology on each switch is shown in the orange curve.
    The top subfigure (a) shows the development of losses over one of the $64$ MAB switches while the bottom subfigure (b) shows the development of the sum of all losses of the MAB as well as the gate driver circuit itself.}
    \label{fig:GDE5}
\end{figure}

Varying the switching frequency of the MAB frequency will decrease the power consumption of the gate driver as seen in Fig. \ref{fig:GDE5}.
However, choosing a frequency too low will result in restrictions of the zero and rise timings which will again yield a decrease in performance.
\section{Conclusion}
This paper presented a novel gate driver model for MAB DC/DC converters. 
To maximize the efficiency of the MAB, a zero-voltage switching scheme was employed within the gate driver. 
Therefore, the topology utilizes a multi-winding transformer to exchange energy stored in the gate-source capacitances of the MOSFETs in the MAB. 
Similarly, the energy within the drain-source capacitance was exchanged with the parasitic inductances in the drain-source capacitance. 
The fundamental working principle was derived with a system model and verified with experiment results. 
The advantages of the topologies are the lower switching losses and the requirement for fewer components.
Furthermore, guidelines for the modulation of the gate driver are given. 
However, it is not possible to implement phase-shifts between the individual active bridges of the MAB which is the major limitation of the proposed topology. 


\ifCLASSOPTIONcaptionsoff
  \newpage
\fi

\bibliographystyle{IEEEtran}
\bibliography{bibtex/bib/References}

\begin{thebibliography}{10}
\providecommand{\url}[1]{#1}
\csname url@samestyle\endcsname
\providecommand{\newblock}{\relax}
\providecommand{\bibinfo}[2]{#2}
\providecommand{\BIBentrySTDinterwordspacing}{\spaceskip=0pt\relax}
\providecommand{\BIBentryALTinterwordstretchfactor}{4}
\providecommand{\BIBentryALTinterwordspacing}{\spaceskip=\fontdimen2\font plus
\BIBentryALTinterwordstretchfactor\fontdimen3\font minus
  \fontdimen4\font\relax}
\providecommand{\BIBforeignlanguage}[2]{{%
\expandafter\ifx\csname l@#1\endcsname\relax
\typeout{** WARNING: IEEEtran.bst: No hyphenation pattern has been}%
\typeout{** loaded for the language `#1'. Using the pattern for}%
\typeout{** the default language instead.}%
\else
\language=\csname l@#1\endcsname
\fi
#2}}
\providecommand{\BIBdecl}{\relax}
\BIBdecl

\bibitem{MABMIMO}
Y.-M. Chen, Y.-C. Liu, and F.-Y. Wu, ``Multi-input dc/dc converter based on the
  multiwinding transformer for renewable energy applications,'' \emph{IEEE
  Transactions on Industry Applications}, vol.~38, no.~4, pp. 1096--1104, 2002.

\bibitem{MABDCDC}
T.~Pereira, F.~Hoffmann, R.~Zhu, and M.~Liserre, ``A comprehensive assessment
  of multiwinding transformer-based dc–dc converters,'' \emph{IEEE
  Transactions on Power Electronics}, vol.~36, no.~9, pp. 10\,020--10\,036,
  2021.

\bibitem{MAB}
C.~Zhao, S.~D. Round, and J.~W. Kolar, ``An isolated three-port bidirectional
  dc-dc converter with decoupled power flow management,'' \emph{IEEE
  Transactions on Power Electronics}, vol.~23, no.~5, pp. 2443--2453, 2008.

\bibitem{MABC}
M.~Liserre, F.~Hoffman, and T.~Pereira, ``Multiwinding-transformer-based dc-dc
  converter solutions for charging stations [technology leaders],'' \emph{IEEE
  Electrification Magazine}, vol.~9, no.~2, pp. 5--9, 2021.

\bibitem{MAB2}
S.~Falcones, R.~Ayyanar, and X.~Mao, ``A dc–dc multiport-converter-based
  solid-state transformer integrating distributed generation and storage,''
  \emph{IEEE Transactions on Power Electronics}, vol.~28, no.~5, pp.
  2192--2203, 2013.

\bibitem{MABBB}
D.-J. Park, S.-Y. Choi, R.-Y. Kim, and D.-S. Kim, ``A novel battery cell
  balancing circuit using an auxiliary circuit for fast equalization,'' in
  \emph{IECON 2014 - 40th Annual Conference of the IEEE Industrial Electronics
  Society}, 2014, pp. 2933--2938.

\bibitem{MABHRE}
M.~A. Rahman, M.~R. Islam, K.~M. Muttaqi, and D.~Sutanto, ``Modeling and design
  of a multiport magnetic bus-based novel wind-wave hybrid ocean energy
  technology,'' \emph{IEEE Transactions on Industry Applications}, pp. 1--1,
  2021.

\bibitem{MABMAB}
F.~Grimm, J.~Wood, and M.~Baghdadi, ``A dc-autotransformer based multilevel
  inverter for automotive applications,'' \emph{arXiv preprint
  arXiv:2011.12164}, 2020.

\bibitem{MABER}
L.~Yin, X.~Weng, K.~Zhang, Z.~Zhao, L.~Yuan, and S.~Yi, ``A new topology of
  energy router with multiple hvac ports for power distribution networks,'' in
  \emph{2016 19th International Conference on Electrical Machines and Systems
  (ICEMS)}, 2016, pp. 1--5.

\bibitem{GD_old_NISO2}
{Jintae Kim}, {Gwanbon Koo}, and {Chung Yuen Won}, ``Loss analysis design of
  charge-pumped voltage supply for floating gate driver circuits in battery
  management system,'' in \emph{2016 IEEE Transportation Electrification
  Conference and Expo, Asia-Pacific (ITEC Asia-Pacific)}, 2016, pp. 061--065.

\bibitem{GD_old_NISO4}
A.~{Soldati}, E.~{Imamovic}, and C.~{Concari}, ``Bidirectional bootstrapped
  gate driver for high-density sic-based automotive dc/dc converters,''
  \emph{IEEE Journal of Emerging and Selected Topics in Power Electronics},
  vol.~8, no.~1, pp. 475--485, 2020.

\bibitem{MMCGD2}
N.~Rouger, Y.~Barazi, M.~Cousineau, and F.~Richardeau, ``Modular multilevel
  soi-cmos active gate driver architecture for sic mosfets,'' in \emph{2020
  32nd International Symposium on Power Semiconductor Devices and ICs (ISPSD)},
  2020, pp. 278--281.

\bibitem{GD_old_NISO1}
D.~{Maksimovic}, ``A mos gate drive with resonant transitions,'' in \emph{PESC
  '91 Record 22nd Annual IEEE Power Electronics Specialists Conference}, 1991,
  pp. 527--532.

\bibitem{GD_old_NISO3}
H.~{Jedi}, T.~{Salvatierra}, A.~{Ayachit}, and M.~K. {Kazimierczuk},
  ``High-frequency single-switch zvs gate driver based on a class $\phi _2$
  resonant inverter,'' \emph{IEEE Transactions on Industrial Electronics},
  vol.~67, no.~6, pp. 4527--4535, 2020.

\bibitem{GD_old_ISO1}
S.~H. {Weinberg}, ``A novel lossless resonant mosfet driver,'' in \emph{PESC
  '92 Record. 23rd Annual IEEE Power Electronics Specialists Conference}, 1992,
  pp. 1003--1010 vol.2.

\bibitem{GD_old_ISO2}
Z.~{Zhang}, F.~{Li}, and Y.~{Liu}, ``A high-frequency dual-channel isolated
  resonant gate driver with low gate drive loss for zvs full-bridge
  converters,'' \emph{IEEE Transactions on Power Electronics}, vol.~29, no.~6,
  pp. 3077--3090, 2014.

\bibitem{GD1}
D.~Rothmund, D.~Bortis, and J.~W. Kolar, ``Highly compact isolated gate driver
  with ultrafast overcurrent protection for 10 kv sic mosfets,'' \emph{CPSS
  Transactions on Power Electronics and Applications}, vol.~3, no.~4, pp.
  278--291, 2018.

\bibitem{GDT2}
M.~Takasaki, Y.~Miura, and T.~Ise, ``Wireless power transfer system for gate
  power supplies of modular multilevel converters,'' in \emph{2016 IEEE 8th
  International Power Electronics and Motion Control Conference (IPEMC-ECCE
  Asia)}, 2016, pp. 3183--3190.

\bibitem{MMCGD1}
V.-S. Nguyen, P.~Lefranc, and J.-C. Crebier, ``Gate driver supply architectures
  for common mode conducted emi reduction in series connection of multiple
  power devices,'' \emph{IEEE Transactions on Power Electronics}, vol.~33,
  no.~12, pp. 10\,265--10\,276, 2018.

\bibitem{MABT}
U.~Khalid, M.~M. Khan, Z.~Xiang, and Y.~Jianyang, ``Bidirectional modular dual
  active bridge (dab) converter using multi-limb-core transformer with
  symmetrical lc series resonant tank based on cascaded converters in solid
  state transformer (sst),'' in \emph{2017 China International Electrical and
  Energy Conference (CIEEC)}, 2017, pp. 627--632.

\bibitem{GD_old_ISOP6}
J.~{Gottschlich}, M.~{Schäfer}, M.~{Neubert}, and R.~W. {De Doncker}, ``A
  galvanically isolated gate driver with low coupling capacitance for medium
  voltage sic mosfets,'' in \emph{2016 18th European Conference on Power
  Electronics and Applications (EPE'16 ECCE Europe)}, 2016, pp. 1--8.

\bibitem{GD_old_ISOP1}
L.~{Zhang}, S.~{Ji}, S.~{Gu}, X.~{Huang}, J.~{Palmer}, W.~{Giewont}, F.~{Wang},
  and L.~M. {Tolbert}, ``Design considerations of high-voltage-insulated gate
  drive power supply for 10 kv sic mosfet in medium-voltage application,'' in
  \emph{2019 IEEE Applied Power Electronics Conference and Exposition (APEC)},
  2019, pp. 425--430.

\bibitem{GD_old_ISOP2}
A.~{Anurag}, S.~{Acharya}, Y.~{Prabowo}, G.~{Gohil}, and S.~{Bhattacharya},
  ``Design considerations and development of an innovative gate driver for
  medium-voltage power devices with high $dv/dt$,'' \emph{IEEE Transactions on
  Power Electronics}, vol.~34, no.~6, pp. 5256--5267, 2019.

\bibitem{GD_old_ISOP4}
A.~{Christe}, M.~{Petkovic}, I.~{Polanco}, M.~{Utvic}, and D.~{Dujic},
  ``Auxiliary submodule power supply for a medium voltage modular multilevel
  converter,'' \emph{CPSS Transactions on Power Electronics and Applications},
  vol.~4, no.~3, pp. 204--218, 2019.

\bibitem{GD_old_ISOP3}
S.~{Fuchs} and J.~{Biela}, ``Output voltage stability of series connected
  transformers for isolated auxiliary supplies in modular medium voltage
  converter systems,'' in \emph{2018 20th European Conference on Power
  Electronics and Applications (EPE'18 ECCE Europe)}, 2018, pp. P.1--P.9.

\bibitem{MABSPECIAL1}
S.~Wei, Z.~Zhao, L.~Yuan, W.~Wen, and K.~Chen, ``Voltage oscillation
  suppression for the high-frequency bus in modular-multiactive-bridge
  converter,'' \emph{IEEE Transactions on Power Electronics}, vol.~36, no.~9,
  pp. 9737--9742, 2021.

\bibitem{MABSPECIAL2}
S.~K. Dam and V.~John, ``A multi-active-half-bridge converter based
  soft-switched fast voltage equalizer for multi-cell to multi-cell charge
  transfer,'' in \emph{2019 IEEE Transportation Electrification Conference
  (ITEC-India)}, 2019, pp. 1--6.

\bibitem{MABSPECIAL3}
P.~Zumel, C.~Fernandez, A.~Lazaro, M.~Sanz, and A.~Barrado, ``Overall analysis
  of a modular multi active bridge converter,'' in \emph{2014 IEEE 15th
  Workshop on Control and Modeling for Power Electronics (COMPEL)}, 2014, pp.
  1--9.

\end{thebibliography}

\end{document}